\documentclass[a4paper,11pt]{article}
\pdfoutput=1 

\usepackage{jheppub} 
                     
\usepackage[T1]{fontenc} 
\usepackage[utf8]{inputenc}
\usepackage{amsfonts}
\usepackage{amsmath}
\usepackage{changepage}

\DeclareMathOperator{\tr}{tr}
\newenvironment{widerequation}{%
    \begin{adjustwidth}{-1cm}{-1cm}\begin{eqnarray}}
    {\end{eqnarray}\end{adjustwidth}}

\usepackage[export]{adjustbox}
\usepackage{subcaption}

\title{Resurgence in 2-dimensional Yang-Mills and a genus altering deformation}

\author[a]{Toshiaki Fujimori,}
\author[a,b,c]{Philip Glass}

\affiliation[a]{Department of Physics, and Research and Education Center for Natural Sciences,
Keio University, 4-1-1 Hiyoshi, Yokohama, Kanagawa 223-8521, Japan}
\affiliation[b]{Department of Mathematical Sciences, Durham University,\\Lower Mountjoy, Stockton Road, Durham, DH1 3LE, UK}
\affiliation[c]{International Research Fellow of Japan Society for the Promotion of Science (Postdoctoral
Fellowships for Research in Japan (Standard))}

\emailAdd{toshiaki.fujimori018@gmail.com}
\emailAdd{philip.glass@outlook.com}

\abstract{We study resurgence in the context of the partition function of 2-dimensional $SU(N)$ and $U(N)$ Yang-Mills theory on a surface of genus $h$. After discussing the properties of the transseries in the undeformed theory, we add a term to the action to deform the theory. The partition function can still be calculated exactly,  and the deformation has the effect of analytically continuing the effective genus parameter in the exact answer to be non-integer. In the deformed theory we find new saddle solutions and study their properties. In this context each saddle contributes an asymptotic series to the transseries which can be analysed using Borel-\`Ecalle resummation. For specific values of the deformation parameter we find Cheshire cat points where the asymptotic series in the transseries truncate to a few terms. We also find new partial differential equations satisfied by the partition function, and a number of applications of these are explained, including low-order/low-order resurgence.}

\begin{document} 
\maketitle
\flushbottom

\section{Introduction}\label{sec:intro}

\`Ecalle's resurgence theory \cite{Ecalle:1981} (see \cite{Dorigoni:2014hea,Aniceto:2018bis,marino_2015www} for nice introductions) is rapidly becoming a standard tool in the toolbox of quantum field theorists. The theory is typically used to make sense of the asymptotic series, with zero radius of convergence, so prevalent in the field. The weakest version of the program (see discussion on p45 of \cite{DiPietro:2021yxb}, also \cite{Marino:2021dzn,Marino:2022ykm,Bajnok:2021zjm,Reis:2022tni}) aims to show that all observables in quantum field theory can be written as ambiguity-free Borel-\`Ecalle resummations of transseries. In this setting a transseries is an object which takes, for example in a theory with weak coupling $g$, the heuristic form
\begin{eqnarray}\label{eq:heuristicTransseries}
\mathcal{O}(g)=\sum\limits_i \sum\limits_j \sigma_i e^{-S_i/g}c_{i,j}g^j\;.
\end{eqnarray}
The series contains a sum of perturbative series $c_{i,j}g^j$, each dressed with a non-perturbative part $e^{-S_i/g}$ with $S_i$ some appropriate constant\footnote{In many examples $S_i$ is the action of a saddle in the semi-classical decomposition, i.e. a finite-action solution to the equations of motion. It remains an open question whether this is always the case.}, and transseries parameter or Stokes constant $\sigma_i$. Note that transseries can be much more complicated than this, containing for example logarithms and multiple couplings \cite{Edgar:2008ga}. For this version of the program there is a large and growing volume of evidence that supports that this is indeed true\footnote{A complete list of references in support of this would be far too long for the present work. The reader is pointed to the comprehensive (but now slightly out of date) bibliography of \cite{Aniceto:2018bis} for a starting point, though many more recent works exist.}.

A stronger version of the program aims to show that the full non-perturbative transseries, up to the values of the transseries parameters can be derived from the perturbative data alone. (There are some caveats to this; for example the contributions to a transseries in a different topological sector are not normally included, which we will address in a moment.) Many positive examples of this exist, but there are also examples that appear to rule this option out (see for example the supersymmetric models discussed in \cite{Pestun:2007rz,Aniceto:2014hoa,Russo:2012kj,Honda:2016mvg,Hama:2012bg,Benini:2012ui,Doroud:2012xw}, and the integrable models discussed in \cite{DiPietro:2021yxb,Marino:2021dzn,Marino:2022ykm,Bajnok:2021zjm,Reis:2022tni}). It is another such supposed counter example of the stronger version of the resurgence program, that being 2-dimensional Yang-Mills theory (2d YM), with which this work in concerned. In this paper we will focus on the resurgence part of the story, and in \cite{Fujimori:2022pld} we will study the Picard-Lefschetz counterpart of the same story.

The Cheshire cat resurgence method, first developed in \cite{Kozcaz:2016wvy}, has in recent years taken a number of theories thought to be counter examples of the stronger version of the resurgence program, and shown them to in fact be more cases where stronger version applies \cite{Dorigoni:2017smz,Dorigoni:2019kux,Dorigoni:2019yoq,Dorigoni:2020oon,Dorigoni:2022bcx}. The work of \cite{Kozcaz:2016wvy} has also recently been made mathematically rigorous in \cite{Kamata:2021jrs} using methods from Exact WKB. In all these cases, the Borel-\`Ecalle resummation procedure applied to the perturbative data alone does not appear to reproduce the full transseries. However, in all these cases, a very slight deformation away from the theory in question renders deformed perturbative data from which the full transseries can be derived, up to the values of the transseries parameters, for the deformed case. At the end of the analysis one can return the deformation to zero, and one is left with the full transseries in the undeformed case.

In the cases studied so far, what is happening is as follows. In the space of possible theories, there exist very special points where there are sufficient cancellations between different contributions to the perturbative data (the bosonic and fermionic contributions in the cases studied so far) that the perturbative series is no longer asymptotic, and a resurgence analysis cannot be used to derive all the non-perturbative data. We call these points Cheshire cat points. However, these points appear to be isolated, and after a small deformation away from them the full transseries can be derived from the perturbative data. It is important to emphasise that the authors are by no means claiming that this is the case for every theory that appears to be a counter example to the stronger version of the resurgence program. Simply, it is interesting to note that this is the case in a growing number of examples, and it is a line of research worth pursuing to see if this is the case in more (or all) theories where it appears at first sight that the full transseries cannot be derived from the perturbative data alone.

To date, the Cheshire cat resurgence phenomenon has only been observed\footnote{At least in physical theories; see \cite{Dorigoni:2019yoq,Dorigoni:2020oon,Dorigoni:2022bcx} for examples in various mathematical functions and series.} in supersymmetric quantum field theories, and supersymmetric or quasi-exact-solvable quantum mechanical systems. In these cases cancellations between bosonic and fermionic contributions are the reason a resurgence analysis of the perturbative data doesn't render the full transseries. For these theories, the Cheshire cat deformation analytically continues the number of bosons, fermions, or both\footnote{Actually, a parameter in an effective theory that appears to describe the boson or fermion number.}, to be slightly non-integer (see \cite{Kozcaz:2016wvy,Dorigoni:2017smz,Dorigoni:2019kux} for details).

In this paper we will discuss a non supersymmetric theory, which as we will see also proves to display the Cheshire cat resurgence phenomenon\footnote{The story is actually slightly more subtle than this. In order to gauge fix, typically ghosts are added with fermion statistics, and this gives the theory BRST symmetry, a supersymmetry. In the 2-dimensional case this can be used to calculate the partition function exactly. However, it must be recalled that these ghosts are not physical, i.e. the physical theory does not actually have this symmetry.}. 2d YM has been studied extensively in various contexts, for example as a solvable QFT providing a toy model for Yang-Mills in higher dimensions (e.g. \cite{Migdal_4090770}), and as a string theory (e.g. \cite{Gross:1992tu,Gross:1993hu}). Many excellent reviews exist, see for example \cite{Blau:1993hj,Cordes:1994fc}, and references therein for a history of the subject.

In fact, in the large $N$ case, the perturbative data is asymptotic with zero radius of convergence. This has been studied in \cite{Okuyama:2018clk,Okuyama:2019rqn}, and in \cite{Okuyama:2018clk} a resurgence analysis was performed in the large $N$ case. The authors studied the theory on a torus, finding a non-Borel summable asymptotic perturbative series from which they could compute the 1-instanton contribution. In this work we will be exclusively interested in small $N$. 

The model can be solved exactly \cite{Migdal_4090770,Rusakov1990MPLA,Fine:1990zz,Witten:1991we,Blau:1991mp}, making a full investigation into its resurgence properties possible. We will mostly consider the partition function in this work, which can be written in various ways. In particular it can be written as a weak or strong coupling transseries. In the weak coupling case the exponentially suppressed terms correspond to semi-classical saddle points. For the case of finite $N$ the perturbative expansion is truncating, both in the strong and weak coupling cases, rendering a resurgence analysis seemingly useless. However, as in the previous cases cited, it turns out that a small deformation of the theory will uncover an asymptotic series in each sector of the transseries on which a resurgent analysis can be performed.

In both the works \cite{Dorigoni:2017smz} and \cite{Dorigoni:2019kux} where Cheshire cat resurgence has been studied in a QFT context, the deformation was applied to an effective description of the partition function after localisation had been performed. To put it another way, a genuine path integral deformation of the UV theory was not found. In the present case, we have been able to find a genuine deformation of the UV theory, i.e. a term we can add to the Lagrangian of the theory which uncovers the Cheshire cat structure. This deformation still allows us to calculate the partition function (and other observables) exactly. In the effective description of the theory, the deformation appears as a deformation of the genus $h$ of the manifold on which the theory lives (i.e. the base space) to be non-integer. The deformation is somewhat similar to that of \cite{Kozcaz:2016wvy}, namely it is a quantum deformation that appears normally upon integrating out certain fields in the theory.

Unlike the previous Cheshire cat examples studied, in 2d YM, we not only have access to the weak coupling representation of the partition function, but also the strong coupling representation. As we will see, the procedure described above will allow us to find a divergent asymptotic transseries in both the weak and strong coupling cases, and from the perturbative data alone (within each topological sector) we will be able to find all the non-perturbative data (within each topological sector), up to the exact value of the transseries parameter. But having access to both a strong and weak coupling transseries description will actually allow us to go even further; by demanding that the strong and weak transseries representations describe the same object, we will be able to determine the transseries parameters exactly, which is not normally possible.

\begin{figure}[ht]
\centering
\includegraphics[width=0.8\textwidth]{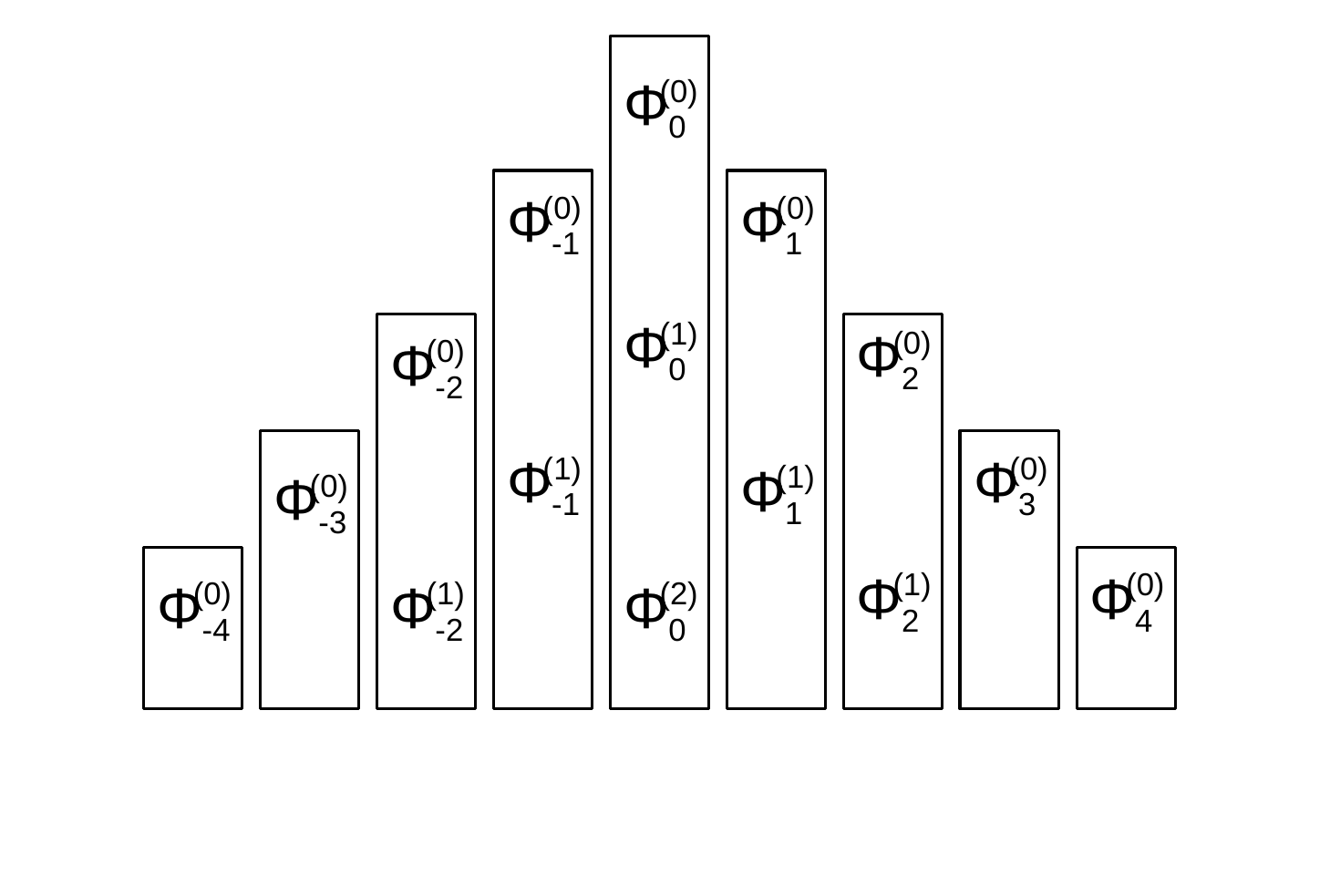} 
\caption{The resurgence triangle. $\Phi^{(k)}_n$ is the contribution from the $k^{th}$ part of the $n^{th}$ topological sector. Using resurgence we can often calculate all of the contents of a column from one entry of that column, but not the contents of any of the other columns. Typically what is observed is that $\Phi^{(0)}_1$ would be the contribution from say an instanton, $\Phi^{(0)}_{-1}$ an anti-instanton, and $\Phi^{(1)}_{0}$ an instanton-anti-instanton. Thus the triangle shape; as we move down the vertical axis the value of the constant $S_i$ (in this example the action of the saddle) in the exponential part of the contribution to the transseries increases.}
\label{fig:resurgenceTriangle}
\end{figure}

As mentioned above, a normal caveat to the stronger version of the resurgence program is that terms in the transseries in a different topological sector can typically not be derived from the perturbative expansion alone, using only a resurgence analysis. Perhaps the best way to see this is through the connection between \`Ecalle resurgence and Picard-Lefschetz theory (see \cite{Dunne:2015eaa} for a nice introduction). In order for us to be able to use Borel-\`Ecalle resummation to generate data for one transseries contribution from another, there must be a Lefschetz thimble connecting the saddles responsible for those contributions (assuming the contributions have saddle explanations). For saddles with different topology, such a smooth thimble cannot occur. Thus non-perturbative contributions with different topology to the perturbative contribution cannot be uncovered using a resurgence analysis of the perturbative part. Thus, in such a theory, the terms in the transseries arrange themselves into what is known as the resurgence triangle. Contributions with the same topology are said to be in the same topological sector. An illustration of this is given in Figure \ref{fig:resurgenceTriangle}.

Determining whether saddles are topologically distinct in general is actually not a simple question. For example, there are cases where there is an emergent topological structure for certain parts of the parameter space, for example in \cite{Dunne:2012ae,Cherman:2014ofa}. As another example, when we are dealing with a gauge theory we need to consider not just the topology of the gauge group, but also the topology of a gauge slice of the fields. There are thus two lines of inquiry we will need to consider in this work. The first line is to determine is whether the truncation of the perturbative series in 2d YM is due to Cheshire cat phenomena, or topological grading of saddles. It turns out we will see examples of both.

The second line is whether there are additional structures that can be combined with resurgence to make it possible to calculate, from the perturbative data, the contributions to the transseries from a different topological sector. We refer to such a calculation as making a sideways step in the resurgence triangle. In \cite{Dunne:2014bca,Dunne:2013ada,Gahramanov:2015yxk} (see also \cite{Ccopa:2000qft,Alvarez:2000aaa,Alvarez:2004aaa}), such a relation was developed for certain quantum mechanical models, known as the Dunne-\"Unsal relation, to allow one to make a sideways step in the resurgence triangle. This is also referred to as low order/low order resurgence, as one can use these relations to calculate the low orders in perturbation theory around one saddle from the low order contributions to a different saddle. This was then further elaborated on in \cite{Dorigoni:2019kux} to explain similar such structures found in $\mathcal{N}=(2,2)$ theories on $S^2$, $\mathcal{N}=2$ theories on squashed $S^3$, and $\mathcal{N}=2$ theories on squashed $S^4$. In all these cases\footnote{The be precise, not including the case of $\mathcal{N}=2$ theories on squashed $S^4$, where a Cheshire cat resurgence analysis has yet to be performed.}, the additional structure allows one (when combined with a Cheshire cat resurgence analysis) to produce the full transseries, with all the non-perturbative data from all topological sectors, from the perturbative data alone.

In the case of 2d YM, when the gauge group is $U(N)$, one can add a topological theta angle to the theory. In this case the normal resurgence triangle structure appears. However it turns out that even without a topological theta angle there is still a non-trivial resurgence triangle structure, graded by monopole numbers. In the first part of this paper we will explore this structure, and the resurgence properties of the partition function within each topological sector, with and without a deformation. We will then consider what happens when we try and apply resurgence unaware of the topological grading of the saddles, having derived the transseries using a method other than saddle decomposition, with some important takeaways. In the latter part of this paper we will discuss various structures, in this case factorisation of the partition function, and various partial differential equations the partition functions satisfy, that can play the role of the above mentioned structures in 2d YM. We will thus be able to produce the full transseries with data from all topological sectors from the perturbative data alone. The partial differential equations will have other applications as well as allowing us to make a sideways step, and we will discuss some of these as well.

The structure of the remaining part of this paper is as follows. In Section \ref{sec:partition_function} we will discuss the exact solution of the partition function, our deformation of the theory, and the saddles of the theory including their topological properties. In Section \ref{sec:Cheshire_cat} we will then perform a resurgence analysis of the weak coupling transseries for gauge groups $U(2)$ and $SU(2)$, within a topological sector. In Section \ref{sec:strongAndConsistency} we will the derive the strong coupling transseries in the deformed case, and analyse its resurgence properties. As we will see, demanding consistency of the weak and strong transseries representations will in fact allow us to completely fix the transseries parameters. Section \ref{sec:WeakCouplingNoTop} will be devoted to studying what happens when we try to analyse the transseries without knowledge of the saddles and their topology. We first derive the perturbative series as an infinite sum of correlators. Without knowledge of the saddles the resurgence structure seems highly unusual, as we will see, but knowledge of the saddles and their topology clarifies what is happening.

In Section \ref{sec:diff_equations_topological_sectors} we will then turn our attention to studying partial differential equations satisfied by the partition function, and discuss various applications including moving sideways in the resurgence triangle. Finally, in Section \ref{sec:conclusion} we will wrap up with some conclusions and future directions. In Appendix \ref{sec:hGeq1} we give some more details about the cases where the genus $h\geq1$. In Appendix \ref{sec:ZagierMethod} we discuss an alternative way of deriving the transseries in the $SU(2)$ case using a method by Zagier\cite{ZagierAppendixTM}.

In the latter stages of completing this work \cite{Griguolo:2022hek} appeared on the arXiv. The authors consider resurgence in 2d YM with $T\Bar{T}$ deformation (not the deformation we have used), and find a number of results similar to our own, in particular the emergence of asymptotic series on which Borel-\`Ecalle resummation can be performed. It would be interesting to consider the relation between the results found here and those of \cite{Griguolo:2022hek}.

\section{2-dimensional Yang-Mills on genus \texorpdfstring{$h$}. surface}\label{sec:partition_function}

Here we begin by recalling some useful facts about 2d YM, its action, saddles and exact partition function. The second half of this section will be dedicated to describing our deformation of the theory, and seeing how various features of the theory are changed once we turn the deformation on. Much of this section follows the excellent reviews \cite{Blau:1993hj} and \cite{Cordes:1994fc}.

In this work we will be concerned with the theory defined on a compact surface (though we will focus mostly on closed surfaces) $\Sigma_h$, where $h$ denotes the genus. The theory will have gauge group $G$ (which for us will be $U(N)$ or $SU(N)$). Depending on the context there are two couplings used in the literature; the Yang-Mills coupling $g_{YM}$, and the string coupling $g$. These are related by
\begin{eqnarray}\label{eq:stringYMcouplingRelation}
g=A\;g_{YM}\;,
\end{eqnarray}
where $A$ is the area of $\Sigma_h$. We will stick with using the string coupling to avoid carrying around an extra factor of $A$ in all our equations (Alternatively of course one may think of this as working with the Yang-Mills coupling whilst setting the area to be 1). For a theory with gauge group $SU(N)$, we cannot include a theta angle term\footnote{Recall $\tr(F)=0$ for all elements of the Lie algebra of $SU(N)$.}, and the action is simply given by
\begin{eqnarray}\label{eq:YM_action_SUN}
S_{SU(N)}(g,h)&=&\frac{1}{2g}\int_{\Sigma_h}\tr(F^2)\;.
\end{eqnarray}
It will also be helpful to write the action as
\begin{eqnarray}\label{eq:YM_action_SUNScalarField}
S_{SU(N)}(g,h)&=&i\int_{\Sigma_h}\tr(\Phi\wedge F)+\frac{g}{2}\int_{\Sigma_h}\tr(\Phi^2)K\;.
\end{eqnarray}
Here $\Phi$ is an auxiliary scalar field taking values in the Lie algebra of $G$, and $K$ is the volume form.

When the gauge group is $U(N)$ we can add a theta angle term. We will work with the action
\begin{eqnarray}\label{eq:YM_action_UN}
S_{U(N)}(g,h,\theta)=\frac{1}{2g}\int_{\Sigma_h}\tr(F^2)+i\frac{\theta}{2\pi}\int_{\Sigma_h}\tr(F)\;.
\end{eqnarray}
However, for the sake of avoiding confusion, it is worth mentioning that there is another popular action found in the literature, given by
\begin{eqnarray}\label{eq:YM_action_TopoUN}
S_{U(N)}(g,h,\theta^\prime)=i\int_{\Sigma_h}\tr(\Phi\wedge F) +i\theta^\prime\int_{\Sigma_h}\tr(\Phi\wedge K)+\frac{g}{2}\int_{\Sigma_h}\tr(\Phi^2)K\;.
\end{eqnarray}
Here again $\Phi$ is a scalar field that can be integrated out. We have put a prime on $\theta^\prime$ to distinguish it from the $\theta$ in \eqref{eq:YM_action_UN}. When one integrates out $\Phi$ one finds the action
\begin{eqnarray}\label{eq:YM_action_TopoUN2}
S_{U(N)}(g,h,\theta^\prime)=\frac{1}{2g}\int_{\Sigma_h}\tr(F^2)+\frac{\theta^\prime}{g}\int_{\Sigma_h}\tr(F)+O\left((\theta^\prime)^2\right)\;,
\end{eqnarray}
where there is a constant term proportional to $(\theta^\prime)^2$. These actions, \eqref{eq:YM_action_UN} and \eqref{eq:YM_action_TopoUN2}, just differ by $\theta=\frac{2\pi\theta^\prime}{ig}$, and removing a constant term from the action. We will focus on \eqref{eq:YM_action_UN} here, which results in our expressions being slightly different from some of those commonly found elsewhere.

\subsection{Saddles points and topology}\label{sec:undeformedSaddles}

The theory possesses non-perturbative saddles, that is non-trivial finite-action solutions to the Euler-Lagrange equations of motion:
\begin{eqnarray}
d\ast F=0\;.
\end{eqnarray}
These solutions are monopoles, and are completely classified by their first Chern class, or monopole number. We will work in the torus gauge (see Section \ref{sec:localisationApproach}) in this paper, which will result in the relevant monopoles being those where $A_\mu$ lies in the Cartan subalgebra. In this case we have that the solutions are given by
\begin{eqnarray}\label{eq:monopoleSolutions}
dA^i=2\pi n_i K\;.
\end{eqnarray}
Here the index $i$ runs over the elements of the Cartan subalgebra. The action for these solutions for $SU(N)$ is given by
\begin{eqnarray}
\frac{1}{2g}\int_{\Sigma_h} \tr( F^2)=\frac{1}{2g}\sum\limits_{i=1}^{N-1} (2\pi n_i)^2\;.
\end{eqnarray}
For $U(N)$ the action is given by
\begin{eqnarray}
\frac{1}{2g}\int_{\Sigma_h}\tr(F^2)+i\frac{\theta}{2\pi}\int_{\Sigma_h}\tr(F)=\sum\limits_{i=1}^{N} \left(\frac{(2\pi n_i)^2
}{2g}+i\theta n_i\right)\;.
\end{eqnarray}

In order to consider if there will be Stokes phenomena between the expansions around each saddle we need to consider the topology of these solutions. For $SU(N)$ we may naively think that there will be resurgence between different saddles, as we cannot associate a theta angle to the monopoles. Likewise for $U(N)$ we would naively expect to find Stokes phenomena occurring between saddles with the same theta angle dependence.

This turns out not to be correct. Although we cannot associate a theta angle to the $SU(N)$ saddle configurations, we can associate a different topological quantity. We have already seen this. These are the monopole numbers. Likewise for saddle configurations in the $U(N)$ case, we can associate $N-1$ more monopole numbers than theta angles ($N$ monopole numbers in total), and thus there exists a much finer topological grading.

The reason for this is as follows. We actually need to consider the topology of a gauge slice of the fields
\begin{eqnarray}
\mathcal{F}: \Sigma_h\rightarrow (A_{\mu}^{gf},\Phi^{gf})\;.
\end{eqnarray}
Here $gf$ means gauged fixed. We will see explicitly shortly that we are dealing with Torus bundles. In this case the $\Phi$ fields are fixed to lie in the Cartan subalgebra, rank $(N-1)$ for $SU(N)$, and rank $N$ for $U(N)$. Torus bundles are completely classified by their first Chern class (i.e. monopole number). Thus we can associate one monopole number to each element of the Cartan subalgebra, and configurations with different sets of monopole numbers will be in different topological sectors.

\begin{figure}
     \centering
     \begin{subfigure}[b]{0.48\textwidth}
         \centering
         \includegraphics[width=\textwidth]{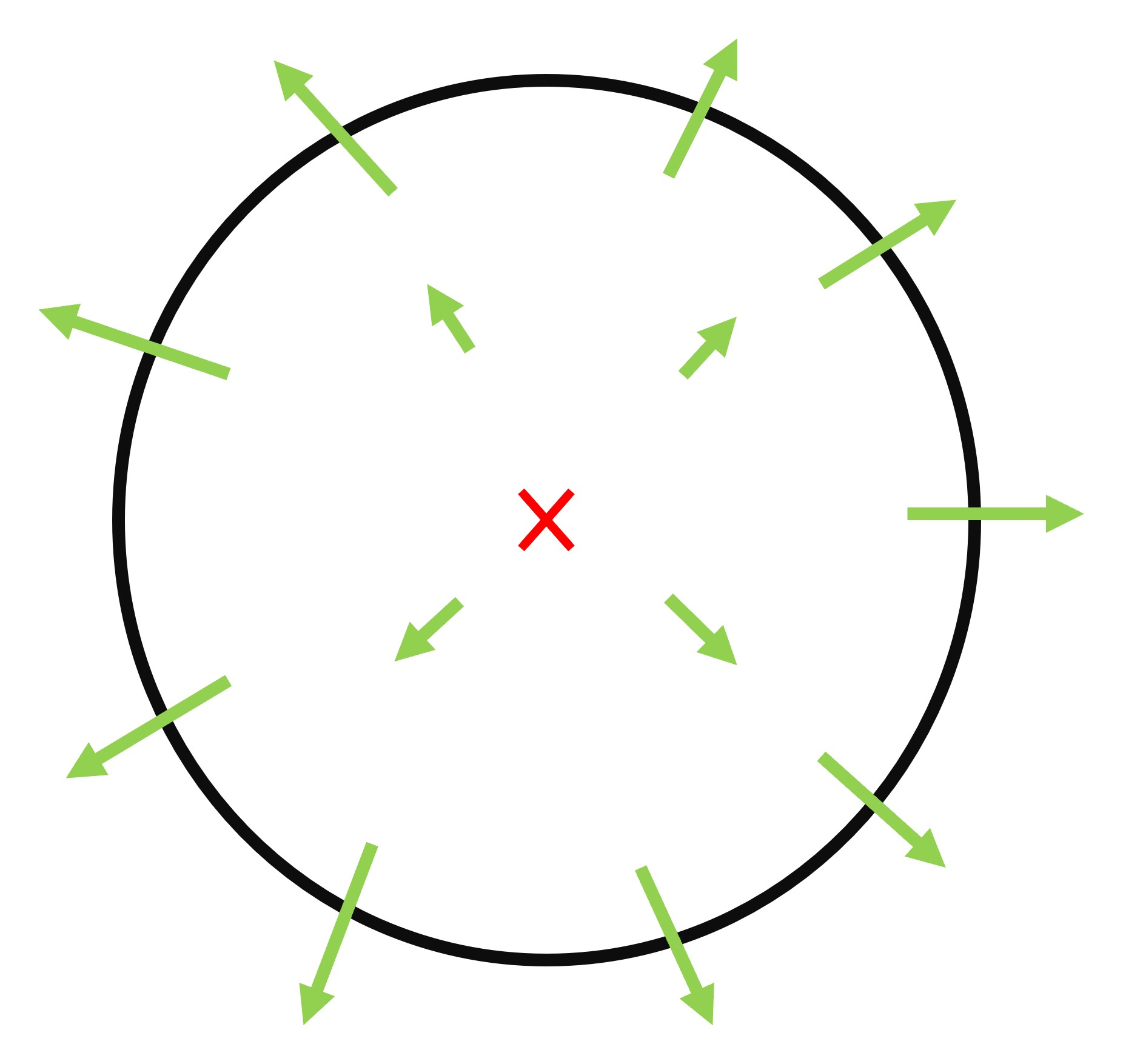}
         \caption{One Monopole. The center is at the center of the sphere when we embed it in 3-dimensions. The monopole is not a lump localized to a point on the surface of the sphere.}
         \label{fig:heuristicMonopoles1}
     \end{subfigure}
     \hfill
     \begin{subfigure}[b]{0.48\textwidth}
         \centering
         \includegraphics[width=\textwidth]{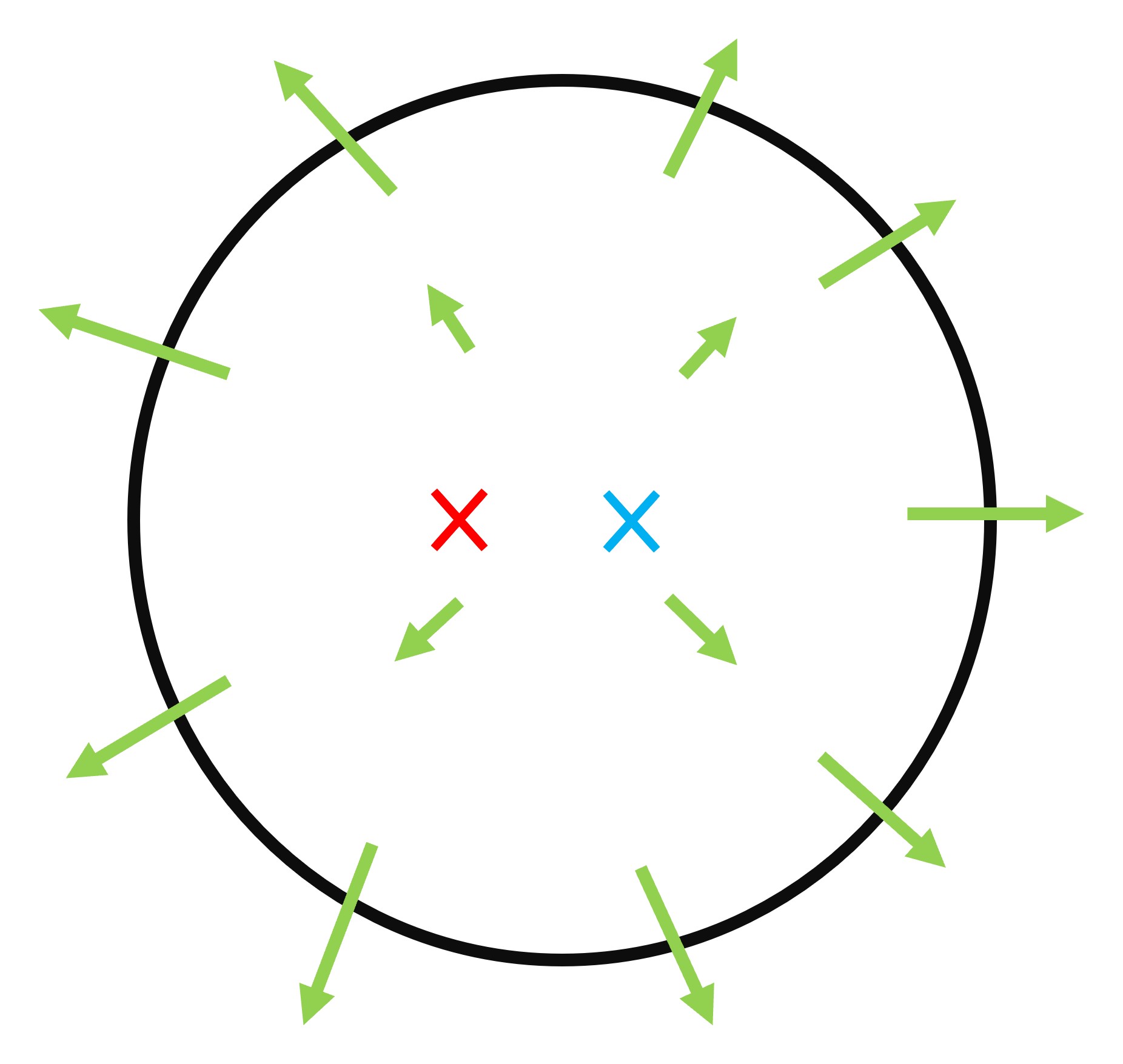}
         \caption{Two monopoles cannot be separated far apart as they are not localized lumps on the surface, and cannot be made small and distant from each other.}
         \label{fig:heuristicMonopoles2}
     \end{subfigure}
        \caption{Heuristic illustration of a monopole configurations when the base space is a sphere. The second plot shows that we can't separate two monopoles.}
        \label{fig:heuristicMonopoles}
\end{figure}

Before concluding that there shouldn't be Stokes phenomena occurring between saddles we need to consider whether we expect composite solutions. In cases where there is topological grading, Stokes phenomena may still occur because of the existence of composite saddles. For example in many theories there is no Stokes phenomenon between the perturbative saddle and an instanton saddle, but there is between the perturbative and instanton-anti-instanton saddles. We say the instanton-anti-instanton saddle is in the same topological sector as the perturbative saddle, as it has the same topology. These saddles may be ``saddle point at infinity'' (see \cite{Behtash:2018voa}), where the approximate saddle point constructed by putting an instanton and an anti-instanton far away from each other becomes exact when one takes the distance between them to infinity. However in many situations these become exact, finite-size saddles in the quantum theory (see for example \cite{Behtash:2015loa,Behtash:2015zha,Behtash:2015kva,Fujimori:2017osz,Behtash:2018voa}).

\begin{figure}[ht]
\centering
\includegraphics[width=0.8\textwidth]{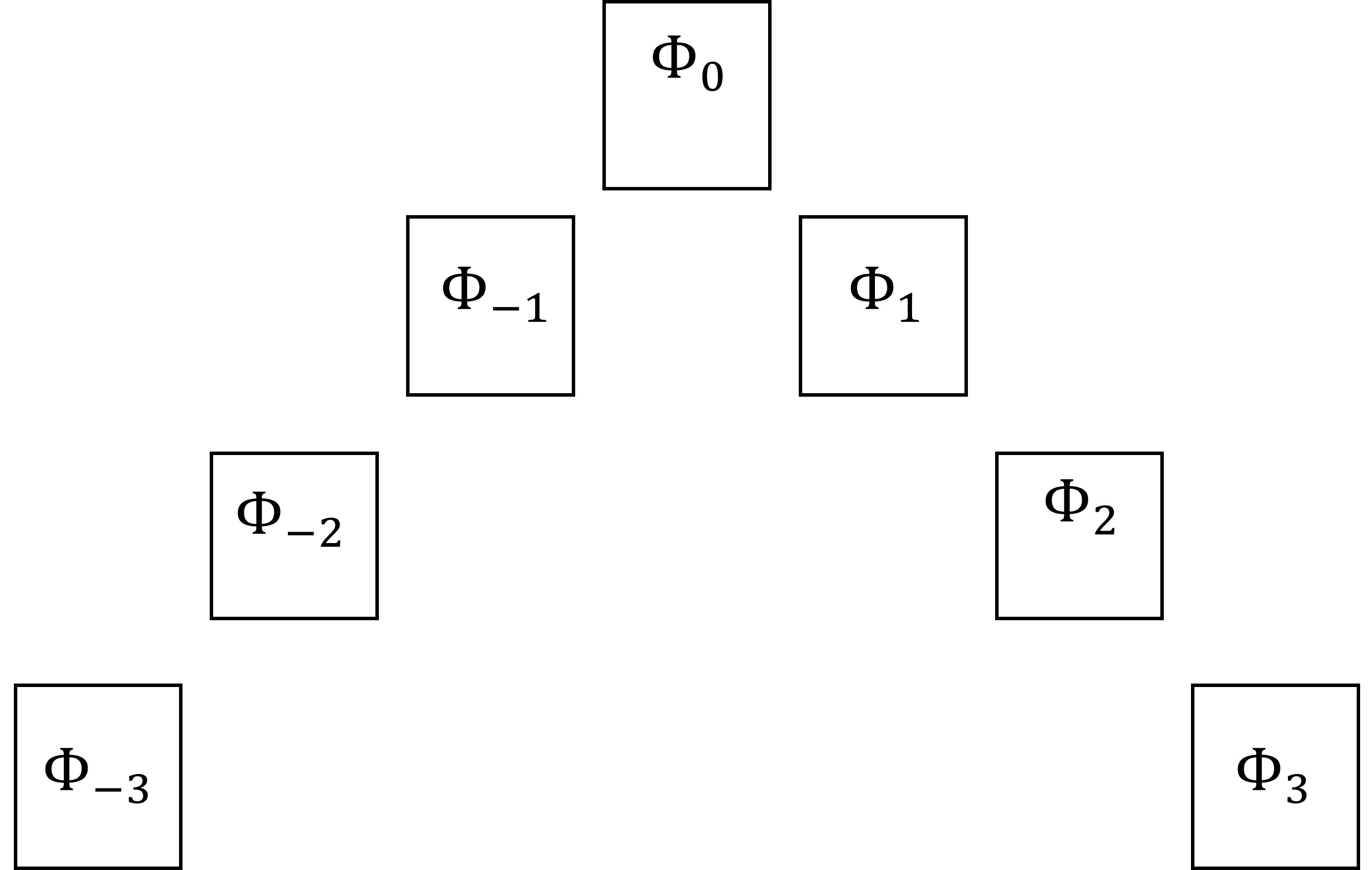} 
\caption{The resurgence triangle for the undeformed $SU(2)$ case. $\Phi_n$ is the contribution from saddle in the $n^{th}$ topological sector. Using resurgence we can often calculate all of the contents of a column from one entry of that column, but not the contents of any of the other columns. In this case this is trivial as there is only one entry in each column. As we move down the vertical axis the value of the action increases.}
\label{fig:resurgenceTriangleUndeformed}
\end{figure}

In the present case however we do not expect such solutions to exist. In order to construct such solutions we typically need the instantons to be localized to some point on the surface of the manifold. In this way we can take two solutions and put them far apart. Our monopoles are not localized to points on the base manifold though (see Figure \ref{fig:heuristicMonopoles}), and thus such a construction is impossible. Thus we do not expect to see any Stokes phenomena between any of the saddles in the theory.

In summary, in the undeformed theory, we have a Resurgence triangle structure as is shown in Figure \ref{fig:resurgenceTriangleUndeformed} for the undeformed $SU(2)$ case. Here we have precisely one saddle in each topological sector (so no column of saddles in each sector). The monopole number(s) parameterises which sector we are in. The triangle shape in Figure \ref{fig:resurgenceTriangleUndeformed} arises due to the action of the saddle in each sector increasing as we move away from the perturbative sector.

\subsection{Calculating the exact partition function}\label{sec:localisationApproach}

The exact partition function for 2d YM has been known for some time and can be calculated in various ways\cite{Migdal_4090770,Rusakov1990MPLA,Fine:1990zz,Witten:1991we,Blau:1991mp}. In this section we will give an outline of one method for calculating the partition function. We will follow here the methods of \cite{Blau:1993hj}. This involves a semi-classical expansion that turns out to be exact. For the reader who is already familiar with this calculation this section can be skipped. There are of course many ways of calculating the partition function for 2d YM theory. This method however allows us to motivate our deformation and make contact with work already conducted on Cheshire cat resurgence\cite{Kozcaz:2016wvy,Dorigoni:2017smz,Dorigoni:2019kux}.

Recall in the $SU(N)$ case the Lagrangian can be rewritten as
\begin{eqnarray}
S_{SU(N)}(g,h)&=&\frac{1}{2g}\int_{\Sigma_h}\tr(F^2)\nonumber\\
&=&i\int_{\Sigma_h}\tr(\Phi\wedge F)+\frac{g}{2}\int_{\Sigma_h}\tr(\Phi^2)K\;.
\end{eqnarray}
Here $K$ is the volume form and $\Phi$ is a scalar field, taking values in the Lie algebra, which can be integrated out to return us to the original Lagrangian.

In order to proceed we now need to gauge fix. We will follow the usual BRST procedure. First we write
\begin{eqnarray}
\Phi_tT^t=\Phi_i T^i+\Phi_\alpha T^\alpha\;,\;\;\;A_tT^t=A_i T^i+A_\alpha T^\alpha\;.
\end{eqnarray}
Here $t$ runs over the whole Lie algebra, $i$ runs over the Cartan sub-algebra, and $\alpha$ over the roots. To fix the gauge to be the torus gauge, we set the off diagonal elements of $\Phi$ to zero:
\begin{eqnarray}
\Phi_\alpha=0\;.
\end{eqnarray}
The situation is actually slightly more complicated than this as this cannot always be done globally. This subtlety will be taken care of for us by summing over monopole backgrounds shortly, but see \cite{Blau:1993hj} for a more detailed explanation of why this is correct. There is still a residual gauge symmetry to be fixed, which we'll return to in a moment. In this gauge the action is given by
\begin{eqnarray}
S_{SU(N)}(g,h)&=&\int_{\Sigma_h}\tr\left(i\Phi^t F^t+i\theta \Phi^t K-\frac{g}{2}\Phi^t\Phi^tK\right)\nonumber\\
&&\;\;\;+\int_{\Sigma_h} K\tr\left(b^\alpha\Phi^\alpha+\Bar{c}^\alpha[\Phi^t,c^\alpha]\right)\;.
\end{eqnarray}
This action is invariant under a BRST symmetry;
\begin{eqnarray}
&&\;\;\;\;Q\phi^\alpha=\alpha(\Phi)c^\alpha\;\;,\;\;\;Q c^\alpha=0\;,\nonumber\\
&&Q\phi^i=0\;\;,\;\;\;Q\Bar{c}^\alpha=b^\alpha\;\;,\;\;\;Qb^\alpha=0\;.
\end{eqnarray}
Here, and in what follows, $\alpha(\Phi)=\alpha(T^i)\Phi^i$. The $b^\alpha$ integral can then be performed, giving a delta function that sets the off-diagonal elements of $\Phi$ to be zero, leaving us with the action
\begin{eqnarray}
S_{SU(N)}(g,h)&=&\sum\limits_{i=1}^{N-1}\int_{\Sigma_h}\left(i\Phi^i dA^i+i\theta \Phi^i K-\frac{g}{2}\Phi^i\Phi^iK\right)\nonumber\\
&&\;\;\;+\sum\limits_{\alpha}\int_{\Sigma_h} \left(\alpha(\Phi)A^\alpha A^{-\alpha}+\alpha(\Phi)\Bar{c}^{-\alpha}c^\alpha K\right)\;.
\end{eqnarray}
With this action we will now find the path integral is quite simple to perform.

BRST symmetry is a supersymmetry, in the sense that it is a symmetry between bosonic and fermionic degrees of freedom. The difference between BRST symmetry and conventional supersymmetry is that here the fermions are ghosts, i.e. they are not physical. There are many cases, starting with the work of \cite{Pestun:2007rz}, where supersymmetry can be utilised via localisation to calculate the partition function exactly. In most settings BRST symmetry does not enable us to calculate observables exactly. This is because BRST-localisation reduces the path integral to integrals over gauge orbits, which is still typically an infinite dimensional path integral. It turns out however that fixing the gauge in 2d YM so drastically reduces the degrees of freedom of the fields that it leads to a finite dimensional integral expression for the partition function. Let us see how this works.

In order to proceed we expand the gauge field around solutions to the Yang-Mills equations of motion;
\begin{eqnarray}
dA^i=2\pi n_i K\;.
\end{eqnarray}
The quantum fluctuations around these solutions (write $A^i=A_c^i+A_q^i$, $c$ for classical and $q$ for quantum) require further gauge fixing, so we demand they satisfy
\begin{eqnarray}
d\ast A_q^i=0\;,
\end{eqnarray}
which is the Landau gauge. At this point we need to add more ghosts to fix the Landau gauge, and the relevant part of the action is then given by
\begin{eqnarray}
\sum\limits_{i=1}^{N-1}\int_{\Sigma_h}2\pi n_i\Phi^i K+\Phi^i dA_q^i+b^id\ast A_q^i+K\Bar{c}^id\ast dc^i\;.
\end{eqnarray}
The integral over $A^i_q$ returns the constraint
\begin{eqnarray}
d\Phi^i+\ast db^i=0\;\;\implies\;\;d\Phi^i=db^i=0\;.
\end{eqnarray}
(To see how the RHS follows from the LHS, square the LHS and integrate.) Thus we now only need to integrate over constant $\Phi^i$ modes. The integrals over $c^i$, $\Bar{c}^i$ and $b^i$ decouple and return an unimportant constant.

The integrals over $A^\alpha$ and $c^\alpha$ can be performed, giving a ratio of 1-loop determinants (this is just the Faddeev-Popov determinant). See \cite{Blau:1993hj} for the details of how this is calculated. The result is
\begin{eqnarray}\label{eq:1loopdetRatio}
\mathrm{det}_\mathbf{k}\left(\mathrm{ad}(\Phi^\mathbf{t})\right)^{\chi(\Sigma_h)/2}\;,\;\;\;\chi(\Sigma_h)=2-2h\;,\;\;\;&\mathrm{if}&\;\;\;\mathrm{det}_\mathbf{k}\left(\mathrm{ad}(\Phi^\mathbf{t})\right)\neq0\;,\nonumber\\
0\;\;\;\;\;\;\;\;\;\;\;\;\;\;\;\;\;\;\;\;\;\;\;\;\;\;\;\;\;\;\;\;\;\;\;\;\;\;\;\;\;\;\;\;\;\;\;\;\;\;\;\;\;\;&\mathrm{if}&\;\;\;\mathrm{det}_\mathbf{k}\left(\mathrm{ad}(\Phi^\mathbf{t})\right)=0\;.
\end{eqnarray}
Here $\chi(\Sigma_h)$ is the Euler characteristic of the manifold, $\mathbf{t}$ is the Cartan subalgebra and $\mathbf{k}$ the roots. If $\chi(\Sigma_h)>0$ the second line above is irrelevant, as we are just removing a point where the integrand is 0. But for $\chi(\Sigma_h)\leq0$ it will mean we need to remove a point where the integrand is singular or 1 from the integration contour. We review one method for achieving this in Appendix \ref{sec:hGeq1}.

It will be important to note here that $h$ can actually be half-integer, not just integer, in the undeformed theory. The reason is that one can calculate the partition function on a surface with appropriate boundary, or equivalently with the insertion of an appropriate Wilson loop. For appropriate loops of boundaries the result is equivalent to increasing the genus of the surface by $\frac{1}{2}$.

Putting this all together then, we are left with the expression for the partition function
\begin{eqnarray}\label{eq:SUNPartitionFunctionIntegralRep}
Z_{SU(N)}(g,h)=\prod\limits_{i=1}^{N-1}\sum\limits_{n_i\in \mathbb{Z}}\int^{'} d\Phi^i e^{-2\pi i n_i\Phi^i -\frac{g}{2}\Phi^i\Phi^i}\mathrm{det}_\mathbf{k}\left(\mathrm{ad}(\Phi^\mathbf{t})\right)^{\chi(\Sigma_h)/2}\;.
\end{eqnarray}
Here we have ignored an unimportant constant multiplicative factor, and $^{'}$ indicates that we are excluding from the integration contour points where $\mathrm{det}_\mathbf{k}\left(\mathrm{ad}(\Phi^\mathbf{t})\right)=0$. As anticipated, gauge fixing has reduced an infinite dimensional integration to a simple finite dimensional integral.

In the case of $U(N)$, the same procedure can be followed in exactly the same way as above. The result in this case is
\begin{eqnarray}\label{eq:UNPartitionFunctionIntegralRep}
Z_{U(N)}(g,h,\theta)=\prod\limits_{i=1}^N\sum\limits_{n_i\in \mathbb{Z}}\int^{'} d\Phi^i e^{-2\pi i n_i\Phi^i -i\theta n_i-\frac{g}{2}\Phi^i\Phi^i}\mathrm{det}_\mathbf{k}\left(\mathrm{ad}(\Phi^\mathbf{t})\right)^{\chi(\Sigma_h)/2}\;.
\end{eqnarray}
Again there is an unimportant constant we have ignored.

One final thing we need here are the 1-loop determinants. The 1-loop determinant for the $U(N)$ case is given by
\begin{eqnarray}\label{eq:UN1loopDet}
\mathrm{det}_\mathbf{k}\left(\mathrm{ad}(\Phi^\mathbf{t})\right)^{\chi(\Sigma_h)/2}=\prod\limits_{1\leq j<i\leq N}(\Phi^i-\Phi^j)^{2-2h}\;.
\end{eqnarray}
For $SU(N)$ things are not quite so neat, but in this paper we will only need explicitly the results for $SU(2)$ and $SU(3)$ which are given by
\begin{eqnarray}
\mathrm{det}_\mathbf{k}\left(\mathrm{ad}(\Phi^\mathbf{t})\right)^{\chi(\Sigma_h)/2}&=&\Phi^{2-2h}\;\;\;\;\;\;\;\;\;\;\;\;\;\;\;\;\;\;\;\;\;\;\;\;\;\;\;\;\;\;\;\;\;\;\;\,\mathrm{for}\;SU(2)\;.\nonumber\\
\mathrm{det}_\mathbf{k}\left(\mathrm{ad}(\Phi^\mathbf{t})\right)^{\chi(\Sigma_h)/2}&=&\Phi_1^{2-2h}\Phi_2^{2-2h}(\Phi_1-\Phi_2)^{2-2h}\;\;\;\;\mathrm{for}\;SU(3)\;.
\end{eqnarray}

From here there are now two standard ways of solving the integral; one which will result in a strong coupling description of the partition function, and the other which will result in a weak coupling description. We will look at each in turn.

\subsubsection{The strong coupling transseries}\label{sec:strongCouplingPF}

To get the strong coupling representation of the partition function we need to use the standard identity
\begin{eqnarray}\label{eq:DeltaFunctionIdentity1}
\sum\limits_{n_i\in \mathbb{Z}}e^{i\frac{n_i \Phi^i}{2 \pi}}=\sum\limits_{m_i\in \mathbb{Z}}\delta(\Phi^i-4\pi^2 m_i)\;.
\end{eqnarray}
Applying this to \eqref{eq:SUNPartitionFunctionIntegralRep} we find
\begin{eqnarray}\label{eq:SUNStrongPF}
Z_{SU(N)}(g,h)=\prod\limits_{i=1}^{N-1}\left(\sum\limits_{m_{i}\in\mathbb{Z}}\right)^{'}\mathrm{det}_\mathbf{k}\left(\mathrm{ad}(m_\mathbf{t})\right)^{2-2h}e^{-g m_i^2/2}\;.
\end{eqnarray}
Here again the $^{'}$ indicates that we are excluding from the sum points where $\mathrm{det}_\mathbf{k}\left(\mathrm{ad}(m_\mathbf{t})\right)=0$. Specialising to $SU(2)$ this is
\begin{eqnarray}\label{eq:SU2localisedPF}
Z_{SU(2)}(g,h)=\sum\limits_{m\in\mathbb{Z}/0}m^{2-2h}e^{-g m^2/2}\;,
\end{eqnarray}
and for $SU(3)$ we have
\begin{eqnarray}
Z_{SU(3)}(g,h)=\left(\sum\limits_{m_1\in\mathbb{Z}}\sum\limits_{m_2\in\mathbb{Z}}\right)^{'}m_1^{2-2h}m_2^{2-2h}(m_1-m_2)^{2-2h}e^{-g (m_1^2+m_2^2)/2}\;.
\end{eqnarray}
For the $U(N)$ case things are much the same. Substituting \eqref{eq:UN1loopDet} into \eqref{eq:UNPartitionFunctionIntegralRep} and applying \eqref{eq:DeltaFunctionIdentity1} the result we get is
\begin{eqnarray}\label{eq:UNStrongPF}
Z_{U(N)}(g,h,\theta)=\left(\sum\limits_{m_1\in\mathbb{Z}}\dots\sum\limits_{m_N\in\mathbb{Z}}\right)^{'}\left(\prod\limits_{1\leq i<j\leq N}(m_i-m_j)^{2-2h}\right)e^{-\sum\limits_{i=1}^{N}g (m_i-\theta/2\pi)^2/2}\;.\;\;\;
\end{eqnarray}
Both of these are clearly transseries representations of the partition function. Moreover, they are also lacking an asymptotic perturbative series (or in fact any series at all) attaching to each exponential. For the weak case this is expected as all the saddles are in different topological sectors. For the strong case we do not have a strong coupling effective action, so we cannot say what saddles there are that may be responsible for the different contributions to the transseries. However, we can see that they do not seem to interact via resurgence and exhibit Stokes phenomena.

A comment is in order here. The expression usually found in the literature is
\begin{eqnarray}\label{eq:PFCasimirs}
Z_G(g,h,\theta)=\sum\limits_R (\dim R)^{2-2h}e^{-\frac{g}{2}C_2(R)+i\theta^\prime C_1(R)}\;.
\end{eqnarray}
Here we are summing over representations $R$ of the Lie algebra of $G$, and $C_1(R)$ and $C_2(R)$ are the first and second Casimirs. We have written $\theta^\prime$ as the difference in $\theta$ dependence between \eqref{eq:UNStrongPF} and \eqref{eq:PFCasimirs} is the difference between \eqref{eq:YM_action_UN} and \eqref{eq:YM_action_TopoUN}, which we have already explained. After this adjustment, to get between our expression and the standard expression found in the literature is then a matter of substituting in expressions for the Casimirs, shifting the dummy variables in the sums appropriately, and absorbing a factor into the multiplicative constant we are suppressing.

For us the most important cases are $SU(2)$ and $U(2)$. For $SU(2)$ we can label representations by a single integer $m$, and the relevant data is then
\begin{eqnarray}\label{eq:SU2Reps}
\dim R &=& m\;,\nonumber\\
C_1(R) &=& 0\;,\nonumber\\
C_2(R) &=& \frac{m^2}{2}-\frac{1}{2}\;,\nonumber\\
\mathrm{for}\;\;m&=&1,2,3,\dots\;.
\end{eqnarray}
To get to $U(2)$ we use the relation
\begin{eqnarray}
U(N)=SU(N)\times U(1)/\mathbb{Z}_N\;.
\end{eqnarray}
In other words we decompose representations of $U(2)$, $\mathcal{R}$, into representations of $SU(2)$, $R$, along with a $U(1)$ charge given by $q=m+2r$, for $r\in\mathbb{Z}$. We then have
\begin{eqnarray}\label{eq:U2Reps}
\dim \mathcal{R} &=& m\;,\nonumber\\
C_1(\mathcal{R}) &=& q\;,\nonumber\\
C_2(\mathcal{R}) &=& \frac{m^2}{2}-\frac{1}{2}+\frac{q^2}{2}\;,\nonumber\\
\mathrm{for}\;\;m&=&1,2,3,\dots\;,\;\;\;q=m+2r\;.
\end{eqnarray}
Using this data it is now a simple exercise to show that for $SU(2)$ and $U(2)$, \eqref{eq:PFCasimirs} reduces to \eqref{eq:SUNStrongPF} and \eqref{eq:UNStrongPF} respectively. See for example \cite{Aganagic_2005} for further explanation of this, including the case of other gauge groups.

\subsubsection{The weak coupling case}\label{sec:weakCouplingPF}

The second way of solving \eqref{eq:SUNPartitionFunctionIntegralRep} and \eqref{eq:UNPartitionFunctionIntegralRep}, first given by Witten \cite{Witten_1992ym2}, is simply solving the Gaussian integrals as they are, without using the identity \eqref{eq:DeltaFunctionIdentity1}. This results in a weak coupling, semi-classical expansion of the partition function.

For the sake of simplicity, we will mostly only consider the case where the genus (actually the effective genus, see next section) is less than 1, to avoid the complication of removing the points where the Faddeev-Popov determinant is zero. However all our conclusions carry over to $h\geq 1$, which we show in Appendix \ref{sec:hGeq1}.

Let us illustrate this here for the $SU(2)$ and $U(2)$ cases, for $h=0$. For $SU(2)$ we have that \eqref{eq:SUNPartitionFunctionIntegralRep} becomes
\begin{eqnarray}\label{eq:SU2weakPFExample}
Z_{SU(2)}(g,h=0)&=&\sum\limits_{n\in\mathbb{Z}}\; \int\limits_{-\infty}^{\infty}d\Phi\;\Phi^{2}e^{-2\pi i n \Phi-\frac{g}{2}\Phi^2}\nonumber\\
&=&\sum\limits_{n\in\mathbb{Z}}\sqrt{2\pi}e^{-\frac{(2\pi n)^2}{2g}}(g^{-3/2}-4n^2\pi^2 g^{-5/2})  \;.
\end{eqnarray}
This is a weak coupling transseries. Again we see that there is no asymptotic series attached to each exponential. Thus there is no Stokes phenomena between saddles, as expected (see Figure \ref{fig:resurgenceTriangleUndeformed}).

In the $U(2)$ case we have
\begin{eqnarray}\label{eq:U2weakPFExample}
Z_{U(2)}(g,h=0,\theta)&=&\sum\limits_{n_1,n_2\in\mathbb{Z}}\int\limits_{-\infty}^\infty d\Phi_1\int\limits_{-\infty}^\infty d\Phi_2 (\Phi_1-\Phi_2)^{2}e^{-2\pi i n_1\Phi_1-2\pi i n_2\Phi_2-\frac{g}{2}(\Phi_1^2+\Phi_2^2)-i\theta(n_1+n_2)}\nonumber\\
&=&\sum\limits_{n_1,n_2\in\mathbb{Z}}4\pi e^{-\frac{1}{2g}((2\pi n_1)^2+(2\pi n_2)^2)-i\theta(n_1+n_2)}(g^{-2}-2\pi^2(n_1-n_2)^2g^{-3})\;.\nonumber\\ \;
\end{eqnarray}
Again, this is a weak coupling transseries, where the perturbative series attached to each exponential is not asymptotic.

In the weak coupling case we have a semi-classical explanation for the non-perturbative contributions to the transseries, i.e. the saddles discussed in Section \ref{sec:undeformedSaddles}. For the strong coupling case, to find such an explanation, we would need to know the strong coupling effective action. For this work, we will content ourselves with working with the transseries without looking for saddle descriptions of the non-perturbative contributions in the strong coupling case.

\subsection{The deformation}

We now introduce the deformation of the theory we will study in this paper. In \cite{Kozcaz:2016wvy,Dorigoni:2017smz,Dorigoni:2019kux} a deformation was shown to work in supersymmetric or quasi-exact-solvable quantum mechanics, $\mathcal{N}=(2,2)$ on $S^2$, and $\mathcal{N}=2$ on a squashed $S^3$,  where the effective parameter for number of bosons or fermions was deformed to be slightly different from each other, such as to slightly break supersymmetry. In the latter two papers, it was also possible to deform the theory such that the effective parameter for the number of bosons and fermions was equal, but non-integer, revealing an asymptotic perturbative series.

In the quantum mechanical case \cite{Kozcaz:2016wvy}, the authors were also able to show how this deformation in the effective parameter could be produced by adding a deformation term to the action in the path integral. This was not possible in \cite{Dorigoni:2017smz,Dorigoni:2019kux}. In \cite{Kozcaz:2016wvy} certain fields, the fermions in their case, could be integrated out, resulting in an additional term in the bosonic effective action. Adding this same term to the action with non-integer coefficient mimics having non-integer number of fermions.

For 2d YM we can also find a bona fide deformation of the UV action of the theory. Like the case of quantum mechanics this can be seen to simply be adding the determinant arising from integrating out certain fields (i.e. that of \eqref{eq:1loopdetRatio}) back into the action, with non-integer coefficient. This happens to mimic analytically continuing the genus of the surface on which the theory lives to be non-integer. Actually, this turns out to be almost identical to deforming the effective parameter for the number of bosonic and fermionic fields to be non-integer as well. Both amount to setting the exponent of the ratio of 1-loop determinants to be non-integer. In this case that exponent is $\chi(\Sigma_h)$. In the cases studied in \cite{Dorigoni:2017smz,Dorigoni:2019kux} the exponent is the effective parameter for the number of chiral multiplets. This naturally prompts the question, can we use the same trick to find a genuine deformation of the UV action in the cases studies in \cite{Dorigoni:2017smz,Dorigoni:2019kux}, but we won't consider this question in this work.

It was noted in \cite{Witten:1991we} that we can consider a family of theories related to 2d YM by adding any symmetric polynomial of $\Phi$ to the action. For example we could add any term of the form $\tr(\Phi^i)^j$ to the action and get another valid theory. Adding such terms gives us a way of deforming the theory.

The particular deformation of the theory we want to consider in this work is given by
\begin{eqnarray}
\delta\int K\log\det\begin{pmatrix}
\tr(\Phi^{2N-2}) &\tr(\Phi^{2N-3})&\dots& \tr(\Phi^{N})&\tr(\Phi^{N-1})\\
\tr(\Phi^{2N-3}) &\tr(\Phi^{2N-4})&\dots& \tr(\Phi^{N-1})&\tr(\Phi^{N-2})\\
\vdots &\vdots&\ddots& \vdots&\vdots\\
\tr(\Phi^{N}) &\tr(\Phi^{N-1})&\dots& \tr(\Phi^{2})&\tr(\Phi)\\
\tr(\Phi^{N-1}) &\tr(\Phi^{N-2})&\dots& \tr(\Phi)&N
\end{pmatrix}\;.
\end{eqnarray}
This may look odd but actually it is nothing but the Faddeev-Popov determinant \eqref{eq:1loopdetRatio}, moved into the action (hence the logarithm), with new coefficient $\delta$.

Calculating the partition function goes through unhindered in the exact same way as achieved in Section \ref{sec:localisationApproach}. The effect on the exact formula for the partition function is to shift $h$:
\begin{eqnarray}
h\rightarrow h+\delta\;.
\end{eqnarray}
For example for the $SU(2)$ case, the deformation is given by
\begin{eqnarray}
\delta\int K\log\left(\tr(\Phi^2)\right)\;.
\end{eqnarray}
The localised partition function is then given by
\begin{eqnarray}\label{eq:SU2DeformedPFIntRep}
Z_{SU(2)}(g,h,\delta)&=&\sum\limits_{n\in\mathbb{Z}}\; \int\limits_{-\infty}^{\infty}d\Phi\;\Phi^{2-2h}e^{-2\pi i n \Phi-\frac{g}{2}\Phi^2-\delta\log(\Phi^2)}\nonumber\\
&=&\sum\limits_{n\in\mathbb{Z}}\; \int\limits_{-\infty}^{\infty}d\Phi\;\Phi^{2-2(h+\delta)}e^{-2\pi i n \Phi-\frac{g}{2}\Phi^2} \;.
\end{eqnarray}
So we have found a genuine deformation of the theory that in the effective description gives rise to a shift in the genus. This will allow us to consider the theory with the effective genus non-integer.

Before moving on two comments are in order. First, in the remainder of this paper, we will denote
\begin{eqnarray}
\tilde{h}=h+\delta\;.
\end{eqnarray}
We will work with $\tilde{h}<1$ to avoid the complications described in Section \ref{sec:weakCouplingPF}, but as explained there, all conclusions carry over simply to the case $\tilde{h}\geq1$. Second, it is worth noting that this deformation is a quantum deformation. The deformation used in the original work on Cheshire cat resurgence \cite{Kozcaz:2016wvy} was also a quantum deformation. With a simple re-scaling of the fields we can rewrite the action as (say for $SU(N)$)
\begin{eqnarray}
S_{SU(N)}&=&\frac{1}{g}\left(i\int_{\Sigma_h}\tr(\Phi\wedge F)+\frac{1}{2}\int_{\Sigma_h}\tr(\Phi^2)K+\phantom{\begin{pmatrix}
\tr(\Phi^{2N-2})\\
\tr(\Phi^{2N-3})\\
\vdots\\
\tr(\Phi^{N})\\
\tr(\Phi^{N-1})
\end{pmatrix}}\right.\\
&&\left.\;\;\;+g\;\delta\int K\log\det\begin{pmatrix}
\tr(\Phi^{2N-2}) &\tr(\Phi^{2N-3})&\dots& \tr(\Phi^{N})&\tr(\Phi^{N-1})\\
\tr(\Phi^{2N-3}) &\tr(\Phi^{2N-4})&\dots& \tr(\Phi^{N-1})&\tr(\Phi^{N-2})\\
\vdots &\vdots&\ddots& \vdots&\vdots\\
\tr(\Phi^{N}) &\tr(\Phi^{N-1})&\dots& \tr(\Phi^{2})&\tr(\Phi)\\
\tr(\Phi^{N-1}) &\tr(\Phi^{N-2})&\dots& \tr(\Phi)&N
\end{pmatrix}\right)\;.\nonumber
\end{eqnarray}
From this expression for the action we can clearly see the additional term we have added is order $g$.

The fact that this is a quantum deformation has important consequences. For starters any saddles will not be ``semi-classical'' saddles in the traditional sense, but rather saddles of a quantum action. We will mostly defer the discussion of this to \cite{Fujimori:2022pld} where we will study the Picard-Lefschetz decomposition of this theory. For now the key thing is that the $\mathcal{O}(g)$ and higher corrections to the saddles we are about to discuss won't appear in the exponent of the exponential factors of the contributions in the transseries (once we have expanded to get perturbative series for each sector). Rather they will contribute to the perturbative series itself in each sector. In other words, the action of these solutions can be expanded as
\begin{eqnarray}
\frac{1}{g}S(g,\tilde{h})=\frac{1}{g}S_0(\tilde{h})+S_1(\tilde{h})+gS_2(\tilde{h})+\dots\;.
\end{eqnarray}
However, the exponent of the exponential factors of the contributions in the transseries will simply be
\begin{eqnarray}
\frac{1}{g}S_0(\tilde{h})\;.
\end{eqnarray}
We could of course send $\delta\rightarrow\frac{\delta}{g}$, and our deformation would no longer be quantum. For a number of reasons though, including making contact with previous work, and the fact that this deformation arises naturally upon integrating out $A^\alpha$ and $c^\alpha$, we will work with the quantum deformation here and in \cite{Fujimori:2022pld}.

\subsection{New saddles}\label{sec:NewSaddles}

\begin{figure}[t]
\centering
\includegraphics[width=0.6\textwidth]{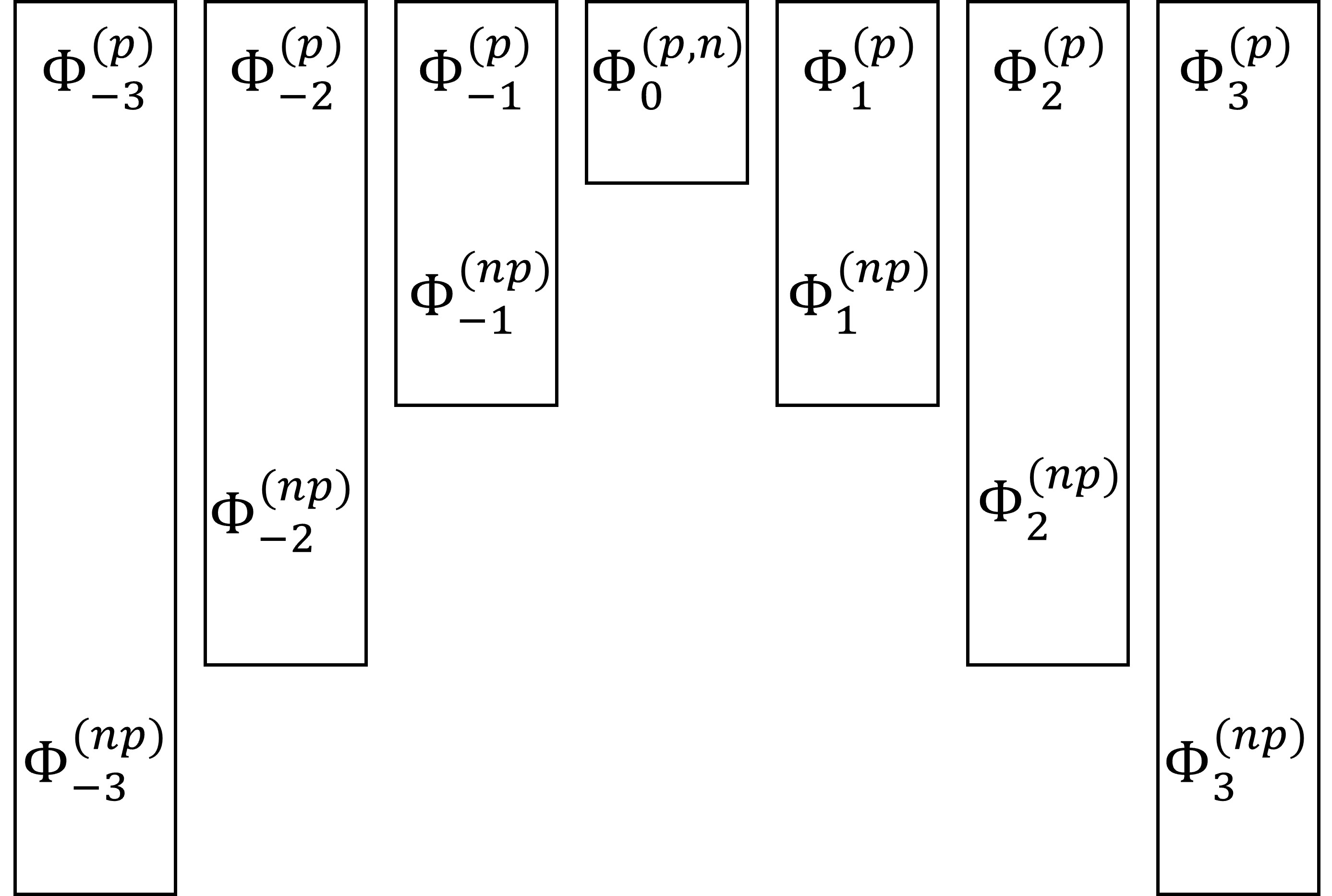} 
\caption{The resurgence triangle for the deformed $SU(2)$ case. $\Phi^{(p)}_n$ is the contribution from the perturbative saddle in the $n^{th}$ topological sector, $\Phi^{(np)}_n$ is the contribution from the non-perturbative saddle in the $n^{th}$ topological sector, and $\Phi^{(n,p)}_0$ is the contribution from the perturbative and non-perturbative saddles in the $0^{th}$ topological sector. Using resurgence we can get all of the contents of the column from one entry of that column, but not the contents of any of the other columns. As we move down the vertical axis the value of the action increases.}
\label{fig:resurgenceTriangleSU2}
\end{figure}

It now turns out that deformation we have just introduced has actually introduced new saddle points into the theory (now saddles of the quantum action rather than saddles of the classical action). To start let us consider the $SU(2)$ theory. The Euler-Lagrange equations of motion after the deformation are given by
\begin{eqnarray}
d\ast \Phi&=&0\;,\nonumber\\
iF+g\Phi K+2\delta\frac{\Phi}{\tr(\Phi^2)}K&=&0\;.
\end{eqnarray}
The first equation means that we need to consider constant $\Phi$. Taking the adjoint exterior derivative of the second equation, and using the first equation to delete the $d\ast \Phi$ terms, we find that in the deformed case we still have at the saddle points
\begin{eqnarray}
d\ast F=0\;.
\end{eqnarray}
I.e our saddles will still be classified by monopole number. Working in the torus gauge again, we restrict $\Phi$ to be in the Cartan subalgebra. For $SU(2)$ this gives us just one scalar field to work with, $\Phi$. Substituting in the monopole solutions \eqref{eq:monopoleSolutions} we are thus left to solve
\begin{eqnarray}
2\pi i n+g\Phi+\frac{2\delta}{\Phi}=0\;.
\end{eqnarray}
This is just a quadratic equation. The saddles are thus given by
\begin{eqnarray}
dA=2\pi n K\;,\;\;\;\Phi&=&\frac{-i}{g}\left(\pi n+\sqrt{\pi^2 n^2+2\delta g}\right)\;,\nonumber\\
dA=2\pi n K\;,\;\;\;\Phi&=&\frac{-i}{g}\left(\pi n-\sqrt{\pi^2 n^2+2\delta g}\right)\;.
\end{eqnarray}
The first of these two solutions is the same as the undeformed solution to order $g^0$. The second is not a solution in the undeformed theory, rather, it makes an appearance in the deformed theory only. It's action is 0 at order $g^0$, i.e. it contributes to the perturbative part of the transseries (but has a distinct contribution for each $n$). In other words, these solutions have respective actions
\begin{eqnarray}
S[A_\mu,\Phi]&=&\frac{(2\pi n)^2+\mathcal{O}(g)}{2g}\;,\nonumber\\
S[A_\mu,\Phi]&=&\frac{0+\mathcal{O}(g)}{2g}\;.
\end{eqnarray}
They thus arrange themselves into the resurgence triangle structure shown in Figure \ref{fig:resurgenceTriangleSU2}. For the remainder of this work we will refer to the first of these solutions as the ``non-perturbative solution'', and the second as the ``perturbative solution'', within each topological sector.

For $U(2)$ things follow through in similar fashion, but now we need two integers to parametrise solutions, as there are two elements of the Cartan subalgebra. The solutions are now given by
\begin{eqnarray}
dA^1=2\pi n_1 K\;,\;\;\;dA^2=2\pi n_2 K\;,\;\;\;\Phi^1&=&\frac{-i}{2g}\left(3\pi n_1+\pi n_2-\sqrt{\pi^2(n_1-n_2)^2 +4\delta g}\right)\;,\nonumber\\
\Phi^2&=&\frac{-i}{2g}\left(\pi n_1+3\pi n_2+\sqrt{\pi^2(n_1-n_2)^2 +4\delta g}\right)\;;\nonumber\\
dA^1=2\pi n_1 K\;,\;\;\;dA^2=2\pi n_2 K\;,\;\;\;\Phi^1&=&\frac{-i}{2g}\left(3\pi n_1+\pi n_2+\sqrt{\pi^2(n_1-n_2)^2 +4\delta g}\right)\;,\nonumber\\
\Phi^2&=&\frac{-i}{2g}\left(\pi n_1+3\pi n_2-\sqrt{\pi^2(n_1-n_2)^2 +4\delta g}\right)\;.\nonumber\\
\;
\end{eqnarray}
Again the first of these saddles in the $\delta\rightarrow 0$ limit is just the monopole saddle, and the second is a new saddle that only exists in the deformed theory. These solutions have respective actions
\begin{eqnarray}\label{eq:U2SaddleValues}
S[A_\mu,\Phi]&=&\frac{(2\pi)^2(n_1^2+n_2^2)+\mathcal{O}(g)}{2g}+i\theta(n_1+n_2)\;,\nonumber\\
S[A_\mu,\Phi]&=&\frac{((n_1+n_2)\pi)^2+\mathcal{O}(g)}{g}+i\theta(n_1+n_2)\;.
\end{eqnarray}
One can draw a resurgence triangle structure similar to that of Figure \ref{fig:resurgenceTriangleSU2}, although now the figure would need to be 3-dimensional.

Thus for both the $SU(2)$ and $U(2)$ cases we have doubled the number of saddles for each $n$ or $(n_1,n_2)$. Also the new saddles have action smaller than the original saddles, so the new saddles will dominate in the transseries. For the new $SU(2)$ saddles, and the new $U(2)$ saddles in the $n_1+n_2=0$ sector, the action of the saddle is 0 (plus order $g$), so it will show up in the transseries as a perturbative contribution.

For higher $N$ things continue in this way. It is easy to see that the vector field part of the equations of motion will always be given by the monopole solutions. The solutions to the scalar field part will be given by solutions to a polynomial in the constant modes of the Cartan-subalgebra elements of the field. For higher $N$ the order of this polynomial grows, and we can no-longer analytically write down all the solutions due to the Abel–Ruffini theorem.

\subsection{Integrals and contours}\label{sec:integralsAndContours}

It will be convenient to introduce the following basis of integrals:
\begin{eqnarray}\label{eq:Zmu}
Z_{\mu}(g,\tilde{h})=\int d\Phi\;\Phi^{2-2\tilde{h}}e^{i \mu\Phi-\frac{g}{2}\Phi^2}\;.
\end{eqnarray}
For the $SU(2)$ and $U(2)$ cases we will be able to write all the relevant integrals in this form.

But we need to be careful about what contour we take. In the undeformed case the contour will just be the real axis. However our deformation has introduced a singularity in the action at the origin, so we now need to use a different contour. For this work we choose to deform the contour so it passes just above the singularity at the origin. The transseries parameters are dependent on this choice. When we consider the Picard-Lefschetz decomposition of the path integral in \cite{Fujimori:2022pld} we will see an explanation of this due to intersection numbers.

Finally let us note that it is now easy to relate this function to the parabolic cylinder function. A standard integral representation of the parabolic cylinder function $U(a,z)$ is \cite{NIST:DLMFCh12}
\begin{eqnarray}\label{eq:UIntegralrep}
U(a,z)=\frac{e^{\frac{1}{4}z^2}}{i\sqrt{2\pi}}\int\limits_{c-i\infty}^{c+i\infty}d\tilde{t}\;e^{-z\tilde{t}+\frac{1}{2}\tilde{t}^2}\tilde{t}^{-a-\frac{1}{2}}\;.
\end{eqnarray}
Here $c>0$. If we make the substitution $\Phi\rightarrow -i\Phi$ in \eqref{eq:Zmu} and do some rearranging, we see that $Z_{\mu}(g,\tilde{h})$ can be written as
\begin{eqnarray}\label{eq:ZmuParabolic}
Z_{\mu}(g,\tilde{h})=-i\sqrt{2\pi}e^{-\frac{\mu^2}{4g}}\left(\frac{i}{\sqrt{g}}\right)^{3-2\tilde{h}} U\left(2\tilde{h}-\frac{5}{2},\frac{\mu}{\sqrt{g}}\right)\;.
\end{eqnarray}
Thus we see that the partition functions in the $SU(2)$ and $U(2)$ cases will be functions of parabolic cylinder functions. The resurgence properties of parabolic cylinder functions are well studied (see e.g. \cite{Dunne:2014bca}), but as we will see the resurgence story of 2d YM is not limited to this.

\section{Resurgence in deformed Yang-Mills theory: Weak coupling semi-classical transseries}\label{sec:Cheshire_cat}

In this section we explore the resurgence structure of the weak coupling semi-classical transseries. We first review the resurgence structure of the parabolic cylinder function $U(a,z)$, and then apply this in the $SU(2)$ and $U(2)$ cases.

\subsection{Review: Resurgence in \texorpdfstring{$U(a,z)$}. parabolic cylinder function}\label{sec:parabolicCFResurgence}

\begin{figure}
     \centering
     \begin{subfigure}[b]{0.48\textwidth}
         \centering
         \includegraphics[width=\textwidth]{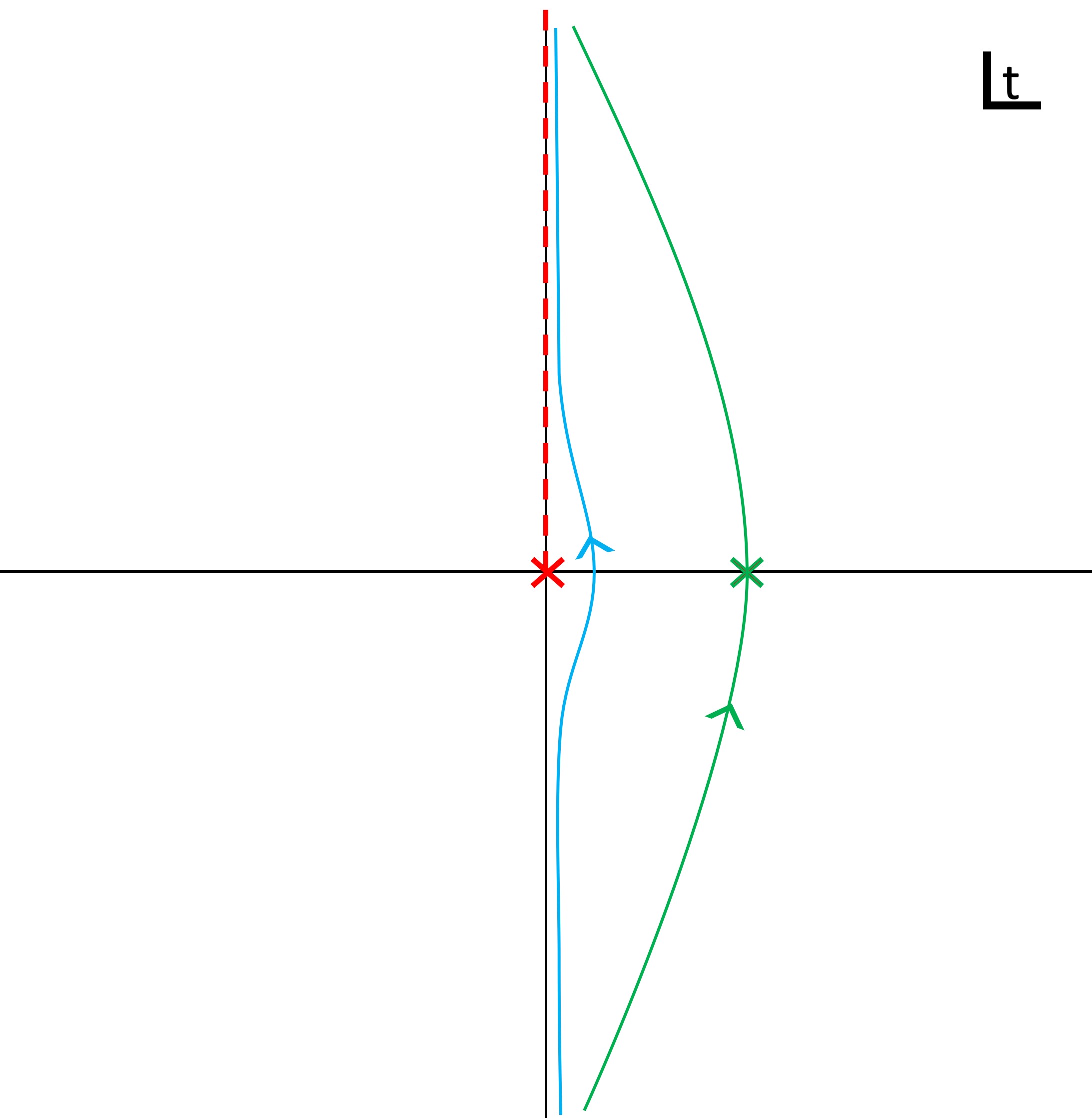}
         \caption{$\Re(z)>0$}
         \label{fig:ParabolicContours1}
     \end{subfigure}
       \hfill
     \begin{subfigure}[b]{0.48\textwidth}
         \centering
         \includegraphics[width=\textwidth]{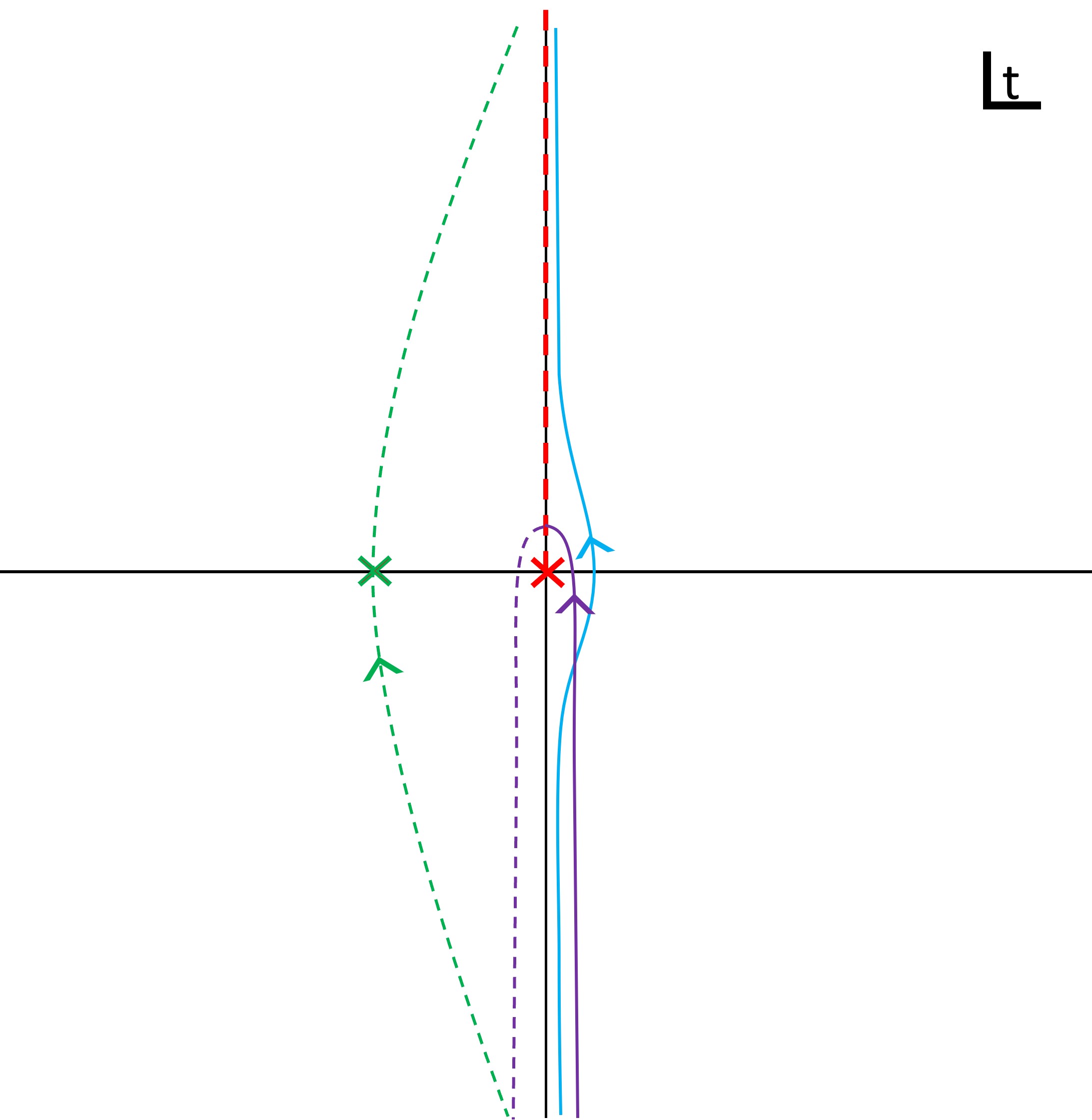}
         \caption{$\Re(z)<0$, $\Im(z)<0$, $\Im(z)\approx 0$}
         \label{fig:ParabolicContours3}
     \end{subfigure}
     \hfill
     \begin{subfigure}[b]{0.48\textwidth}
         \centering
         \includegraphics[width=\textwidth]{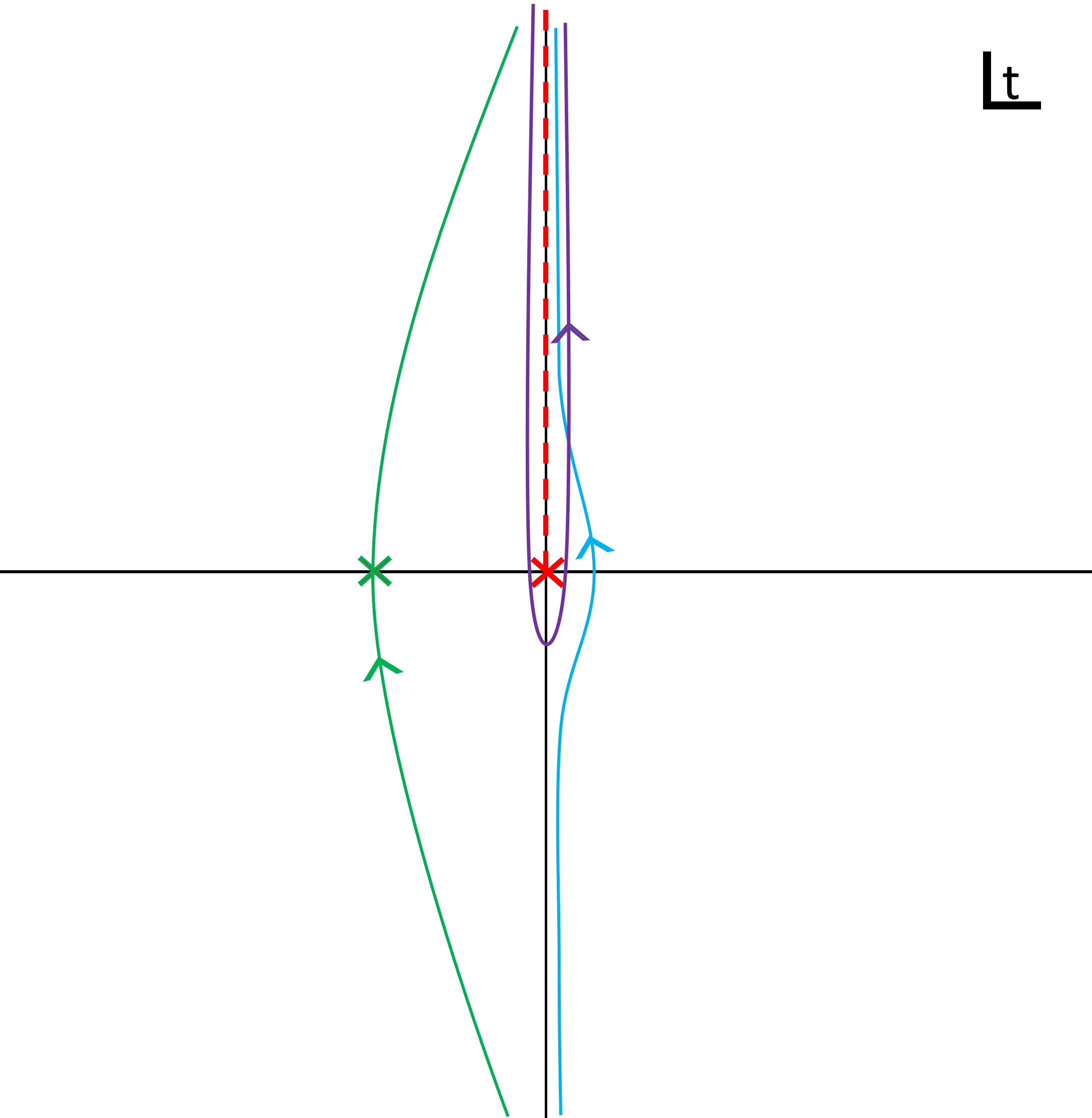}
         \caption{$\Re(z)<0$, $\Im(z)>0$, $\Im(z)\approx 0$}
         \label{fig:ParabolicContours2}
     \end{subfigure}
        \caption{Plots of the contours for the parabolic cylinder function integral for real part of $z$ being $\Re(z)>0$ and $\Re(z)<0$. For $\Re(z)<0$ we also distinguish the cases for imaginary part of $z$ being $\Im(z)>0$ and $\Im(z)<0$, which we have plotted for infinitesimal imaginary part. The blue line shows the original contour. The green lines and purple lines show the contours we integrate over. The green line passes through the saddle point, and the purple around the branch cut. Solid lines are on the same sheet as the original contour, and the dashed lines on the neighbouring sheet. The green cross represents the saddle point, and the red cross the branch point, with the red dashed line representing the branch cut.}
        \label{fig:ParabolicContours}
\end{figure}

As we have discussed, in many cases we will be able to write the partition function in terms of the parabolic cylinder function $U(a,z)$. It will therefore be efficient to review the resurgent properties of this function here. Later we will be able to make use of this in a number of contexts.

Consider again the integral representation of $U(a,z)$ presented in \eqref{eq:UIntegralrep}. We have two cases to consider; the case when $\Re(z)>0$ and the case when $\Re(z)<0$. The exponent in the integrand of \eqref{eq:UIntegralrep} has a saddle at $\tilde{t}=z$. For $\Re(z)>0$ the contour can be deformed to pass through the saddle with no added complications. For $\Re(z)<0$, in deforming the contour we pick up a contribution from the branch point at the origin. Thus in the first of these cases we only have one contribution to the transseries for the parabolic function, and in the latter case we have two. These are plotted in Figure \ref{fig:ParabolicContours}

For the case $\Re(z)>0$ we first deform the contour to pass through the saddle. Changing variables so $\tilde{t}=0$ is at the saddle point we are left with the integral
\begin{eqnarray}
U(a,z)=\frac{e^{-\frac{1}{4}z^2}}{i\sqrt{2\pi}}\int\limits_{-i\infty}^{i\infty}d\tilde{t}\;(\tilde{t}+z)^{-a-\frac{1}{2}}e^{\frac{1}{2}\tilde{t}^2}\;.
\end{eqnarray}
Taylor expanding $(\tilde{t}+z)^{-a-\frac{1}{2}}$ for large z, and then performing the remaining Gaussian integral, we find the standard asymptotic expansion of $U(a,z)$ for large z:
\begin{eqnarray}\label{eq:parabolicCFexpansionPveZ}
U(a,z)\sim e^{-\frac{1}{4}z^2}z^{-a-\frac{1}{2}}\sum_{j=0}^\infty(-1)^j\frac{(a+1/2)_{2j}}{j!(2z^2)^j}, \quad \quad (x)_n = \frac{\Gamma(x+n)}{\Gamma (x)} \;.
\end{eqnarray}
This is a factorially divergent, alternating sum. For us this will correspond to the series expansion around the non-perturbative term of the transseries in each sector. Thus, the series being alternating and factorially divergent is good news; we expect to find a branch cut on the negative real axis of the Borel plane corresponding to the contribution from the perturbative contribution to the transseries.

For $\Re(z)<0$ things differ as follows. As already stated, when deforming the integral to pass through the saddle we pick up a contribution from the branch cut originating at the origin. The contribution coming from the saddle is identical to the case $\Re(z)>0$. So we just need to calculate the contribution from the branch cut.

In order to calculate the perturbative series in large $z$ for the contribution coming from the branch-cut, we first Taylor expand the $e^{\frac{1}{2}\tilde{t}^2}$ term in the integral:
\begin{eqnarray}
U(a,z)\sim\sum\limits_{j=0}^\infty \frac{(1/2)^j}{j!}\frac{e^{\frac{1}{4}z^2}}{i\sqrt{2\pi}}\int\limits_{c-i\infty}^{c+i\infty} d\tilde{t} \, e^{-z\tilde{t}}\tilde{t}^{2j-a-\frac{1}{2}}\;.
\end{eqnarray}
As we have negative $z$ we can now close the contour around the branch-point at the origin and negative real axis, leaving us with a simple integral which returns a Gamma function. Thus we find the following transseries for $U(a,z)$:
\begin{eqnarray}\label{eq:parabolicCFexpansionNveZ}
U(a,z)&\sim& \frac{\sqrt{2\pi}}{\Gamma(a+1/2)}e^{\frac{1}{4}z^2}z^{a-1/2}\sum_{j=0}^\infty\frac{(1/2-a)_{2j}}{j!(2z^2)^j}\nonumber\\
&&\;\;\;\mp i e^{\pm i\pi a}e^{-\frac{1}{4}z^2}z^{-a-\frac{1}{2}}\sum_{j=0}^\infty(-1)^j\frac{(a+1/2)_{2j}}{j!(2z^2)^j}\;.
\end{eqnarray}
The upper sign of the Stokes constant will be for $\Im(z)>0$ and the lower sign for $\Im(z)<0$. For us this new contribution will correspond to that of the perturbative saddle. It is a  non-alternating factorially divergent asymptotic series, as expected. This will allow us to calculate the non-perturbative data from the perturbative data.

Let us now study how such a resurgence analysis of the above perturbative series works. We first focus on the series in what will become our perturbative sector:
\begin{eqnarray}
U_p(a,z)=\frac{\sqrt{2\pi}}{\Gamma(a+1/2)}e^{\frac{1}{4}z^2}z^{a-1/2}\sum_{j=0}^\infty\frac{(1/2-a)_{2j}}{j!(2z^2)^j}\;.
\end{eqnarray}
We choose a Borel transform that divides each term by $\Gamma(j-a/2+1/4)$;
\begin{eqnarray}\label{eq:ParabolBorelPlane1}
\sum_{j=0}^\infty\frac{(1/2-a)_{2j}}{j!(2z^2)^j}&=&\left(2z^2\right)^{\frac{1}{4}-\frac{a}{2}}\int\limits_0^\infty dt\sum_{j=0}^\infty\frac{(1/2-a)_{2j}}{j!\Gamma(j-\frac{a}{2}+\frac{1}{4})}t^{j-\frac{a}{2}-\frac{3}{4}}e^{-2z^2 t}\nonumber\\
&=&\left(2z^2\right)^{\frac{1}{4}-\frac{a}{2}}\int\limits_0^\infty dt\;\frac{t^{-\frac{a}{2}-\frac{3}{4}}(1-4t)^{\frac{a}{2}-\frac{3}{4}}}{\Gamma(\frac{1}{4}-\frac{a}{2})}e^{-2z^2 t}\;.
\end{eqnarray}
The Borel plane has a cut starting at $t=\frac{1}{4}$. Thus we can see from the Borel plane that there is a non-perturbative part with exponential part $e^{-\frac{1}{4}z^2}$. We can calculate the imaginary part of the jump as we cross the Stokes line in the usual way, by calculating the discontinuity across this cut:
\begin{eqnarray}\label{eq:ParaResurgenceDisc}
\mathrm{Disc}_{0}(a,z)&=& \frac{\sqrt{2\pi}}{\Gamma(a+1/2)}e^{\frac{1}{4}z^2}z^{a-1/2}\left(2z^2\right)^{\frac{1}{4}-\frac{a}{2}}\\
&&\;\;\;\;\;\;\;\;\;\;\;\;\;\;\;\;\;\;\;\;\;\;\;\;\;\;\;\;\;\;\;\;\;\;\;\;\;\;\left(\int\limits_0^{\infty+i\epsilon}-\int\limits_0^{\infty-i\epsilon}\;\right) dt\;\frac{t^{-\frac{a}{2}-\frac{3}{4}}(1-4t)^{\frac{a}{2}-\frac{3}{4}}}{\Gamma(\frac{1}{4}-\frac{a}{2})}e^{-2z^2 t}\nonumber\\
&=&\mp\left(e^{i\pi(\frac{a}{2}-\frac{3}{4})}-e^{-i\pi(\frac{a}{2}-\frac{3}{4})}\right)\frac{\sqrt{2\pi}}{\Gamma(a+1/2)}e^{-\frac{1}{4}z^2}z^{a-1/2}\left(2z^2\right)^{\frac{1}{4}-\frac{a}{2}}\nonumber\\
&&\;\;\;\;\;\;\;\;\;\;\;\;\;\;\;\;\;\;\;\;\;\;\;\;\;\;\;\;\;\;\;\;\;\;\;\;\;\;\int\limits_0^{\infty\pm i\epsilon} dt\;\frac{(t+\frac{1}{4})^{-\frac{a}{2}-\frac{3}{4}}(4t)^{\frac{a}{2}-\frac{3}{4}}}{\Gamma(\frac{1}{4}-\frac{a}{2})}e^{-2z^2 t}\;.\nonumber
\end{eqnarray}
In the second line we have written the discontinuity in terms of the integral either above or below the cut, with an appropriate jump in the imaginary part of the transseries parameter dressing this. Note, the discontinuity in the first line begins at $t=1/4$, so the integrals exactly cancel till then, i.e in the first line we really only have the integrals $\int\limits_\frac{1}{4}^{\infty\pm i\epsilon}$. To get to the second line we have then changed variables, $t\rightarrow t+1/4$, so the lower limit returns to 0 and we pull out the exponential term that dresses the non-perturbative contribution. We will see momentarily that this ambiguity is exactly cancelled by the ambiguity in the non-perturbative sector.

We can turn this discontinuity into a perturbative series expansion around the non-perturbative contribution by expanding the $(t+\frac{1}{4})^{-\frac{a}{2}}$ term in the integrand and performing the integral, which just returns a Gamma function. The result is, up to fixing the transseries parameter, identical to \eqref{eq:parabolicCFexpansionPveZ}. In other words, the Borel-\`Ecalle resummation procedure has allowed us to calculate the non-perturbative part of the transseries from the perturbative part, up to transseries parameter.

More briefly let us now consider a resurgence analysis of the non-perturbative contribution to the transseries. Starting from \eqref{eq:parabolicCFexpansionPveZ} we choose a Borel transformation that divides each term by $\Gamma(j+a/2+1/4)$. In this way we have
\begin{eqnarray}\label{eq:ParaResurgeNPBorel}
U(a,z)&\sim& e^{-\frac{1}{4}z^2}z^{-a-\frac{1}{2}}\sum_{j=0}^\infty(-1)^j\frac{(a+1/2)_{2j}}{j!(2z^2)^j}\nonumber\\
&=&e^{-\frac{1}{4}z^2}z^{-a-\frac{1}{2}}\left(2z^2\right)^{\frac{a}{2}+\frac{1}{4}}\int\limits_0^\infty dt\;\sum_{j=0}^\infty(-1)^j\frac{(a+1/2)_{2j}}{j!\Gamma\left(j+\frac{a}{2}+\frac{1}{4}\right)}t^{j+\frac{a}{2}-\frac{3}{4}}e^{-2z^2 t}\nonumber\\
&=&e^{-\frac{1}{4}z^2}z^{-a-\frac{1}{2}}\left(2z^2\right)^{\frac{a}{2}+\frac{1}{4}}\int\limits_0^\infty dt\;\frac{t^{\frac{a}{2}-\frac{3}{4}}(1+4t)^{-\frac{3}{4}-\frac{a}{2}}}{\Gamma\left(\frac{a}{2}+\frac{1}{4}\right)}e^{-2z^2 t}\;.
\end{eqnarray}
We have a cut beginning at $t=-\frac{1}{4}$ corresponding to the perturbative contribution to the transseries. Indeed, after a change of coordinates for $t$, we see we have the same integrand in the Borel plane as before. As above we can calculate the discontinuity across the cut to find the perturbative contribution to the transseries up to transseries parameter.

From \eqref{eq:ParaResurgeNPBorel} it is also simple to see that the ambiguity in the Borel resummation of the perturbative sector is exactly cancelled by the jump in the transseries parameters in the non-perturbative sector. Applying the duplication formula for the Gamma function, we see that the jump \eqref{eq:ParaResurgenceDisc} is exactly canceled by \eqref{eq:ParaResurgeNPBorel} once we have dressed the latter with the transseries parameters from \eqref{eq:parabolicCFexpansionNveZ}.

\subsection{\texorpdfstring{$SU(2)$}. resurgence analysis}\label{sec:SU(2)resurgenceAnalysis}

We have seen that we can write the partition function of undeformed 2d YM as transseries in two ways; as a strong coupling transseries and as a weak coupling transseries. The weak coupling case has an interpretation as a sum over contributions from saddle points. In both cases the perturbative series encountered in each contribution, in the undeformed case, are not asymptotic but rather truncating. In the strong case this truncation is rather severe, to just one term (i.e a number). This is fine, as we were not expecting to find resurgence phenomena occurring due to topological grading.

In this section will begin to discuss how this changes when we deform the theory. Here we focus on the gauge group $SU(2)$, and in the next subsection we will tackle the case of $U(2)$. With the deformation we have introduced new saddles into the theory. We will see that we now have non-truncating asymptotically divergent series in each contribution to the transseries. We also have the manifestation of Cheshire cat phenomena occurring for specific values of the deformation parameter. As discussed in the introduction, there are multiple approaches to deriving the weak coupling expansions we can study. In this section we will be concerned with the semi-classical expansion around each of the saddles.

Thanks to the results of Sections \ref{sec:integralsAndContours} and \ref{sec:parabolicCFResurgence} most of the hard work is now over. We can write the $SU(2)$ partition function in terms of the parabolic cylinder functions $U(a,z)$ as follows:
\begin{eqnarray}\label{eq:SU2WeakTransSumOvern}
Z_{SU(2)}(g,\tilde{h})&=&\sum\limits_{n=-\infty}^\infty Z_{-2\pi n}(g,\tilde{h})\nonumber\\
&=&\sum\limits_{n=-\infty}^\infty-i\sqrt{2\pi}e^{-\frac{(2\pi n)^2}{4g}}\left(\frac{i}{\sqrt{g}}\right)^{3-2\tilde{h}} U\left(2\tilde{h}-\frac{5}{2},-\frac{2\pi n}{\sqrt{g}}\right)\;.
\end{eqnarray}
For the semi-classical expansion of the partition function we can restrict ourselves to working within a single topological sector, $n$. We can then use \eqref{eq:parabolicCFexpansionPveZ} and \eqref{eq:parabolicCFexpansionNveZ} to derive the perturbative series and non-perturbative series in each topological sector. Working with $g$ real and positive for the moment, for $n>0$ we thus have
\begin{eqnarray}\label{eq:SU2PvenTransseries}
 Z_{-2\pi n}(g,\tilde{h})&=&- ie^{\pi i\tilde{h}}\frac{(2\pi)^{2\tilde{h}-2}n^{2\tilde{h}-3}}{\Gamma(2\tilde{h}-2)} \sum_{j=0}^\infty\frac{(3-2\tilde{h})_{2j}}{j!}\left(\frac{g}{8\pi^2n^2}\right)^j\\
&&\;\;\;-e^{\pm 2\pi i\tilde{h}}\frac{e^{2\pi \tilde{h}}}{\sqrt{n}} \left(\frac{2\pi n}{g}\right)^{\frac{5}{2}-2\tilde{h}} e^{-\frac{(2\pi n)^2}{2g}} \sum_{j=0}^\infty (-1)^j \frac{(2\tilde{h}-2)_{2j}}{j!} \left(\frac{g}{8\pi^2n^2}\right)^j\;.\nonumber
\end{eqnarray}
Here we have the non-perturbative contribution coming from the monopole, now dressed with a divergent asymptotic series, and also a contribution coming from the new perturbative saddle. For $n<0$ we have
\begin{eqnarray}\label{eq:SU2NvenTransseries}
 Z_{-2\pi n}(g,\tilde{h})&=&-\frac{e^{2\pi \tilde{h}}}{\sqrt{n}} \left(\frac{2\pi n}{g}\right)^{\frac{5}{2}-2\tilde{h}} e^{-\frac{(2\pi n)^2}{2g}} \sum_{j=0}^\infty (-1)^j \frac{(2\tilde{h}-2)_{2j}}{j!} \left(\frac{g}{8\pi^2n^2}\right)^j\;.
\end{eqnarray}
In this case we only have the non-perturbative saddle, as discussed in Section \ref{sec:parabolicCFResurgence}. For generic values of the deformation parameter we can perform a Borel-\`Ecalle analysis of each of these series to find the other contributions to the transseries, up to the transseries parameters, following the steps in Section \ref{sec:parabolicCFResurgence}.

Here a comment is in order. The above transseries are for real and positive $g$. As we vary the phase of $g$, the second argument of the parabolic cylinder $U(a,z)$ function varies in phase. As the real part of $z$ crosses over from positive to negative values, we again get Stokes phenomena. This jump in the transseries parameters effectively swaps the $\Re(n)>0$ and $\Re(n)<0$ transseries with each other.

Let us now consider what happens as we vary the deformation parameter. The above series truncate for any value of $2-2\tilde{h}$ which is non-negative integer. This is due to the factor of $\Gamma(2\tilde{h}-2)$ in the denominator of both the contributions to the transseries (in the non-perturbative contribution this is coming from the $(2\tilde{h}-2)_{2j}$ factor). At these points the non-perturbative series truncate to a finite number of terms, and the perturbative contribution vanishes entirely (not the saddle itself, just it's contribution).

We also have that when $2\tilde{h}-3$ is a non-negative integer, the perturbative series in the perturbative sector truncates to few terms. These are the terms we have shown how to calculate in Appendix \ref{sec:hGeq1}. In this case the non-perturbative sector doesn't truncate, and we can still derive the perturbative data from the non-perturbative data, but not vice-versa. Note also that at these points the jump in the transseries parameters vanishes. This is to be expected as there is no ambiguity in the Borel-resumption of the perturbative sector.

We can make use of the following formula to calculate the truncated perturbative series:
\begin{eqnarray}
\lim\limits_{\epsilon\rightarrow 0}\frac{\Gamma(-m+\epsilon)}{\Gamma(-n+\epsilon)}=(-1)^{(m-n)}\frac{\Gamma(n+1)}{\Gamma(m+1)}\;\;.
\end{eqnarray}
Thus, for example, for the case $\tilde{h}=0$ we have no perturbative contribution, and the non-perturbative contribution is given by
\begin{eqnarray}
Z_{-2\pi n}(g,0)&=&\sqrt{2\pi}e^{-\frac{(2\pi n)^2}{2g}}(g^{-3/2}-4n^2\pi^2 g^{-5/2})\;.
\end{eqnarray}
This is exactly what we saw in \eqref{eq:SU2weakPFExample}.

In summary, in the $SU(2)$ case we have found that upon deforming the theory we both introduced new saddles into the theory, and rendered the previously truncating weak-coupling perturbative series in each sector of the transseries now divergent asymptotic. For generic values of the deformation parameter this leads to us being able to analyse the transseries using Borel-\`Ecalle resummation. Indeed, for generic values of the deformation parameter we can calculate all the data in a given topological sector of the transseries from the contribution of a single saddle in that sector.

However, at specific values of the deformation parameter we land on a Cheshire cat point. Here the series truncate, and Borel-\`Ecalle resummation is trivial. At these points the effective genus is an integer or half-integer, i.e. the partition function is identical to that of undeformed YM on a different genus surface, perhaps with boundaries or Wilson loop insertions.

Thus we have two different descriptions of the points where the perturbative series truncate. In undeformed 2d YM (for example $\tilde{h}=1$ with $h=1$ and $\delta=0$) the extra saddles in our deformed theory don't exist, and there is only one saddle in each sector, thus no non-trivial resurgence structure. However they are still present in our deformed theory (for example $\tilde{h}=1$ with $h=0$ and $\delta=1$), even at the Cheshire cat points, but at the Cheshire cat points they don't contribute. The latter of these two descriptions is Cheshire Cat resurgence. There is a nice geometrical reason for the perturbative saddles not contributing at the Cheshire cat points, which will be presented in \cite{Fujimori:2022pld}.

\subsection{\texorpdfstring{$U(2)$}. resurgence analysis}

We now turn to the $U(2)$ case. In this case we have an explicit topological angle, which splits the partition function into topological sectors as usual, but also the topological grading discussed in Section \ref{sec:undeformedSaddles}. Upon deforming we have new saddles, now two saddles in each topological sector.

We have that the $U(2)$ partition function can be written as:
\begin{eqnarray}
Z_{U(2)}(g,h,\theta)&=&\sum\limits_{n_1,n_2\in\mathbb{Z}}\int\limits_{-\infty}^\infty d\Phi_1\int\limits_{-\infty}^\infty d\Phi_2 (\Phi_1-\Phi_2)^{2-2\tilde{h}}e^{-2\pi i n_1\Phi_1-2\pi i n_2\Phi_2-\frac{g}{2}(\Phi_1^2+\Phi_2^2)-i\theta(n_1+n_2)}\;.\nonumber\\
\end{eqnarray}
We can analyse the resurgence structure of this by making the following change of coordinates:
\begin{eqnarray}\label{eq:u2Subs}
x=\Phi_1-\Phi_2\;,\;\;\;y=\Phi_1+\Phi_2\;,
\end{eqnarray}
which gets us to the expression
\begin{eqnarray}\label{eq:1topoStartingPoint}
Z_{U(2)}(g,h,\theta)&=&\frac{1}{2}\sum\limits_{n_1,n_2\in\mathbb{Z}}\int\limits_{-\infty}^\infty dx\int\limits_{-\infty}^\infty dy\; x^{2-2\tilde{h}}e^{-\pi ix (n_1-n_2)-\pi iy(n_1+n_2)-\frac{g}{4}(x^2+y^2)-i\theta (n_1+n_2) }\nonumber\\
&=&\sqrt{\frac{\pi}{g}}\sum\limits_{n_1,n_2\in\mathbb{Z}}e^{-\frac{\left(\pi(n_1+n_2)\right)^2}{g}-i\theta(n_1+n_2)}\int\limits_{-\infty}^\infty dx\; x^{2-2\tilde{h}}e^{-\pi ix (n_1-n_2)-\frac{g}{2}x^2}\nonumber\\
&=&\sqrt{\frac{\pi}{g}}\sum\limits_{n_1,n_2\in\mathbb{Z}}e^{-\frac{\left(\pi(n_1+n_2)\right)^2}{g}-i\theta(n_1+n_2)}Z_{(n_2-n_1)\pi}\left(\frac{g}{2},\tilde{h}\right)\;.
\end{eqnarray}
In the last line we have managed to write the partition function in terms of $Z_\mu (g,\tilde{h})$ integrals, which as we have seen can be written in terms of parabolic cylinder functions.

Thus from here the resurgence story for gauge group $U(2)$ in each topological sector is almost identical to that of $SU(2)$, the difference being a different pre-factor and a change of arguments in the $Z_\mu (g,\tilde{h})$ functions. Within each topological sector of the \eqref{eq:1topoStartingPoint}, that is each choice of $(n_1,n_2)$, we find two contributions, with the exponent of the exponential factor given by
\begin{eqnarray}
S(g,\theta)&=&\frac{\left(\pi(n_1+n_2)\right)^2}{g}+i\theta(n_1+n_2)\\
S(g,\theta)&=&\frac{\left(\pi(n_1+n_2)\right)^2}{g}+\frac{\left(\pi(n_1-n_2)\right)^2}{g}+i\theta(n_1+n_2)=\frac{2\left((n_1^2+n_2^2)\right)\pi^2}{g}+i\theta(n_1+n_2)\;.\nonumber
\end{eqnarray}
This is exactly as we expected from \eqref{eq:U2SaddleValues}.

Moreover in the $U(2)$ case we have Cheshire cat phenomena, as in the $SU(2)$ case, for the same values of the deformation parameter. At each of these points we have two descriptions, one being undeformed Yang-Mills with no new saddles, on a different genus surface, perhaps with Wilson loop insertions or boundaries. The other description is that of our deformed theory with new saddles included but exhibiting Cheshire cat resurgence phenomena. Again the disappearance of the additional saddles in the deformed theory has a nice geometrical explanation, which we will present in \cite{Fujimori:2022pld}.

\section{Resurgence in deformed Yang-Mills theory: Strong coupling transseries}\label{sec:strongAndConsistency}

In this section we now turn our attention to the strong coupling transseries for deformed Yang-Mills. We will focus here just on the $SU(2)$ gauge group. Making the substitutions as in \eqref{eq:u2Subs} things carry over to $U(2)$ just as before. We will first start by deriving the transseries and analysing its resurgence properties. It turns out that in the case of 2d YM, we can use the consistency of the strong and weak transseries representations of the partition function to completely determine the transseries parameters, which resurgence on its own will normally not accomplish. Our weapon of choice to achieve this is Poisson resummation. We will explore this after we have explored the strong coupling transseries.

\subsection{Strong coupling}\label{sec:strongResurgenceAnalysis}

Let us now turn our attention to the strong coupling transseries in the case of $SU(2)$. The partition function \eqref{eq:SU2localisedPF} looks, even for non-integer genus, like there is a truncating expansion in every sector. However, the story is slightly more complicated than this. Let us begin with the integral representation of the partition function and begin by performing the sum in a different manner to before:
\begin{eqnarray}\label{eq:strongCouplingPoisson}
Z_{SU(2)}(g,\tilde{h})&=&\sum\limits_{n\in\mathbb{Z}}\int\limits_{-\infty}^\infty d\Phi\;\Phi^{2-2\tilde{h}}e^{-2\pi i n\Phi-\frac{g}{2}\Phi^2}\nonumber\\
&=&(1+e^{-2\pi i\tilde{h}})\left(2^{1/2-\tilde{h}}\Gamma(3/2-\tilde{h})g^{-3/2+\tilde{h}}+\sum\limits_{n=1}^\infty\int\limits_{-\infty}^\infty d\Phi\;\Phi^{2-2\tilde{h}}e^{-2\pi i n\Phi-\frac{g}{2}\Phi^2}\right)\nonumber\\
&=&(1+e^{-2\pi i\tilde{h}})\left(2^{1/2-\tilde{h}}\Gamma(3/2-\tilde{h})g^{-3/2+\tilde{h}}+\int\limits_{-\infty}^\infty d\Phi\;\Phi^{2-2\tilde{h}} \frac{e^{-\frac{g}{2}\Phi^2}}{e^{2\pi i \Phi}-1}\right)\;.
\end{eqnarray}
Here we have separated out the $n=0$ case, then turned the remainder of the sum into a sum over positive $n$ only by making a substitution $\Phi\rightarrow -\Phi$ for the case of negative $n$, and then we have performed the sum using the geometric series summation formula.

Next we focus on the remaining integral, and expand the integrand as follows:
\begin{eqnarray}\label{eq:strongseries}
\int\limits_{-\infty}^\infty d\Phi\;\Phi^{2-2\tilde{h}}\frac{e^{-\frac{g}{2}\Phi^2}}{e^{2\pi i \Phi}-1}&=&\int\limits_{-\infty}^\infty d\Phi\;e^{-\frac{g}{2}\Phi^2}\Phi^{2-2\tilde{h}}\left(-\frac{1}{2}+\sum\limits_{j=0}^\infty\frac{B_{2j}(2\pi i\Phi)^{2j-1}}{(2j)!}\right)\nonumber\\
&=&-\frac{(1+e^{-2\pi i\tilde{h}})}{2}2^{1/2-\tilde{h}}\Gamma(3/2-\tilde{h})g^{-3/2+\tilde{h}}\\
&&\;\;\;\;\;+(1-e^{-2\pi i\tilde{h}})\frac{1}{2^{1+\tilde{h}}\pi i}\sum\limits_{j=0}^\infty\frac{B_{2j}(-8\pi^2)^{j}}{(2j)!}\Gamma(1+j-\tilde{h})g^{\tilde{h}-1-j}\nonumber\;.
\end{eqnarray}
Substituting this into \eqref{eq:strongCouplingPoisson} we find
\begin{eqnarray}\label{eq:strongCouplingPert}
Z_{SU(2)}(g,\tilde{h})&=&(1-e^{-4\pi i\tilde{h}})\left(2^{-1/2-\tilde{h}}\Gamma(3/2-\tilde{h})g^{-3/2+\tilde{h}}\phantom{\sum\limits_0^0}\right.\\
&&\left.\;\;\;\;\;\;\;\;\;\;\;\;\;\;\;\;\;\;\;\;\;\;\;\;\;\;\;\;\;\;\;\;+\frac{1}{2^{1+\tilde{h}}\pi i}\sum\limits_{j=0}^\infty\frac{B_{2j}(-8\pi^2)^{j}}{(2j)!}\Gamma(1+j-\tilde{h})g^{\tilde{h}-1-j}\right)\;.\nonumber
\end{eqnarray}
We have found a perturbative expansion of the partition function in the strong coupling limit.

As $j\rightarrow\infty$ we have
\begin{eqnarray}
B_{2j}\sim (-1)^{j+1}\frac{2(2j)!}{(2\pi)^{2j}}\;\;.
\end{eqnarray}
Thus this perturbative series is non-alternating, and thus non-Borel summable. We have an asymptotic divergent series. Thus we expect to be able to apply Borel-\`Ecalle resummation to be able to determine the other contributions to the transseries.

The Cheshire cat points are again half-integers; $\tilde{h}\in\mathbb{Z}/2$. In these limits $(1-e^{-4\pi i\tilde{h}})\rightarrow 0$. Thus the deformations where the partition function is identical to that of the undeformed theory on different genus surface, perhaps with boundaries or Wilson loop insertions, are Cheshire cat points.

Let us now apply Borel-\`Ecalle resummation to the perturbative series to see how we derive the non-perturbative data in the strong coupling case. We can focus on the sum in \eqref{eq:strongCouplingPert}, and Borel transform as follows:
\begin{eqnarray}
\sum\limits_{j=0}^\infty\frac{B_{2j}(-8\pi^2 )^{j}}{(2j)!}\Gamma(1+j-\tilde{h})g^{\tilde{h}-1-j}&=&\int\limits_{0}^\infty dt\;e^{-gt}\sum\limits_{j=0}^\infty\frac{B_{2j}(-8\pi^2)^{j}}{(2j)!}t^{j-\tilde{h}}\nonumber\\
&=&\sqrt{2}\pi \int\limits_{0}^\infty dt\;e^{-gt}t^{1/2-\tilde{h}}\cot(\pi\sqrt{2t})\;.
\end{eqnarray}
The integrand has singularities at $\frac{m^2}{2}$ on the Borel plane, for $m=0,1,2,\dots$, precisely corresponding to the non-perturbative contributions to the transseries. Inserting this into \eqref{eq:strongCouplingPert} we can write the partition function as
\begin{eqnarray}\label{eq:SU2StrongDefPFPert}
Z_{SU(2)}(g,\tilde{h})&=&(1-e^{-4\pi i\tilde{h}})\left(2^{-1/2-\tilde{h}}\Gamma(3/2-\tilde{h})g^{-3/2+\tilde{h}}\phantom{\sum\limits_0^0}\right.\\
&&\left.\;\;\;\;\;\;\;\;\;\;\;\;\;\;\;\;\;\;\;\;\;\;\;\;\;\;\;\;\;\;\;\;+\frac{1}{2^{1/2+\tilde{h}} i}\int\limits_{0}^\infty dt\;e^{-gt}t^{1/2-\tilde{h}}\cot(\pi\sqrt{2t})\right)\;.\nonumber
\end{eqnarray}

We can now use the discontinuity across the poles to calculate the non-perturbative contributions to the transseries. In this case the discontinuity of the above integral can be written as
\begin{eqnarray}
\mathrm{Disc}_{0}(g,\tilde{h})&=&\frac{(1-e^{-4\pi i \tilde{h}})}{2^{1/2+\tilde{h}}i}\left(\int\limits_{0}^{\infty+i\epsilon}-\int\limits_{0}^{\infty-i\epsilon}\;\right) dt\;e^{-gt}t^{1/2-\tilde{h}}\cot(\pi\sqrt{2t})\nonumber\\
&=&-(1-e^{-4\pi i \tilde{h}})\sum\limits_{m=1}^\infty m^{2-2\tilde{h}}e^{-gm^2/2}\;.
\end{eqnarray}
This looks a lot like the non-perturbative contribution we are expecting. We do have a complication though. This jump clearly vanishes when $\tilde{h}$ is an integer. But we have no way of determining the real part of the transseries parameter by applying only resurgence, and thus cannot find the real contribution that remains as we approach these Cheshire cat points. For now we write
\begin{eqnarray}\label{eq:strongNonPertTrans}
Z_{SU(2)}^{(m)}(g,\tilde{h})&=&\sigma_s(m)m^{2-2\tilde{h}}e^{-gm^2/2}\;.
\end{eqnarray}
Here $\sigma(m)$ is the undetermined transseries parameter. We next turn to look at how we can calculate this parameter using consistency of the strong and weak coupling transseries representations.

\subsection{Using Weak-strong consistency to determine the transseries parameters}\label{sec:weakStrongConsistency}

Thus far we have considered Borel-\`Ecalle resummation of the strong and weak coupling expansions of the partition function. As noted, this process only allows us to calculate the jump in the imaginary part of the transseries parameters, but does not allow us to calculate the transseries parameters exactly. Of course in this setting we can calculate the transseries parameters (and indeed the full non-perturbative contributions to the transseries) directly from \eqref{eq:SU2DeformedPFIntRep}. But here there is another method that doesn't require access to \eqref{eq:SU2DeformedPFIntRep} at all, just access to the strong and weak coupling transseries, and a single transseries parameter.

Our weapon of choice to achieve this is Poisson resummation:
\begin{eqnarray}
\sum\limits_{m\in\mathbb{Z}}f(m)=\sum\limits_{n\in\mathbb{Z}}\int\limits_0^\infty dm\;e^{-2\pi i m n}f(m)\;.
\end{eqnarray}
Applying Poisson resummation to say the strong coupling transseries will allow us to derive a weak coupling transseries. But the weak coupling transseries we get will be dependent on the transseries parameters of the strong coupling transseries. Thus by demanding consistency we can derive the transseries parameters.

As a simply way of introducing the method, let us consider the undeformed case, and suppose that by some means (perhaps Cheshire cat resurgence coupled with the methods of Section \ref{sec:diff_equations_topological_sectors}) we had obtained the strong and weak transseries representations of the partition function up to transseries parameter. Here let us also consider $\tilde{h}=0$ for simplicity. We now have the following two representations of the partition function:
\begin{eqnarray}\label{eq:su2StrongWeakComp}
Z_{SU(2)}(g,\tilde{h}=0)&=&\sum\limits_{m\in\mathbb{Z}}\sigma_s(m)m^2e^{-\frac{g}{2}m^2}\;,\nonumber\\
Z_{SU(2)}(g,\tilde{h}=0)&=&\sum\limits_{n\in\mathbb{Z}}\sigma_w(n)\sqrt{2\pi}e^{-\frac{(2\pi n)^2}{2g}}(g^{-3/2}-4n^2\pi^2 g^{-5/2})  \;.
\end{eqnarray}
Here the top line is the strong coupling representation, and the bottom line the weak coupling representation. $\sigma_s(n)$ are the undetermined transseries parameters in the strong case, and $\sigma_w(m)$ are the undetermined transseries parameters in the weak case.

Note that upon summation in \eqref{eq:su2StrongWeakComp}, any odd part of either $\sigma_s(n)$ or $\sigma_w(m)$ will give no contribution to the partition function. We will thus assume the odd parts of $\sigma_s(n)$ and $\sigma_w(m)$ are both 0.

We of course also know what the transseries parameters are in the perturbative sectors of each transseries, i.e. we know that
\begin{eqnarray}
\sigma_w(0)=\sigma_s(0)=1\;.
\end{eqnarray}
This fact allows us to use consistency of these expressions to calculate all the other transseries parameters.

Taking the Poisson summation of the strong coupling representation we have
\begin{eqnarray}\label{eq:strongWeakConstistancy1}
Z_{SU(2)}(g,\tilde{h}=0)&=&\sum\limits_{\tilde{n}\in\mathbb{Z}}\int\limits_{-\infty}^\infty dm\;e^{-2\pi i m \tilde{n}}\sigma_s(m)m^2e^{-\frac{g}{2}m^2}\;.
\end{eqnarray}
We now equate this with the second line of \eqref{eq:su2StrongWeakComp}. Equating coefficients of the exponentials, we see we have $n=\tilde{n}$. For the $n=0$ case, using our knowledge that $\sigma_w(0)=1$, we have
\begin{eqnarray}
\frac{\sqrt{2\pi}}{g^{3/2}}=\int\limits\limits_{-\infty}^\infty dm\;\sigma_s(m)m^2e^{-\frac{g}{2}m^2}\;.
\end{eqnarray}
From this we can read off (from the $g$ dependence of both sides, and assuming that the odd part of $\sigma_s(m)$ is 0)
\begin{eqnarray}
\sigma_s(m)=1\;.
\end{eqnarray}
Then from equating the coefficients of the exponentials in \eqref{eq:strongWeakConstistancy1} for the $n\neq0$ cases we find
\begin{eqnarray}
\sigma_w(n)=1\;.
\end{eqnarray}
Thus, once we have returned to the undeformed case, it is very simple to use consistency of the strong and weak versions of the transseries to calculate the transseries parameters exactly. In fact we didn't even need to use our knowledge of $\sigma_s(0)$. The knowledge of just one transseries parameter in just one of the transseries representations was enough to determine the all of the transseries parameters in both transseries representations.

Let us consider the deformed case where $\tilde{h}\neq0$. Here things are conceptually the same, though the equations are more complicated.

We have for the strong case that the perturbative part is given by \eqref{eq:strongCouplingPert}, and the non-perturbative part by \eqref{eq:strongNonPertTrans}. In order to use the Poisson resummation method outlined above we need to have the transseries in the form
\begin{eqnarray}
Z_{SU(2)}(g,\tilde{h})=\sum\limits_{n\in\mathbb{Z}}f(n)\;.
\end{eqnarray}
The non-perturbative part is already in this form. For the perturbative part we can use the identity
\begin{eqnarray}
B_{2j}=-2j\zeta(-2j+1)\;,
\end{eqnarray}
and then write $\zeta(-2j+1)$ as a sum over the integers:
\begin{eqnarray}\label{eq:zetaSumStrong}
\zeta(-2j+1)=\sum\limits_{n=1}^\infty n^{2j-1}.
\end{eqnarray}
Applying these identities we get \eqref{eq:strongCouplingPert} into the form of a sum over the integers. In this way we can write the strong partition function in the form
\begin{eqnarray}
Z_{SU(2)}(g,\tilde{h})=\sum\limits_{m\in\mathbb{Z}}\left(\sigma_s^{p}(m)g^{\tilde{h}-1}\sum\limits_{j=0}^\infty c_j m^{2j-1} g^{-j}  +\sigma_s^{np}(m)m^{2-2\tilde{h}}e^{-gm^2/2}\right)\;.
\end{eqnarray}
The coefficients $c_a$ can easily be read from \eqref{eq:strongCouplingPert}, and because the sum in \eqref{eq:zetaSumStrong} starts at $n=1$ we have
\begin{eqnarray}
\sigma_s^{p}(m)=0\;\;\mathrm{for}\;\;m<1\;.
\end{eqnarray}

For the sake of clarity of presentation, we write the weak transseries as a sum of integrals as follows:
\begin{eqnarray}
Z_{SU(2)}(g,\tilde{h})=\sum\limits_{n\in\mathbb{Z}}\left(\sigma_w^{p}(n)\int_{J_p}\Phi^{2-2\tilde{h}}e^{-2\pi in\Phi-\frac{g}{2}\Phi^2} + \sigma_w^{np}(n)\int_{J_{np}}\Phi^{2-2\tilde{h}}e^{-2\pi in\Phi-\frac{g}{2}\Phi^2}\right)\;.\nonumber\\ \;
\end{eqnarray}
Here $J_p$ is the perturbative contour that circles round the branch point at the origin, and $J_{np}$ is the non-perturbative contour that goes from negative infinity to positive infinity passing through the saddle (see Figure \ref{fig:ParabolicContours}). For the $n=0$ case we know what the transseries parameters are from our input (Section \ref{sec:integralsAndContours}), and let us suppose that this is all we know.

We are now in a position to do Poisson resummation and compare our two representations of the transseries. We choose to resum the strong series. The result is
\begin{eqnarray}\label{eq:PoissonResumStrongTrans}
Z_{SU(2)}(g,\tilde{h})=\sum\limits_{n\in\mathbb{Z}}\int\limits_{-\infty}^\infty dm\left(\sigma_s^{p}(m)g^{\tilde{h}-1}\sum\limits_{j=0}^\infty c_j m^{2j-1} g^{-j}  +\sigma_s^{np}(m)m^{2-2\tilde{h}}e^{-gm^2/2}\right)e^{-2\pi i mn}\;.\nonumber\\ \;
\end{eqnarray}
From the $n=0$ part of the weak transseries that we know, we can determine that
\begin{eqnarray}
\sigma_s^{p}(m)=0\;\;,\;\;\;\sigma_s^{np}(m)=1\;.
\end{eqnarray}
(We can also determine the ambiguity in the contour of the integral after Poisson resummation, which passes through a singularity). We can then substitute these into \eqref{eq:PoissonResumStrongTrans}, and from here determining the weak coupling transseries parameters is identical to the contour decomposition we discussed in Section \ref{sec:parabolicCFResurgence}. Thus we can calculate all the transseries parameters in both the strong and weak coupling transseries just from knowledge of the weak $n=0$ transseries parameters.

This sheds some light on the strong coupling transseries. We see that the additional perturbative series we found in the deformed case arises due to the ambiguity in the contour in the integral representation of the partition function. Fixing the ambiguity, here just by fixing one of the transseries parameters in the weak transseries, has caused this strong perturbative series to have zero contribution, as we expected from \eqref{eq:SU2localisedPF}.

\section{Resurgence in deformed Yang-Mills theory: More on weak-coupling \\ transseries}\label{sec:WeakCouplingNoTop}

Thus far, for weak coupling, we have examined the perturbative series associated to a saddle expansion around the various saddles in the theory. In this section we examine the weak coupling expansion produced from expanding the partition function as a sum of (nearly) topological correlators\footnote{The correlators depend on the area of the base space, but aside from this depend only on topology.}.

Let us see what we mean explicitly in the $SU(2)$ case. Starting from the path integral expression for the partition function, we can expand it as
\begin{eqnarray}
Z_{SU(2)}(g,\tilde{h})&=&\int\mathcal{D}\Phi\mathcal{D}A\;e^{i\int_{\Sigma_h}\tr(\Phi\wedge F)+\frac{g}{2}\int_{\Sigma_h}\tr(\Phi^2)K+\delta\int_{\Sigma_h}\log(\tr(\Phi^2))K}\nonumber\\
&=&\int\mathcal{D}\Phi\mathcal{D}A\sum\limits_{j=0}^\infty\frac{\left(\frac{g}{2}\int_{\Sigma_h}\tr(\Phi^2)K\right)^j}{j!}\;e^{i\int_{\Sigma_h}\tr(\Phi\wedge F)+\delta\int_{\Sigma_h}\log(\tr(\Phi^2))K}\;.
\end{eqnarray}
We write this as
\begin{eqnarray}
Z_{SU(2)}(g,\tilde{h})&=&Z_{SU(2)}(g=0,\tilde{h})+\sum\limits_{j=1}^\infty\frac{\left(\frac{g}{2}\right)^j}{j!}\left\langle\left(\int_{\Sigma_h}\tr(\Phi^2)K\right)^j\right\rangle_{g=0}\;.
\end{eqnarray}
Here we have defined
\begin{eqnarray}
\left\langle\mathcal{O}\right\rangle_{g=0}=\int\mathcal{D}\Phi\mathcal{D}A\;\mathcal{O}\;e^{-i\int_{\Sigma_h}\tr(\Phi\wedge F)-\delta\int_{\Sigma_h}\log(\tr(\Phi^2))K}\;,
\end{eqnarray}
i.e. it is a correlation function which hasn't been normalised by dividing out the partition function. Thus we see that we can write the partition function as a perturbative series in weak coupling, summing over a particular set of correlator-like objects. Let us look at the undeformed and deformed cases in turn.

\subsection{Undeformed \texorpdfstring{$SU(2)$}. theory}

For the undeformed theory, the remaining action once we have expanded out the $\frac{g}{2}\int_{\Sigma_h}\tr(\Phi^2)K$ term, is given by
\begin{eqnarray}
S_{\mathrm{top}}=i\int_{\Sigma_h}\tr(\Phi\wedge F)\;.
\end{eqnarray}
This is a well studied topological theory (hence the ``top'' subscript on the above action), i.e. it has no metric dependence. See \cite{Blau:1993hj} for a nice review. The partition function is in fact the symplectic volume of the space of flat connections, and the correlators are observables in the topological theory that lie in the $4^{th}$ cohomology class of the space of flat connections. In summary
\begin{eqnarray}
Z_{SU(2)}(g=0,h)&=&\mathrm{Vol}\left(\mathcal{M}_{\mathcal{F}}(\Sigma_h,G)\right)\;,\nonumber\\
\int_{\Sigma_h}\tr(\Phi^2)K&\in& H^{4}\left(\mathcal{M}_{\mathcal{F}}(\Sigma_h,G)\right)\;.
\end{eqnarray}
Thus in the undeformed case, our perturbative series has an interpretation in terms of a sum over topological correlators of a particular topological theory. Note here we have written $h$ rather than $\tilde{h}$ as $\delta=0$.

However, as the reader has probably noticed, in the undeformed case, this is not a particularly useful series (for resurgence). We have that
\begin{eqnarray}
\left\langle\left(\int_{\Sigma_h}\tr(\Phi^2)K\right)^j\right\rangle_{g=0}=\sum\limits_{n\in\mathbb{Z}}\int\limits_{-\infty}^\infty d\Phi\;\Phi^{2-2h+2j}e^{-2\pi i n \Phi}\;.
\end{eqnarray}
For $h=0$ the integral is 0 for all $n$ (recall $2-2h+2j$ is an integer in this case), apart from $n=0$ where it diverges for all $j$. In this case we can regulate the divergence, say by regulating the integral limits so the integral goes from $-\frac{\beta}{2}$ to $\frac{\beta}{2}$. We now get a finite answer, for which we can do the $j$ summation. At the end we can take $\beta\rightarrow\infty$, and the result is
\begin{eqnarray}
Z_{SU(2)}(g,h)&\sim&\frac{2\sqrt{\pi}}{g^{3/2}}\;.
\end{eqnarray}
This is just the perturbative contribution to the transseries we found in the exact result in \eqref{eq:SU2weakPFExample}.

For $h>0$ we pick up extra terms coming from the pole that now exists at the origin. These terms are exactly the extra terms we derive in Appendix \ref{sec:hGeq1} for the case $h\geq 1$. But again this series is truncating, i.e. not an asymptotic divergent series, so Borel-\`Ecalle resummation is trivial.

Of course this is all we ever could have expected in the undeformed case. However, once we apply the deformation to the theory, things begin to get more interesting.

\subsection{Deformed \texorpdfstring{$SU(2)$}. theory}\label{sec:deformedSU2WeakZeta}

In the deformed case we now have
\begin{eqnarray}
\left\langle\left(\int_{\Sigma_h}\tr(\Phi^2)K\right)^j\right\rangle_{g=0}=\sum\limits_{n\in\mathbb{Z}}\int\limits_{-\infty}^\infty d\Phi\;\Phi^{2-2\tilde{h}+2j}e^{-2\pi i n \Phi}\;.
\end{eqnarray}
To calculate this we need to consider three cases; $n=0$, $n$ positive and $n$ negative. For $n=0$ the integrals diverge. We again need to regulate as in the previous subsection, and the result after summing over $j$ is
\begin{eqnarray}
(1+e^{-2\pi i \tilde{h}})2^{1/2-2\tilde{h}}\Gamma(3/2-\tilde{h})g^{-3/2+\tilde{h}}\;\;.
\end{eqnarray}
This is the contribution to the perturbative and non-perturbative saddles in the $n=0$ topological sector.

Recall we are taking the contour to pass over the branch point at the origin. For negative $n$ we can close the contour in the upper half plane. The contour encloses no singularities or branch points, and this for positive $n$ all the integrals are 0.

For positive $n$ however, we must close the contour in the negative half plane, around the branch-cut. In this case the contribution to the correlator is
\begin{eqnarray}
&&\sum\limits_{n=1}^\infty e^{\pi i \tilde{h}}\left(1-e^{-4\pi i \tilde{h}}\right)\left(\frac{i}{2\pi n } \right)^{3-2\tilde{h}}\frac{(g/(8\pi^2n^2))^j}{j!}\Gamma(3+2j-2\tilde{h})\nonumber\\
&=&e^{\pi i \tilde{h}}\left(1-e^{-4\pi i \tilde{h}}\right)\left(\frac{i}{2\pi } \right)^{3-2\tilde{h}}\frac{(g/(8\pi^2))^j}{j!}\zeta(3+2j-2\tilde{h})\Gamma(3+2j-2\tilde{h})\;.
\end{eqnarray}
Putting this all together, we have the perturbative series
\begin{eqnarray}\label{eq:deformedSU2PF}
Z_{SU(2)}^{\mathrm{pert}}(g,\tilde{h})&=&(1+e^{-2\pi i \tilde{h}})2^{1/2-2\tilde{h}}\Gamma(3/2-\tilde{h})g^{-3/2+\tilde{h}} \\
&&\;\;\;\;+e^{\pi i \tilde{h}}\left(1-e^{-4\pi i \tilde{h}}\right)\left(\frac{i}{2\pi }\right)^{3-2\tilde{h}}\nonumber\\
&&\;\;\;\;\;\;\;\;\;\;\;\;\times\sum\limits_{j=0}^{\infty}\frac{(g/(8\pi^2))^j}{j!}\zeta(3+2j-2\tilde{h})\Gamma(3+2j-2\tilde{h})\;.\nonumber
\end{eqnarray}
Thus we have an alternative perturbative asymptotic expansion for the partition function. In fact, this is just the sum of the perturbative series associated to each of the perturbative saddles in all the topological sectors with positive $n$. But of course, the transseries, unlike the saddle decomposition, doesn't distinguish contributions with identical exponential part. Thus all the perturbative saddles are included in the transseries as just one contribution.

\subsubsection{Weak coupling resurgence analysis I: Non-perturbative data from perturbative data}\label{sec:SU2weakresurgenceanalysis}

We now perform a resurgence analysis of \eqref{eq:deformedSU2PF}. Here we want to ask whether we can recover the non-perturbative part of the transseries from this perturbative part alone, for which the answer will be yes. In the next section we will ask the converse question, can the perturbative data be derived from the non-perturbative data, for which the answer is more complicated.

There are various ways of applying resurgence to \eqref{eq:deformedSU2PF}. The most basic is just to use the standard Borel resummation, dividing out by the factorial part, and then applying the Laplace transformation to get to a resummed result. We will look at this way in a moment. However, for example, one can divide our by something more complicated, and change the measure of the Laplace transform appropriately to get the resummed result (see \cite{Russo:2012kj,Hatsuda:2015owa}). We mention this here as, in this case, it is quite tempting to divide out by $\Gamma(3+2j-2\tilde{h})$, or $\zeta(3+2j-2\tilde{h})\Gamma(3+2j-2\tilde{h})$. However calculating the non-perturbative contributions to the transseries using either of these methods turns out to be somewhat non-standard\footnote{To be more precise, if one divides out by too much, rather than a function on the Borel plane with singularities, one ends up with an entire function with exponential part that decays faster than the Borel measure. In this case the non-perturbative contributions manifest themselves through thimble decomposition of the Borel inverse transform, rather through discontinuities on the Borel plane.}.

Let us now perform a resurgence analysis of the weak perturbative series. First we need to Borel resum \eqref{eq:deformedSU2PF}. We focus on the contents of the sum, so write
\begin{eqnarray}
Z_{SU(2)}^{\mathrm{pert}}(g,\tilde{h})&=&(1+e^{-2\pi i \tilde{h}})2^{1/2-2\tilde{h}}\Gamma(3/2-\tilde{h})g^{-3/2+\tilde{h}} \nonumber\\
&&\;\;\;\;+e^{\pi i \tilde{h}}\left(1-e^{-4\pi i \tilde{h}}\right)\left(\frac{i}{2\pi }\right)^{3-2\tilde{h}}\Tilde{Z}_{SU(2)}^{\mathrm{pert}}(g,\tilde{h})\;.
\end{eqnarray}
Our chosen way to Borel resum is
\begin{eqnarray}\label{eq:AsymptoticSum1}
\Tilde{Z}_{SU(2)}^{\mathrm{pert}}(g,\tilde{h})&=&\sum\limits_{j=0}^{\infty}\frac{(g/(8\pi^2))^j}{j!}\zeta(3+2j-2\tilde{h})\Gamma(3+2j-2\tilde{h})\\
&=&\left(\frac{8\pi^2}{g}\right)^{\frac{3}{2}}\sum\limits_{j=0}^{\infty}\int\limits_0^\infty dt\;e^{-\frac{8\pi^2t}{g}}t^{j+1/2}\frac{
\zeta(3+2j-2\tilde{h})\Gamma(3+2j-2\tilde{h})}{\Gamma(j+1)\Gamma(j+3/2)}\;.\nonumber
\end{eqnarray}
Here we have chosen a Borel summation that divides each term in the series by $\Gamma(j+3/2)$. We can then use the following representation of the zeta function:
\begin{eqnarray}\label{eq:zetaIntegralRep}
\zeta(z)\Gamma(z)=\int\limits_0^\infty dx\;\frac{x^{z-1}}{e^x-1}\;.
\end{eqnarray}
This allows us to write \eqref{eq:AsymptoticSum1} as
\begin{eqnarray}\label{eq:AsymptoticSum2}
\Tilde{Z}_{SU(2)}^{\mathrm{pert}}(g,\tilde{h})&=&\left(\frac{8\pi^2}{g}\right)^{\frac{3}{2}}\sum\limits_{j=0}^{\infty}\int\limits_0^\infty dt\;e^{-\frac{8\pi^2t}{g}}\frac{
t^{j+1/2}}{\Gamma(j+1)\Gamma(j+3/2)}\int\limits_0^\infty dx\;\frac{x^{2+2j-2\tilde{h}}}{e^x-1}\nonumber\\
&=&\left(\frac{8\pi^2}{g}\right)^{\frac{3}{2}}\int\limits_0^\infty dt\;e^{-\frac{8\pi^2t}{g}} \int\limits_0^\infty dx\;\frac{x^{1-2\tilde{h}}\sinh(2\sqrt{t}x)}{\sqrt{\pi}(e^x-1)}\;.
\end{eqnarray}
Here we have used the identity
\begin{eqnarray}
\sum\limits_{j=0}^{\infty}\frac{x^j}{\Gamma(j+1)\Gamma(j+3/2)}=\frac{\sinh(2\sqrt{x})}{\sqrt{x}\sqrt{\pi}}\;.
\end{eqnarray}
We also now make use of the integral representation of the Hurwitz zeta function:
\begin{eqnarray}\label{eq:HzetaIntegralRep}
\Gamma(s)\zeta(s,a)=\int\limits_0^\infty dx \;\frac{x^{s-1}e^{-ax}}{1-e^{-x}}\;.
\end{eqnarray}
We can use this identity, after writing the $\sinh(2\sqrt{t}x)$ function in \eqref{eq:AsymptoticSum2} in terms of exponentials, to write \eqref{eq:AsymptoticSum2} as
\begin{eqnarray}\label{eq:AsymptoticSum3}
\Tilde{Z}_{SU(2)}^{\mathrm{pert}}(g,\tilde{h})&=&\left(\frac{8\pi^2}{g}\right)^{\frac{3}{2}}\frac{\Gamma(2-2\tilde{h})}{2\sqrt{\pi}}\int\limits_0^\infty dt\;e^{-\frac{8\pi^2t}{g}}\left(\zeta(2-2\tilde{h},1-2\sqrt{t})-\zeta(2-2\tilde{h},1+2\sqrt{t})\right)\nonumber\\
&=&\left(\frac{8\pi^2}{g}\right)^{\frac{3}{2}}\frac{\Gamma(2-2\tilde{h})}{8\sqrt{\pi}}\int\limits_0^\infty dt\;e^{-\frac{2\pi^2t}{g}}\left(\zeta(2-2\tilde{h},1-\sqrt{t})-\zeta(2-2\tilde{h},1+\sqrt{t})\right)\;.\nonumber\\
\;
\end{eqnarray}
Thus, using Borel resummation, we can write our perturbative series as the Laplace transformation of a particular function, and as desired, this function has branch cuts starting at the locations of the non-perturbative contributions. In particular, $\zeta(s,t)$ has cuts on in the $t$ plane at $t=0,-1,-2,-3,\dots$ .

In order to retrieve the non-perturbative data contained in perturbative series we now need to calculate the ambiguities in the Laplace transform. To do this we need let us first use the standard formula for the Hurwitz zeta function
\begin{eqnarray}\label{eq:HurZetaSum}
\zeta(s,a)=\sum\limits_{n=0}^\infty \frac{1}{(n+a)^s}\;.
\end{eqnarray}
Substituting this sum into \eqref{eq:AsymptoticSum3} we see that we can write the inverse Borel-transform as an infinite sum of inverse Borel-transforms, one corresponding to each non-perturbative contribution.

The calculation of the ambiguity is now elementary. The discontinuity across the cuts on the positive real axis is given by
\begin{eqnarray}
\mathrm{Disc}_0(g,\tilde{h})&=&\left(\frac{8\pi^2}{g}\right)^{\frac{3}{2}}\frac{\Gamma(2-2\tilde{h})}{2\sqrt{\pi}}\left(\int\limits_0^{\infty+i\epsilon}-\int\limits_0^{\infty-i\epsilon}\;\right) dt\;e^{-\frac{2\pi^2t}{g}}\sum\limits_{n=1}^\infty\frac{1}{(n-\sqrt{t})^{2-2\tilde{h}}}\nonumber\\
&=&(1-e^{-4\pi i\tilde{h}})\left(\frac{8\pi^2}{g}\right)^{\frac{3}{2}}\frac{\Gamma(2-2\tilde{h})}{2\sqrt{\pi}}\sum\limits_{n=1}^\infty\int\limits_{n^2}^{\infty+i\epsilon} dt\;e^{-\frac{2\pi^2t}{g}}\frac{1}{(n-\sqrt{t})^{2-2\tilde{h}}}\;.\nonumber\\
\;
\end{eqnarray}
By a series of involved transformations the integrand of this integral can be shown to be of the form of \eqref{eq:ParabolBorelPlane1}. In other words, as one would expect, the Borel transform of the perturbative series \eqref{eq:deformedSU2PF} is that of the infinite sum of the Borel-transforms over each sector of \eqref{eq:SU2WeakTransSumOvern}.

In order to extract the perturbative series for each non-perturbative part we shift $t\rightarrow t+n$ and use the expansion
\begin{eqnarray}
\left(\frac{1}{n-\sqrt{t+n^2}}\right)^{2-2\tilde{h}}=e^{-2\pi i \tilde{h}} (2\tilde{h}-2)\left(\frac{2n}{t}\right)^{2-2\tilde{h}} \sum\limits_{j=0}^\infty \frac{\Gamma(2j-2+2\tilde{h})}{\Gamma(j+1)\Gamma(j-1+2\tilde{h})}\left(-\frac{t}{(2n)^2}\right)^j\;.\nonumber\\
\;
\end{eqnarray}
After performing the integral and manipulating the Gamma functions we end up with the non-perturbative part of \eqref{eq:SU2PvenTransseries} up to the transseries parameter.

From here things are as before. In order to determine the transseries parameter we need to use something like the strong-weak consistency already discussed, or an analysis of the original integral as discussed in Section \ref{sec:parabolicCFResurgence}. Having determined them we will be able to analytically continue the deformation back to 0, retaining the non-perturbative data, as we did in Section \ref{sec:SU(2)resurgenceAnalysis}.

\subsubsection{Weak coupling resurgence analysis I: Perturbative data from non-perturbative data}\label{sec:pertfromnonpert}

We now briefly comment on what happens when we now try and go the other way round, calculating the perturbative contribution to the transseries from the non-perturbative part alone. As should be clear, if we try and calculate the perturbative contribution from the $n^{th}$ non-perturbative contribution we will not find the full perturbative contribution in \eqref{eq:deformedSU2PF}. Instead we will find only part of it. To be more precise, we will find the contribution to the perturbative series \eqref{eq:deformedSU2PF} that is coming from the perturbative saddle in the $n^{th}$ topological sector. From the discussion of saddles and topology in Sections \ref{sec:undeformedSaddles} and \ref{sec:NewSaddles} this is obvious, but it is not obvious if we were only to have access to the perturbative series \eqref{eq:deformedSU2PF}.

The takeaway lesson for this section is that a knowledge of the saddles and their topology really does help to untangle some of the mysteries of the transseries. In this case we have an infinite number of saddles with action equal to 0 up to quantum corrections, which all contribute to the same term in the transseries. Without knowledge of the saddles and their topology we have a strange phenomenon where we can calculate all the non-perturbative data from the perturbative data, but the other way round we can only calculate part of the perturbative data. But from studying the saddles and their topology, we can see the reason is that there are in fact an infinite number of perturbative saddles, one for each topological sector. Thus the perturbative contribution to the transseries contains information in this case about all the topological sectors, whilst each non-perturbative saddle only contains information about only one topological sector. Hence one way the analysis is possible, but not the other.

\subsection{Deformed \texorpdfstring{$U(2)$}. theory}

Here we briefly consider the case of $U(2)$, focusing on the contributions with $n_1=1-n_2$. Without knowledge of the saddles and their topology, one might be tempted to think that these contributions are all in the same topological sector, after all they all have the same theta angle dependence. The saddles in this subset have action (excluding quantum corrections)
\begin{eqnarray}
S(g,\theta)&=&\frac{((n_1+n_2)\pi)^2}{g}-i\theta(n_1+n_2)=\frac{\pi^2}{g}-i\theta\;,\nonumber\\
S(g,\theta)&=&\frac{(2\pi)^2(n_1^2+n_2^2)}{2g}-i\theta(n_1+n_2)=\frac{(2\pi)^2(2n^2-2n+1)}{2g}-i\theta\;.
\end{eqnarray}
In the second line we have briefly defined $n=n_1$. Thus we see the saddles with the smallest action are of the first type, and all the saddles of the first type have the same action. Thus we find very the same phenomena as we found in the $SU(2)$ theory in other topological sectors of $U(2)$ as well. In fact, it is easy to see that the first kind of saddle will have identical action (up to quantum corrections) for all contributions with the same theta angle dependence, that is $n_1=c-n_2$, for any $c$.

In the case of $U(2)$, when analysing the transseries without knowledge of the saddles and their topology, one may be tempted to think that topological sectors are graded only by theta angle dependence. Within the set of saddles with given theta angle dependence, one may again be tempted to think that the contribution to transseries with action of the first kind of saddle above is coming from only one saddle. In this case, one would be surprised to find that from this contribution one could calculate the contributions to the transseries from all the other sectors with equal theta angle dependence, but not the other way around. Moreover, one cannot calculate the contribution from one of these other non-minimal non-perturbative contributions with a given theta angle dependence from a different one with the same theta angle dependence. The answer again lies in the topology of the saddles. There is a finer topological grading than theta angle dependence, and for a given theta angle dependence there are an infinite number of saddles, one for each topological sector, that have the same action.

Let us briefly see this for the $n_1=1-n_2$ sector. Starting from \eqref{eq:1topoStartingPoint} we see that the $n_1=1-n_2$ part is given by
\begin{eqnarray}
Z_{U(2)}\vert_{n_1=1-n_2}(g,h,\theta)&=&\sqrt{\frac{\pi}{g}}e^{-\frac{\pi^2}{g}-i\theta} \sum\limits_{n\in\mathbb{Z}} \int\limits_{-\infty}^\infty dx\; x^{2-2\tilde{h}}e^{-\pi ix (2n-1)-\frac{g}{2}x^2}\;.
\end{eqnarray}
The sum and integral here are very similar to what we have already calculated in the $SU(2)$ case. Shifting $n$ by $\frac{1}{2}$ (thus dropping the $n=0$ contribution), and substituting $g\rightarrow\frac{g}{2}$ into the $SU(2)$ case we can extract the sum and integral we need for the $U(2)$ case. The result is
\begin{eqnarray}
Z_{U(2)}\vert_{n_1=1-n_2}(g,h,\theta)&=&\sqrt{\frac{\pi}{g}}e^{-\frac{\pi^2}{g}-i\theta}e^{\pi i \tilde{h}}\left(1-e^{-4\pi i \tilde{h}}\right)\left(\frac{i}{2\pi }\right)^{3-2\tilde{h}}\\
&&\;\;\;\;\;\;\;\;\;\;\;\;\times\sum\limits_{j=0}^{\infty}\frac{(g/(16\pi^2))^j}{j!}\zeta(3+2j-2\tilde{h},1/2)\Gamma(3+2j-2\tilde{h})\;.\nonumber
\end{eqnarray}
Writing this as
\begin{eqnarray}
Z_{U(2)}\vert_{n_1=1-n_2}(g,h,\theta)&=&\sqrt{\frac{\pi}{g}}e^{-\frac{\pi^2}{g}-i\theta}e^{\pi i \tilde{h}}\left(1-e^{-4\pi i \tilde{h}}\right)\left(\frac{i}{2\pi }\right)^{3-2\tilde{h}}\Tilde{Z}_{U(2)}^{\mathrm{pert}}\vert_{n_1=1-n_2}(g,h)\;,\nonumber\\
\;
\end{eqnarray}
we can perform a Borel-\`Ecalle resummation as before. The result is
\begin{eqnarray}\label{eq:1toposectorBorelSum}
\Tilde{Z}_{U(2)}^{\mathrm{pert}}\vert_{n_1=1-n_2}(g,h)&=&\left(\frac{16\pi^2}{g}\right)^{\frac{3}{2}}\frac{\Gamma(2-2h)}{8\sqrt{\pi}}\\
&&\;\;\;\;\;\;\times\int\limits_0^\infty dt\;e^{-\frac{4\pi^2t}{g}}\left(\zeta(2-2h,1/2-\sqrt{t})-\zeta(2-2h,\sqrt{t}+1/2)\right)\;.\nonumber
\;
\end{eqnarray}
This has singularities in the right location and from it we can calculate all the contributions to the transseries with the same theta angle dependence. But as before, this does not work the other way around.

\section{Analysis for higher \texorpdfstring{$N$}. }\label{sec:higherNAnalysis}

For higher $N$ in principal things things are much the same as for $N=2$. However our partition function (\eqref{eq:SUNPartitionFunctionIntegralRep} or \eqref{eq:UNPartitionFunctionIntegralRep}) now involves multiple infinite sums, and higher dimensional integrals, which practically complicates things greatly. The higher dimensional integrals turn out to be very tricky to solve, and regulating multiple-dimensional infinite sums is difficult to do. Here we will briefly look at the strong coupling $SU(3)$ case, where we can see some of these issues come into play.

The integral representation for the $SU(3)$ partition function is given by
\begin{eqnarray}
Z_{SU(3)}(g,\tilde{h})&=&\sum\limits_{n_1,n_2\in\mathbb{Z}}\int d\Phi_1\,d\Phi_2\; \Phi_1^{2-2\tilde{h}}\Phi_2^{2-2\tilde{h}}\left(\Phi_1-\Phi_2\right)^{2-2\tilde{h}}e^{2\pi i(n_1\Phi_1+n_2\Phi_2)-\frac{g}{2}\left(\Phi_1^2+\Phi_2^2\right)}\;.\nonumber\\
&\;&
\end{eqnarray}
We will find the strong coupling transseries representation by a direct analysis of the integral.

First we slit the sums up, and make appropriate substitutions such that they are over positive integers, so that the results of the sums will give zeta functions. This gives us
\begin{eqnarray}\label{eq:SU3PFIntegralRep}
&&\;\;\;Z_{SU(3)}(g,\tilde{h})=\int d\Phi_1\,d\Phi_2\; \Phi_1^{2-2\tilde{h}}\Phi_2^{2-2\tilde{h}}\left(\Phi_1-\Phi_2\right)^{2-2\tilde{h}}e^{-\frac{g}{2}\left(\Phi_1^2+\Phi_2^2\right)}\\
&&+(1+e^{-4\pi i \tilde{h}})\sum\limits_{n=1}^\infty \int d\Phi_1\,d\Phi_2\; \Phi_1^{2-2\tilde{h}}\Phi_2^{2-2\tilde{h}}\left(\Phi_1-\Phi_2\right)^{2-2\tilde{h}}e^{2\pi in_1\Phi_1-\frac{g}{2}\left(\Phi_1^2+\Phi_2^2\right)}\nonumber\\
&&+(e^{-2\pi i \tilde{h}}+e^{-4\pi i \tilde{h}})\sum\limits_{n=1}^\infty \int d\Phi_1\,d\Phi_2\; \Phi_1^{2-2\tilde{h}}\Phi_2^{2-2\tilde{h}}\left(\Phi_1+\Phi_2\right)^{2-2\tilde{h}}e^{2\pi in_1\Phi_1-\frac{g}{2}\left(\Phi_1^2+\Phi_2^2\right)}\nonumber\\
&&+(1+e^{-6\pi i \tilde{h}})\sum\limits_{n_1=1}^\infty\sum\limits_{n_2=1}^\infty\int d\Phi_1\,d\Phi_2\; \Phi_1^{2-2\tilde{h}}\Phi_2^{2-2\tilde{h}}\left(\Phi_1-\Phi_2\right)^{2-2\tilde{h}}e^{2\pi i(n_1\Phi_1+n_2\Phi_2)-\frac{g}{2}\left(\Phi_1^2+\Phi_2^2\right)}\;.\nonumber
\end{eqnarray}
These terms can now be expanded to produce a perturbative series, using the identity
\begin{eqnarray}
\sum\limits_{n=1}^\infty e^{2\pi i n \Phi}=1+\sum\limits_{m=1}^\infty\frac{(2\pi i\Phi)^m\zeta(-m)}{m!}\;,
\end{eqnarray}
and the result of the integral
\begin{eqnarray}
&&\int\limits_{-\infty}^\infty dx\int\limits_{-\infty}^\infty dy\;\; x^a y^b (x-y)^c e^{-g/4(x^2+y^2)}\\
&&\;\;\;\;\;\;\;\;\;\;\;\;\;\;\;\;\;\;\;\;\;\;\;\;\;=\pi ^{3/2} 2^{a-1} e^{-i \pi  b} g^{\frac{1}{2} (-a-b-c-2)}\left(A(x)+2^{b+c}(B(x)+C(x))\right)\;.\nonumber
\end{eqnarray}
Here we have
\small{
\begin{eqnarray}
A(x)&=&\left(-\frac{2 \left(-1+e^{2 i \pi  b}\right) \Gamma (b+1) \left((-1)^{a+b+c}-1\right) \csc (\pi  (b+c)) \Gamma \left(\frac{1}{2} (a+b+c+2)\right) }{\Gamma (-c)}\right.\nonumber \\
&&\left.\;\;\;\;\;\;\;\;\;\;\;\;\;\;\times\, _3\tilde{F}_2\left(\frac{b+1}{2},\frac{b+2}{2},\frac{1}{2} (a+b+c+2);\frac{1}{2} (b+c+2),\frac{1}{2}
   (b+c+3);-1\right)\right)\;,\nonumber\\
B(x)&=&\left((-1)^{a+1}+1\right) c \Gamma \left(\frac{a}{2}+1\right) \left(e^{2 i \pi  b}-e^{i \pi  (b+c)}\right) \csc \left(\frac{1}{2} \pi  (b+c)\right) \nonumber\\ &&\;\;\;\;\;\;\;\;\;\;\;\;\;\;\times\,
   _3\tilde{F}_2\left(\frac{a}{2}+1,\frac{1}{2}-\frac{c}{2},1-\frac{c}{2};\frac{3}{2},-\frac{b}{2}-\frac{c}{2}+1;-1\right)\;,\\
   C(x)&=&2 \left((-1)^a+1\right) \Gamma \left(\frac{a+1}{2}\right) \left(e^{i \pi  (b+c)}+e^{2 i \pi  b}\right) \sec \left(\frac{1}{2} \pi  (b+c)\right) \nonumber\\ &&\;\;\;\;\;\;\;\;\;\;\;\;\;\;\times\, _3\tilde{F}_2\left(\frac{a+1}{2},\frac{1-c}{2},-\frac{c}{2};\frac{1}{2},\frac{1}{2} (-b-c+1);-1\right)\;.\nonumber
\end{eqnarray}}
\normalsize

It's a simple matter of applying these formula to \eqref{eq:SU3PFIntegralRep} to get a perturbation series in $\frac{1}{g}$. The result however is a very long formula which we shall not write down here as it is not particularly illuminating. Importantly the terms do diverge factorially and are non-alternating, so we can apply a resurgence analysis to it.

In summary, with higher $N$ the deformation does indeed introduce new saddles and render the perturbative series in each sector divergent asymptotic. However it reintroduces the problem of the complexity involved in deriving the asymptotic series themselves. Whilst for $N=3$ we still have access to the explicit perturbation series, for higher $N$ we would probably need to turn to numerical methods. The takeaway lesson is that, whilst the deformation renders the truncating perturbative series asymptotic and divergent, thus uncovering a resurgence structure, perturbation theory is generally hard.

\section{Factorisation and partial differential equations}\label{sec:diff_equations_topological_sectors}

In this section we now turn our attention away from Borel-\`Ecalle resummation and towards structures that are not restricted to a single column of the resurgence triangle. As explained in the introduction, and as we have seen throughout this paper, the normal caveat to the stronger version of the resurgence program is that one cannot expect to be able to derive contributions to the transseries in different topological sectors from the perturbative data alone. However, as we explained, in \cite{Dunne:2014bca,Dunne:2013ada,Gahramanov:2015yxk,Dorigoni:2019kux} various additional structures have been looked at that can be combined with resurgence to take such a sideways step in the resurgence triangle. In this section we will unpack three such structures that are present in 2d YM. These are factorisation, various partial differential equations for the $N=2$ case, and a way of writing higher $N$ partition functions in terms of the $N=1$ partition function. We will look at each of these in turn, and consider some applications including ways of taking a sideways step, and low-order/low-order resurgence.

\subsection{Factorisation}\label{sec:Factorisation}

At large $N$ the 2d YM partition function factorises (or at least is conjectured to), satisfying the OSV conjecture \cite{Ooguri_2004}:
\begin{eqnarray}
Z_{YM}=\vert\Psi^{\mathrm{top}}\vert^2\;.
\end{eqnarray}
Here $\Psi^{\mathrm{top}}$ is the partition function of a topological string on a particular local Calabi-Yau threefold. In this case we have a factorisation formula very similar to the cases of 3-dimensional $\mathcal{N}=2$ and 4-dimensional $\mathcal{N}=2$ supersymmetric Yang-Mills studied in \cite{Dorigoni:2019kux}. In that work it was demonstrated how to use such a factorisation formula to take a sideways step in the resurgence triangle.

For finite $N$ such factorisation is not possible in general, but for $U(2)$ we can find similar factorization equations that our partition functions satisfy that will allow us to move sideways in the resurgence triangle. Let us start with $U(2)$.

We work with the strong coupling representation of the partition function to derive the factorization equation for $\tilde{h}=0$. Starting from \eqref{eq:UNPartitionFunctionIntegralRep} we have
\begin{eqnarray}\label{eq:dunneUnsalU2}
Z_{U(2)}(g,0,\theta)&=&\sum\limits_{m_1,m_2\in\mathbb{Z}}(m_1-m_2)^2 e^{-\frac{g}{2}((m_1-\theta/2\pi)^2+(m_2-\theta/2\pi)^2)}\nonumber\\
&=&\left(-4\frac{\partial}{\partial g}-\frac{4\pi^2}{g^2}\frac{\partial^2}{\partial\theta^2}-\frac{2}{g}\right)\left(\sum\limits_{m_1,m_2\in\mathbb{Z}} e^{-\frac{g}{2}((m_1-\theta/2\pi)^2+(m_2-\theta/2\pi)^2)}\right)\nonumber\\
&=&\left(-4\frac{\partial}{\partial g}-\frac{4\pi^2}{g^2}\frac{\partial^2}{\partial\theta^2}-\frac{2}{g}\right)\left(\sum\limits_{n\in\mathbb{Z}}e^{-\frac{g}{2}((n-\theta/2\pi)^2)}\right)^2\;.
\end{eqnarray}
We have found a way of writing the $U(2)$ partition function in terms of a differential operator acting on a factorised object, similar to the $\tilde{h}=1$ partition function. Generalising to $\tilde{h}=1$ is obviously trivial. For higher $\tilde{h}$ we need an integral operator rather than a differential operator, but this is quite easy to do. We can now use this to take a sideways step in the resurgence triangle.

Let us demonstrate this by calculating the contribution to the $(1,0)$ and $(0,1)$ topological sectors from the contribution in the $(1,1)$ topological sector, in the weak coupling case (using notation $(n_1,n_2)$). From our knowledge of the saddle point of the action we can write \eqref{eq:dunneUnsalU2} in the form
\begin{eqnarray}\label{eq:dunneUnsalCalc}
Z_{U(2)}(g,0,\theta)&=&\frac{4\pi}{g^2}+4\pi e^{-\frac{4\pi^2}{g}} \left(\frac{1}{g^2}-\frac{8\pi^2}{g^3} \right)+Z^{(1,0)}(g)e^{-\frac{2\pi^2}{g}+i\theta}+Z^{(0,1)}(g)e^{-\frac{2\pi^2}{g}-i\theta}+\dots \\
&=&\left(-4\frac{\partial}{\partial g}-\frac{4\pi^2}{g^2}\frac{\partial^2}{\partial\theta^2}-\frac{2}{g}\right)\left[\left(Z_0(g)+Z_1(g)e^{-\frac{2\pi^2}{g}+i\theta}+Z_{-1}(g)e^{-\frac{2\pi^2}{g}-i\theta}+\dots\right)^2\right]\;.\nonumber
\end{eqnarray}
Here $Z_n(g)$ is the $n^{th}$ component of the sum in the final line of \eqref{eq:dunneUnsalU2}. We want to calculate $Z^{(1,0)}(g)$ and $Z^{(0,1)}(g)$. The tactic is to calculate $Z_0(g)$, $Z_1(g)$ and $Z_{-1}(g)$ by separating out the coefficients of the exponentials in the equation, which gives us
\begin{eqnarray}
Z_0(g)&=&\sqrt{\frac{2\pi}{g}}\;,\nonumber \\
Z_1(g)&=&Z_{-1}(g)=\sqrt{\frac{\pi}{g}}\;.
\end{eqnarray}
Then we can substitute these back into \eqref{eq:dunneUnsalCalc} to find
\begin{eqnarray}
Z^{(1,0)}(g)=Z^{(0,1)}(g)=4\pi \left(\frac{1}{g^2}-\frac{2\pi^2}{g^3}\right)\;.
\end{eqnarray}
This is exactly the contribution we expected from \eqref{eq:U2weakPFExample}.

We have found our first additional structure that can allow us to make a sideways step in the resurgence triangle. Unfortunately we have not found a way of extending this procedure to other $U(N)$ or $SU(N)$ gauge groups (except for infinite $N$ by the OSV conjecture).

\subsection{Differential equations the partition function satisfies}\label{sec:DifferentialEquations}

In \cite{Dorigoni:2019kux} it was noted that the $tt^\ast$ equations in 2-dimensional $\mathcal{N}=(2,2)$ theories provide a relation in that context to take a sideways step in the resurgence triangle. The $tt^\ast$ equations are partial differential equations satisfied by the partition functions of such theories. We now turn to look at partial differential equations that are satisfied by the $N=2$ partition functions of 2d YM.

We have found multiple such equations; a number are satisfied by the $SU(2)$ partition function as well as the $U(2)$. In the next subsection we'll discuss the equations obeyed by the $SU(2)$ partition function, equivalent to the sum of the $(n,-n)$ topological sectors of the $U(2)$ partition function, and their applications. Then in Section \ref{sec:u2diffeq} we'll discuss the $U(2)$ specific partial differential equations.

\subsubsection{\texorpdfstring{$SU(2)$}. partition function}

We first look for a differential operator $\mathcal{F}$ such that we can write a formula of the form
\begin{eqnarray}\label{eq:recursionform1}
Z_{SU(2)}(g,\tilde{h})=\mathcal{F}[Z_{SU(2)}^{n=0}(g,\tilde{h})]\;.
\end{eqnarray}
In other words we are looking for a differential operator that we can apply to the perturbative data to get the whole partition function. Here we are working with $SU(2)$, or equivalently, the sum of the $(n,-n)$ topological sectors of $U(2)$.

We can derive such an operator and formula in the following way. We start with the integral representation of the $SU(2)$ partition function \eqref{eq:SUNPartitionFunctionIntegralRep} and write
\begin{eqnarray}\label{eq:recursionPert}
Z_{SU(2)}(g,\tilde{h})&=&\sum\limits_{n\in\mathbb{Z}}\int\limits_{-\infty}^\infty d\Phi\;\Phi^{2-2\tilde{h}}e^{-2\pi in\Phi-\frac{g}{2}\Phi^2}\nonumber\\
&=&\int\limits_{-\infty}^\infty d\Phi\;\Phi^{2-2\tilde{h}}e^{-\frac{g}{2}\Phi^2}+\sum\limits_{n\neq0}\int\limits_{-\infty}^\infty d\Phi\;\Phi^{2-2\tilde{h}}e^{-2\pi in\Phi-\frac{g}{2}\Phi^2}\nonumber\\
&=&\int\limits_{-\infty}^\infty d\Phi\;\Phi^{2-2\tilde{h}}e^{-\frac{g}{2}\Phi^2}+\sum\limits_{n\neq0}\sum\limits_{j=0}^\infty\int\limits_{-\infty}^\infty d\Phi\;\Phi^{2-2\tilde{h}}\frac{(-2\pi in\Phi)^j}{j!}e^{-\frac{g}{2}\Phi^2}\\
&=&\int\limits_{-\infty}^\infty d\Phi\;\Phi^{2-2\tilde{h}}e^{-\frac{g}{2}\Phi^2}+\sum\limits_{n=1}^\infty\sum\limits_{j=0}^\infty(1+(-1)^j)\int\limits_{-\infty}^\infty d\Phi\;\Phi^{2-2\tilde{h}}\frac{(2\pi in\Phi)^j}{j!}e^{-\frac{g}{2}\Phi^2}\;.\nonumber
\end{eqnarray}
Here we have Taylor expanded $e^{2\pi in\Phi}$, and to get to the final line have used that the sum over negative integers is the sum over the positive integers with a factor of $-1$ inserted appropriately. Now the factor $(1+(-1)^j)$ will only survive for even $j$, so we can substitute $j\rightarrow 2j$. Proceeding we have
\begin{eqnarray}\label{eq:recursionPert2}
Z_{SU(2)}(g,\tilde{h})&=&\int\limits_{-\infty}^\infty d\Phi\;\Phi^{2-2\tilde{h}}e^{-\frac{g}{2}\Phi^2}+2\sum\limits_{n=1}^\infty\sum\limits_{j=0}^\infty\int\limits_{-\infty}^\infty d\Phi\;\Phi^{2-2\tilde{h}}\frac{(2\pi in\Phi)^{2j}}{(2j)!}e^{-\frac{g}{2}\Phi^2}\nonumber\\
&=&\int\limits_{-\infty}^\infty d\Phi\;\Phi^{2-2\tilde{h}}e^{-\frac{g}{2}\Phi^2}+2\sum\limits_{n=1}^\infty\sum\limits_{j=0}^\infty\frac{\left(8\pi^2n^2\frac{\partial}{\partial g}\right)^{j}}{(2j)!}\int\limits_{-\infty}^\infty d\Phi\;\Phi^{2-2\tilde{h}}e^{-\frac{g}{2}\Phi^2}\nonumber\\
&=&\left(1+2\sum\limits_{n=1}^\infty\cosh\left(2\sqrt{2}\pi n\sqrt{\frac{\partial}{\partial g}}\right)\right)[Z_{SU(2)}^{n=0}(g,\tilde{h})]\nonumber\\
&=&\sum\limits_{n\in\mathbb{Z}}^\infty\cosh\left(2\sqrt{2}\pi n\sqrt{\frac{\partial}{\partial g}}\right)[Z_{SU(2)}^{n=0}(g,\tilde{h})]\;.
\end{eqnarray}
Here we have used the summation formula
\begin{eqnarray}
\sum\limits_{j=0}^\infty\frac{x^j}{(2j)!}=\cosh(\sqrt{x})\;.
\end{eqnarray}
Thus we have a formula of the form \eqref{eq:recursionform1}.

Let us note here that this formula is actually much more general than just a formula for the full partition function from the $n=0$ part alone. Writing
\begin{eqnarray}
Z_{SU(2)}(g,\tilde{h})&=&\sum\limits_{n\in\mathbb{Z}}\int\limits_{-\infty}^\infty d\Phi\;\Phi^{2-2\tilde{h}}e^{-2\pi in\Phi-2\pi i\Phi-\frac{g}{2}\Phi^2}\;,
\end{eqnarray}
and then expanding $e^{-2\pi in\Phi}$ as before and following the same steps as above, we would end up with almost the same formula:
\begin{eqnarray}
Z_{SU(2)}(g,\tilde{h})&=&\sum\limits_{n\in\mathbb{Z}}^\infty\cosh\left(2\sqrt{2}\pi n\sqrt{\frac{\partial}{\partial g}}\right)[Z_{SU(2)}^{n=1}(g,\tilde{h})]\;.
\end{eqnarray}
One can see that this is in fact a formula for the full partition function in terms of any sector you like:
\begin{eqnarray}
Z_{SU(2)}(g,\tilde{h})&=&\sum\limits_{n\in\mathbb{Z}}^\infty\cosh\left(2\sqrt{2}\pi n\sqrt{\frac{\partial}{\partial g}}\right)[Z_{SU(2)}^{(n)}(g,\tilde{h})]\;.
\end{eqnarray}
Taking a more careful look at the derivation above we can also see that $\cosh\left(2\sqrt{2}\pi n\sqrt{\frac{\partial}{\partial g}}\right)$ is a sort of shift operator, acting as
\begin{eqnarray}\label{eq:SU2ShiftCosh}
\cosh\left(2\sqrt{2}\pi m\sqrt{\frac{\partial}{\partial g}}\right)[Z_{SU(2)}^{(n)}(g,\tilde{h})]=\frac{1}{2}\left(Z_{SU(2)}^{(n+m)}(g,\tilde{h})+Z_{SU(2)}^{(n-m)}(g,\tilde{h})\right)\;.
\end{eqnarray}
It is now tempting to look for an operator that looks like $\sinh$ rather than $\cosh$ as a complimentary shift operator. This is possible, but we have a problem in defining odd powers of $\sqrt{\frac{\partial}{\partial g}}$. We can circumnavigate this by shifting $\tilde{h}$ by $\frac{1}{2}$ using the shift operator. In this way we have
\begin{eqnarray}\label{eq:SU2ShiftSinh}
2\pi i m e^{-\frac{1}{2}\partial_{\tilde{h}}}\frac{\sinh\left(2\sqrt{2}\pi m\sqrt{\frac{\partial}{\partial g}}\right)}{2\sqrt{2}\pi m\sqrt{\frac{\partial}{\partial g}}}[Z_{SU(2)}^{(n)}(g,\tilde{h})]=\frac{1}{2}\left(Z_{SU(2)}^{(n+m)}(g,\tilde{h})-Z_{SU(2)}^{(n-m)}(g,\tilde{h})\right)\;.\nonumber\\
\;
\end{eqnarray}
One can now see how we can combine these to get any one sector of the transseries from any other sector. For example we have
\begin{eqnarray}\label{eq:ShiftOperator}
\left(\cosh\left(2\sqrt{2}\pi m\sqrt{\frac{\partial}{\partial g}}\right)+2\pi i m e^{-\frac{1}{2}\partial_{\tilde{h}}}\frac{\sinh\left(2\sqrt{2}\pi m\sqrt{\frac{\partial}{\partial g}}\right)}{2\sqrt{2}\pi m\sqrt{\frac{\partial}{\partial g}}}\right)[Z_{SU(2)}^{(n)}(g,\tilde{h})]=Z_{SU(2)}^{(n+m)}(g,\tilde{h})\;.\nonumber\\
\;
\end{eqnarray}
We have found a shift operator for contributions to the transseries from individual topological sectors.

Let's see three different ways \eqref{eq:recursionPert2} comes in handy. First, for the case where $\tilde{h}=0$, let's check we can indeed get all the non-perturbative data from the perturbative part only. In this case we have
\begin{eqnarray}
Z_{SU(2)}^{n=0}(g,0)=\frac{\sqrt{2\pi}}{g^{3/2}}\;.
\end{eqnarray}
Thus, applying \eqref{eq:recursionPert2} we have
\begin{eqnarray}\label{eq:FullFromPertSU2}
Z_{SU(2)}(g,0)&=&\sum\limits_{n\in\mathbb{Z}}^\infty\cosh\left(2\sqrt{2}\pi n\sqrt{\frac{\partial}{\partial g}}\right)\left[\frac{\sqrt{2\pi}}{g^{3/2}}\right]\nonumber\\
&=&\sum\limits_{n\in\mathbb{Z}}^\infty\sum\limits_{j=0}^\infty\frac{\left(8\pi^2n^2\frac{\partial}{\partial g}\right)^{j}}{(2ja)!}\left[\frac{\sqrt{2\pi}}{g^{3/2}}\right]\nonumber\\
&=&\sqrt{2\pi}\sum\limits_{n\in\mathbb{Z}}^\infty\sum\limits_{j=0}^\infty\frac{\left(8\pi^2n^2\right)^{j}\Gamma(-1/2)}{(2j)!\Gamma(-1/2-j)g^{3/2+j}}\nonumber\\
&=&\sum\limits_{n\in\mathbb{Z}}\sqrt{2\pi}e^{-\frac{(2\pi n)^2}{2g}}(g^{-3/2}-4n^2\pi^2 g^{-5/2})  \;.
\end{eqnarray}
Excellent!

Of course, we can also do the above calculation with $\tilde{h}$ not integer. Performing this calculation will give us the perturbative series in all the sectors contributing to the transseries, and also allow us to compute the transseries parameters exactly.

A second thing we can do is use \eqref{eq:recursionPert2} to re-derive the asymptotic perturbation series \eqref{eq:deformedSU2PF} in a different way, starting from $Z_{SU(2)}^{n=0}(g,\tilde{h})$ as defined above. Let's see this. We now have
\begin{eqnarray}
Z_{SU(2)}^{n=0}(g,\tilde{h})=(1+e^{-2\pi i \tilde{h}})2^{\frac{1}{2}-\tilde{h}}\Gamma(3/2-\tilde{h})g^{-3/2+\tilde{h}}\;.
\end{eqnarray}
Applying the part of \eqref{eq:recursionPert2} with $n>0$ (recalling the discussion of Section \ref{sec:deformedSU2WeakZeta} as to why we don't include negative $n$), let's see how we can get the rest of the perturbative expansion in the deformed case. We have
\begin{eqnarray}
&&2\sum\limits_{n=1}^\infty\cosh\left(2\sqrt{2}\pi n\sqrt{\frac{\partial}{\partial g}}\right)[Z_{SU(2)}^{n=0}(g,\tilde{h})]\nonumber\\
&&\;\;\;\;\;=(1+e^{-2\pi i \tilde{h}})2^{\frac{3}{2}-\tilde{h}}\Gamma(3/2-\tilde{h})\sum\limits_{n=1}^\infty\sum\limits_{j=0}^\infty\frac{\left(8\pi^2n^2\frac{\partial}{\partial g}\right)^{j}}{(2j)!}[g^{-3/2+\tilde{h}}]\nonumber\\
&&\;\;\;\;\;=(1+e^{-2\pi i \tilde{h}})2^{\frac{3}{2}-\tilde{h}}\Gamma(3/2-\tilde{h})\sum\limits_{n=1}^\infty\sum\limits_{j=0}^\infty \frac{\left(8\pi^2n^2\right)^{j}\Gamma(-1/2+\tilde{h})}{(2j)!\Gamma(-1/2+\tilde{h}-j)g^{3/2-\tilde{h}+j}}\nonumber\\
&&\;\;\;\;\;=(1+e^{-2\pi i \tilde{h}})2^{\frac{3}{2}-\tilde{h}}\sum\limits_{n=1}^\infty\sum\limits_{j=0}^\infty \frac{\left(-8\pi^2n^2\right)^{j}\Gamma(3/2-\tilde{h}+j)}{(2j)! g^{3/2-\tilde{h}+j}}\;.
\end{eqnarray}
To get to the last line we have used the formula
\begin{eqnarray}
\Gamma(z)\Gamma(1-z)=\frac{\pi}{\sin(\pi z)}\;.
\end{eqnarray}
We now use the definition of the Gamma function to write the above equation in the form
\begin{eqnarray}
&=&(1+e^{-2\pi i \tilde{h}})2^{\frac{3}{2}-\tilde{h}}\sum\limits_{n=1}^\infty\sum\limits_{j=0}^\infty \frac{\left(-8\pi^2n^2\right)^{j}}{(2j)!}\int\limits_0^\infty dt\;e^{-gt}t^{1/2-\tilde{h}+j}\nonumber\\
&=&(1+e^{-2\pi i \tilde{h}})2^{\frac{3}{2}-\tilde{h}}\sum\limits_{n=1}^\infty\int\limits_0^\infty dt\;e^{-gt}t^{1/2-\tilde{h}}\cos\left(2\sqrt{2}\pi n\sqrt{t}\right)\\
&=&(1+e^{-2\pi i \tilde{h}})2^{\frac{1}{2}-\tilde{h}}\sum\limits_{n=1}^\infty\left(\int\limits_0^\infty dt\;e^{-gt}t^{1/2-\tilde{h}}e^{2\sqrt{2}\pi in\sqrt{t}}+\int\limits_0^\infty dt\;e^{-gt}t^{1/2-\tilde{h}}e^{-2\sqrt{2}\pi in\sqrt{t}}\right)\;.\nonumber
\end{eqnarray}
Note that keeping the order of the sums thus far has been very important. Doing the sum over $n$ before the sum over $j$ returns a $\zeta(-2j)$ which is zero except for $a=0$.

We can now expand the $e^{- gt}$ factor in each of these integrals as a Taylor series. The contour of the first (second) integral can then be deformed to be from 0 to positive (negative) imaginary infinity, and then the integrals performed, returning the usual gamma function factor. Finally the sum over $n$ will give us the zeta function factor for each term. Putting this all together we get
\begin{eqnarray}
Z_{SU(2)}^{\mathrm{pert}}(g,\tilde{h})&=&(1+e^{-2\pi i \tilde{h}})2^{1/2-2\tilde{h}}\Gamma(3/2-\tilde{h})g^{-3/2+\tilde{h}} \\
&&\;\;\;\;+e^{\pi i \tilde{h}}\left(1-e^{-4\pi i \tilde{h}}\right)\left(\frac{i}{2\pi }\right)^{3-2\tilde{h}}\nonumber\\
&&\;\;\;\;\;\;\;\;\;\;\;\;\times\sum\limits_{j=0}^{\infty}\frac{(g/(8\pi^2))^j}{j!}\zeta(3+2j-2\tilde{h})\Gamma(3+2j-2\tilde{h})\;.\nonumber
\end{eqnarray}
Comparing this with \eqref{eq:deformedSU2PF} we see we have recovered the asymptotic perturbative expansion we wanted.

A third thing we can do is low-order/low-order resurgence. The process here is almost the same as what we did in \eqref{eq:FullFromPertSU2}, but rather than summing over $n$ we choose a single shift operator \eqref{eq:ShiftOperator}. If we apply this to the contribution from a particular saddle, we will get the contribution from another saddle. If we apply it to only the low-order contributions to a particular saddle, we will be able to find the low-order contributions to a different saddle.

Before moving onto $U(2)$, one final thing to note is that we can also find a formula of the form
\begin{eqnarray}\label{eq:Su2intermsSU2}
Z_{SU(2)}(g,\tilde{h})=\mathcal{F}[Z_{SU(2)}(g,\tilde{h})]\;.
\end{eqnarray}
That is to say, we can find a differential operator that acts on the full transseries and returns itself. This just follows from the fact that we have found shift operators for the contributions to the transseries. As the transseries is just a sum over all the contributions from the different sectors, an operator that shifts the sectors, so long as it acts uniformly on all the sectors, will leave the transseries unchanged. In other words we can write down formula like
\begin{eqnarray}\label{eq:Su2intermsSU22}
\cosh\left(2\sqrt{2}\pi \sqrt{\frac{\partial}{\partial g}}\right)[Z_{SU(2)}(g,\tilde{h})]&=&Z_{SU(2)}(g,\tilde{h})\;.
\end{eqnarray}
This is a formula of the form \eqref{eq:Su2intermsSU2}. We could indeed have written any combination of the operators \eqref{eq:SU2ShiftCosh} and \eqref{eq:SU2ShiftSinh} here. Applications of these formulas are much the same, for example we can compute the non-perturbative data and transseries parameters from the perturbative data by demanding that \eqref{eq:Su2intermsSU22} hold.

\subsubsection{\texorpdfstring{$U(2)$}. case}\label{sec:u2diffeq}

Now we turn to look at various partial differential equations that the $U(2)$ partition function satisfies, that will similarly enable us to make sideways steps in the resurgence triangle. It is first useful to recall \eqref{eq:1topoStartingPoint} such that we can write the $U(2)$ action in terms of the $SU(2)$ action. We write this here again for convenience.
\begin{eqnarray}
Z_{U(2)}(g,h,\theta)&=&\frac{1}{2}\sum\limits_{n_1,n_2\in\mathbb{Z}}\int\limits_{-\infty}^\infty dx\int\limits_{-\infty}^\infty dy\; x^{2-2\tilde{h}}e^{-\pi ix (n_1-n_2)-\pi iy(n_1+n_2)-\frac{g}{4}(x^2+y^2)-i\theta (n_1+n_2) }\nonumber\\
&=&\sqrt{\frac{\pi}{g}}\sum\limits_{n_1,n_2\in\mathbb{Z}}e^{-\frac{\left(\pi(n_1+n_2)\right)^2}{g}-i\theta(n_1+n_2)}\int\limits_{-\infty}^\infty dx\; x^{2-2\tilde{h}}e^{-\pi ix (n_1-n_2)-\frac{g}{2}x^2}\nonumber\\
&=&\sqrt{\frac{\pi}{g}}\sum\limits_{n_1,n_2\in\mathbb{Z}}e^{-\frac{\left(\pi(n_1+n_2)\right)^2}{g}-i\theta(n_1+n_2)}Z_{(n_2-n_1)\pi}\left(\frac{g}{2},\tilde{h}\right)\;.
\end{eqnarray}
We first want to see how we can apply our formulas for the $SU(2)$ partition function to the above formula. We have
\begin{eqnarray}
&&\left(\frac{\partial}{\partial g}-\frac{\pi^2}{g^2}\frac{\partial^2}{\partial\theta^2}+\frac{1}{2g}\right)[Z_{U(2)}(g,h,\theta)]\\
&&\;\;\;\;\;\;\;\;\;\;\;\;\;\;\;\;\;\;\;\;\;\;\;\;\;\;\;\;\;\;\;\;=\sqrt{\frac{\pi}{g}}\sum\limits_{n_1,n_2\in\mathbb{Z}}e^{-\frac{\left(\pi(n_1+n_2)\right)^2}{g}-i\theta(n_1+n_2)}\frac{\partial}{\partial g}\left[Z_{(n_2-n_1)\pi}\left(\frac{g}{2},\tilde{h}\right)\right]\;.\nonumber
\end{eqnarray}
We thus see that by replacing $\frac{\partial}{\partial g}$ by the operator $\left(\frac{\partial}{\partial g}-\frac{\pi^2}{g^2}\frac{\partial^2}{\partial\theta^2}+\frac{1}{2g}\right)$ in \eqref{eq:SU2ShiftCosh} and \eqref{eq:SU2ShiftSinh}, we can apply all the operators in the previous subsection to the $U(2)$ case, which will act by shifting the $(n_1-n_2)$ in the $SU(2)$ factor of the $U(2)$ partition function.

Thus, in order to get any topological sector from any other topological sector in the $U(2)$ case, all we need to find in addition to the $SU(2)$ case is a way of shifting $(n_1+n_2)$. But from the above formula it is easy to see how this is done using shift operators for $\theta$. We have
\begin{eqnarray}\label{eq:U2Shift}
e^{-\frac{\pi^2}{g}-2i\frac{\pi^2}{g}\partial_\theta-i\theta}[Z_{U(2)}(g,h,\theta)]=\sqrt{\frac{\pi}{g}}\sum\limits_{n_1,n_2\in\mathbb{Z}}e^{-\frac{\left(\pi(n_1+n_2+1)\right)^2}{g}-i\theta(n_1+n_2+1)}Z_{(n_2-n_1)\pi}\left(\frac{g}{2},\tilde{h}\right)\;.\nonumber\\
\;
\end{eqnarray}

In summary we can use the shift operators of \eqref{eq:SU2ShiftCosh} and \eqref{eq:SU2ShiftSinh}, with the above substitution for $\frac{\partial}{\partial g}$, to shift the $(n_1-n_2)$ argument in the above $SU(2)$ contribution to the $U(2)$ partition function. We can then use \eqref{eq:U2Shift} to shift the $(n_1+n_2)$ argument in the pre-factor. Combining these we can shift $n_1$ and $n_2$ individually, as much as we like. In this way we can write down operators to get any contribution from a particular sector of the $U(2)$ partition function transseries from any other, and formulas for the partition function in terms of itself etc. Application of these formulas are much the same as discussed for $SU(2)$ in the previous section.

\subsection{\texorpdfstring{$U(N)$}. in terms of \texorpdfstring{$U(1)$}.}

Finally for this section we will briefly look at a partial differential equation for the partition function, but this time the formula will relate the partition function for higher $N$ in terms of $N=1$. Whilst everything we have looked at so far only applies to $N=2$, this applies to higher $N$. There is one setback though, which is that this formula only applies for integer $\tilde{h}$.

Consider the $U(1)$ theory with the strong coupling representation of the partition function given by
\begin{eqnarray}
Z_{U(1)}(g,\tilde{h},\theta)=\sum\limits_{n\in\mathbb{Z}/0}e^{\frac{g(n-\theta/2\pi)^2}{2}}\;.
\end{eqnarray}
We can use this as building block to  build partition functions for higher $N$. Lets work with $\tilde{h}=0$. For $\tilde{h}=1$ things are trivial, and for higher $\tilde{h}$ we would need to replace the differentials by integrals. The formula is for the $U(N)$ partition function in terms of the $N=1$ partition function is then given by 
\small{\begin{widerequation}\label{eq:Nlower1}
Z_{U(N)}=\det\begin{pmatrix}
1 & \left(-\frac{2\pi}{g}\frac{\partial}{\partial \theta_1}+\frac{g\theta_1}{2\pi}\right) & \dots & \left(-\frac{2\pi}{g}\frac{\partial}{\partial \theta_1}+\frac{g\theta_1}{2\pi}\right)^{N-1} \\
1 & \left(-\frac{2\pi}{g}\frac{\partial}{\partial \theta_2}+\frac{g\theta_2}{2\pi}\right) & \dots & \left(-\frac{2\pi}{g}\frac{\partial}{\partial \theta_2}+\frac{g\theta_2}{2\pi}\right)^{N-1} \\
\vdots & \vdots & \vdots & \vdots \\
1 & \left(-\frac{2\pi}{g}\frac{\partial}{\partial \theta_N}+\frac{g\theta_N}{2\pi}\right) & \dots & \left(-\frac{2\pi}{g}\frac{\partial}{\partial \theta_N}+\frac{g\theta_N}{2\pi}\right)^{N-1} \\
\end{pmatrix}^2\left(Z_{U(1)}(g,0,\theta_1)\dots Z_{U(1)}(g,0,\theta_N)\right)\vert_{\theta_1=\dots=\theta_N=\theta} \;.\nonumber\\ \;
\end{widerequation}}
\normalsize
For example, we can write $Z_{U(2)}$ as
\small{\begin{widerequation}
Z_{U(2)}&=&\det\begin{pmatrix}
1 & \left(-\frac{2\pi}{g}\frac{\partial}{\partial \theta_1}+\frac{g\theta_1}{2\pi}\right)\\
1 & \left(-\frac{2\pi}{g}\frac{\partial}{\partial \theta_2}+\frac{g\theta_2}{2\pi}\right)
\end{pmatrix}^2\left(Z_{U(1)}(g,0,\theta_1) Z_{U(1)}(g,0,\theta_2)\right)\vert_{\theta_1=\theta_2=\theta}\nonumber\\
&=&\left(\left(-\frac{2\pi}{g}\frac{\partial}{\partial \theta_2}+\frac{g\theta_2}{2\pi}\right)-\left(-\frac{2\pi}{g}\frac{\partial}{\partial \theta_1}+\frac{g\theta_1}{2\pi}\right)\right)^2\left(\sum\limits_{n_1,n_2\in\mathbb{Z}}e^{-\frac{g}{2}((n_1-\theta_1/2\pi)^2+(n_2-\theta_2/2\pi)^2)}\right)\vert_{\theta_1=\theta_2=\theta}\nonumber\\
&=&\sum\limits_{n_1,n_2\in\mathbb{Z}}(n_1-n_2)^2e^{-\frac{g}{2}((n_1-\theta/2\pi)^2+(n_2-\theta/2\pi)^2)}\;.
\end{widerequation}}
\normalsize
We can see that this kind of formula can't be used for non-integer $\tilde{h}$, without perhaps utilizing fractional calculus or something similar. However for the integer $\tilde{h}$ case, i.e. once the deformation has been returned to zero, it again gives us a structure by which we can move sideways in the resurgence triangle.

This sideways step can be done in almost exactly the same way as in the factorisation case. In particular, we have something very similar to \eqref{eq:dunneUnsalCalc}, but with a different differential operator acting on a different ansatz. Thus, in this way we can perform a sideways step using \eqref{eq:Nlower1}. Of course, the most obvious further application of \eqref{eq:Nlower1} is calculating the full partition function directly from the $N=1$ theory, for higher $N$ in the undeformed theories.

\section{Conclusion}\label{sec:conclusion}

In this paper we have analysed the partition function of 2d YM in order to explore its resurgence structure. For the undeformed theory both a weak coupling and strong coupling transseries representation of the partition function for general gauge group have been known for some time. In this case the series in each sector is truncating, so Borel-\`Ecalle resummation cannot be applied to determine contributions from different sectors to the transseries from other contributions. We have explained that this is due to the topology of the contributions.

We have been able to find a deformation of the UV theory where the partition function is still calculable, and the contributions in each sector are no longer truncating but asymptotic divergent. In this case, this is due to appearance of new saddles. The deformation results in an effective theory describing 2d YM on a non-integer genus surface. There are now multiple saddles within each topological sector, and within a topological sector Borel-\`Ecalle resummation can be applied to determine contributions from different sectors to the transseries from other contributions.

Moreover, for certain values of the deformation parameter, these new saddles still exist, but the perturbative series associated to them truncate. The values of the deformation parameter are the values where the effective genus is an integer, or perhaps half-integer, imitating a Wilson loop insertion or boundary. These are Cheshire cat points of the theory. There are still multiple saddles within a topological sector, but for a precise value of the deformation parameter there are substantial cancellations within the transseries rendering the perturbative series associated with the saddles no longer asymptotic divergent. There is a nice geometric reason behind this which will be explored in more detail in \cite{Fujimori:2022pld}.

A further phenomena we have been able to study relates to what happens when we have multiple saddles in the transseries with identical action. In our case we have seen two examples, the $SU(2)$ and $U(2)$ gauge groups, where we have infinite saddles with equal action all contributing to the transseries. With prior knowledge of the saddles points, calculating the transseries via saddle decomposition, one can distinguish them in the transseries. However if one is just handed the transseries, calculated via some other means, one can't distinguish them. This leads to phenomena where from one term in the transseries we can calculate the contributions in all the other sectors, but not vice-versa. The reason is that the one contribution contains contributions from saddles in every topological sector, but the other contributions are from a single saddle in a single topological sector. This is another example of an unusual phenomena in the transseries with a topological underlying reason, hidden if one doesn't have knowledge of the saddles and their topology.

An additional calculation we have managed to achieve in this work, which is not normally possible, is to determine the transseries parameters exactly. Whilst this is normally not possible, having access to both the strong and weak coupling perturbative data makes it possible in this case, by demanding that the strong and weak coupling transseries are describing the same object.

In the case of 2d YM we have been able to extend the observations of \cite{Dorigoni:2019kux} in finding additional structures that allow us to calculate contributions to transseries from different topological sectors from the perturbative sector. Such structures allow us to calculate the whole transseries from the perturbative part alone, circumventing the standard caveat that we can only calculate terms in the transseries in the same topological sector as the perturbative data.

This work provides some evidence (but by no means conclusive evidence!) that the strong version of the resurgence program may be right, if Cheshire cat points and topology are taken into account. This is done by adding 2d YM to a growing list of theories thought to have been counter examples to the strong version of the resurgence program, but are in fact examples of it.

Regarding 2d YM, one important follow-up which will be presented in \cite{Fujimori:2022pld} is the Picard-Lefschetz decomposition of the partition function. This will give us a clear explanation for the Stokes phenomena in terms of thimble decomposition of the path integral. We can perform this decomposition exactly once we have integrated out certain modes that only contribute Gaussian terms to the action, even in the deformed case, which is interesting in itself. But in this case, it will further allow us to provide a geometrical reason for why the series truncate at Cheshire cat points, shining more light on the importance of imaginary quantum contributions to saddles in transseries.

Another interesting direction for future research is whether techniques can be found that would allow us to calculate the perturbative series in the deformed case for higher $N$. This would allow us to test more carefully whether the Cheshire cat resurgence procedure can be applied in these cases. A further direction would be to see if we can apply the procedure to different observables in the theory.

Many other questions remain when we consider theories other than 2d YM. The most notable counter example to the strong version of the resurgence program that remains is that of 4-dimensional $\mathcal{N}=2$ supersymmetric Yang-Mills. Here a Cheshire cat resurgence analysis is technically very challenging, and has yet to be performed. There are also counter examples studied in the works of \cite{DiPietro:2021yxb,Marino:2021dzn,Marino:2022ykm,Bajnok:2021zjm,Reis:2022tni} in the context of integrable theories. Beyond this, there is the general question of whether Cheshire cat points are responsible for all theories where at first sight it appears the strong version of the resurgence program cannot apply, and there is no topological reason for the truncation. Finally, related to this, is the question of whether there is always some addition structure that allows us to move sideways in the resurgence triangle, thus (in cases where the strong version of the program does apply) allowing us to compute the full transseries from the perturbative part alone.

\acknowledgments

The authors are particularly grateful to Daniele Dorigoni for making P.G. aware of 2d YM being a potentially fruitful ground for the study of Cheshire cat resurgence, and for early collaboration on this project. This work is supported by JSPS KAKENHI Grant Numbers JP20F20787, JP21K03558.

\appendix

\section{\texorpdfstring{$h\geq 1$}. }\label{sec:hGeq1}

In the bulk of this work we have specialised to $h<1$ for simplicity. Let us now briefly comment on what happens when $h\geq 1$. Here we need to be careful about the locations where $\mathrm{det}_\mathbf{k}\left(\mathrm{ad}(\Phi^\mathbf{t})\right)=0$. First let us see one way of removing these points from the integrals. Focusing on $SU(N)$ we have
\begin{eqnarray}
Z_{SU(N)}(g,h,\theta)&=&\prod\limits_{i=1}^{N-1}\sum\limits_{n_i\in \mathbb{Z}}\int^{'} d\Phi^i e^{-2\pi i n_i\Phi^i -\frac{g}{2}\Phi^i\Phi^i}\mathrm{det}_\mathbf{k}\left(\mathrm{ad}(\Phi^\mathbf{t})\right)^{\chi(\Sigma_h)/2}\nonumber\\
&=&\prod\limits_{i=1}^{N-1}\sum\limits_{m_i\in \mathbb{Z}}\int^{'} d\Phi^i \delta(\Phi^i-m_i) e^{-\frac{g}{2}\Phi^i\Phi^i} \mathrm{det}_\mathbf{k}\left(\mathrm{ad}(\Phi^\mathbf{t})\right)^{\chi(\Sigma_h)/2}\nonumber\\
&=&\prod\limits_{i=1}^{N-1}\left(\sum\limits_{m_i\in \mathbb{Z}}\right)^{'}\int d\Phi^i \delta(\Phi^i-m_i) e^{-\frac{g}{2}\Phi^i\Phi^i} \mathrm{det}_\mathbf{k}\left(\mathrm{ad}(\Phi^\mathbf{t})\right)^{\chi(\Sigma_h)/2}\;.
\end{eqnarray}
To get to the final line here we switched how we remove the singular points by removing them from the sum rather than the integral. We can now write the sum as
\begin{eqnarray}
\prod\limits_{i=1}^{N-1}\left(\sum\limits_{m_i\in \mathbb{Z}}\right)^{'}\delta(\Phi^i-m_i)&=&\prod\limits_{i=1}^{N-1}\sum\limits_{m_i\in \mathbb{Z}}\delta(\Phi^i-m_i)-\prod\limits_{i=1}^{N-1}\sum\limits_{m_i\in m_\mathrm{cr}}\delta(\Phi^i-m_i)\nonumber\\
\mathrm{for}\;\;\;m_\mathrm{cr}&=&\lbrace m_i:\mathrm{det}_\mathbf{k}\left(\mathrm{ad}(m_\mathbf{t})\right)=0\rbrace\;.
\end{eqnarray}
Reversing the logic, we can write the first of these sums as a sum over exponentials again. We thus can write the partition function as
\begin{flalign}
&Z_{SU(N)}(g,h,\theta)\\
&\;\;\;\;\;\;\;=\int d\Phi^i e^{-\frac{g}{2}\Phi^i\Phi^i} \mathrm{det}_\mathbf{k}\left(\mathrm{ad}(\Phi^\mathbf{t})\right)^{\chi(\Sigma_h)/2}\left(\prod\limits_{i=1}^{N-1}\sum\limits_{n_i\in \mathbb{Z}}e^{-2\pi i n_i\Phi^i}-\prod\limits_{i=1}^{N-1}\sum\limits_{m_i\in m_\mathrm{cr}}\delta(\Phi^i-m_i)\right)\;.\nonumber
\end{flalign}
Importantly we have been able to remove the $^{'}$ from the integral. For $h<1$ the extra delta functions make no difference, but we need to consider it when $h\geq1$.

Let us demonstrate how we deal with this for the $SU(2)$ case for simplicity. In this case, with $h\geq 1$, we have
\begin{eqnarray}
Z_{SU(2)}(g,h)&=&\int\limits_{-\infty}^{\infty}d\Phi\;\Phi^{2-2h}e^{-\frac{g}{2}\Phi^2}\left(\sum\limits_{n\in\mathbb{Z}}\;e^{-2\pi i n \Phi}-\delta(\Phi)\right)\;.
\end{eqnarray}
Differentiating this $h-1$ times with respect to $g$ we have
\begin{eqnarray}
\frac{\partial^{h-1}}{\partial g^{h-1}}Z_{SU(2)}(g,h)&=&\left(\frac{-1}{2}\right)^{h-1}\int\limits_{-\infty}^{\infty}d\Phi\;e^{-\frac{g}{2}\Phi^2}\left(\sum\limits_{n\in\mathbb{Z}}\;e^{-2\pi i n \Phi}-\delta(\Phi)\right)\nonumber\\
&=&\left(\frac{-1}{2}\right)^{h-1}\left(\sum\limits_{n\in\mathbb{Z}}\frac{\sqrt{2\pi}}{\sqrt{g}}e^{-\frac{(2\pi n)^2}{2g}}-1\right)\;.
\end{eqnarray}
We can now integrate $h-1$ times with respect to $g$ to get the transseries for the partition function. We thus get an extra $h$ terms in the perturbative series for $h\geq 1$ that are not present in the $h<1$ case. This will make no difference to the resurgence structure of the partition function.

\section{Deriving transseries via Zagier's method}\label{sec:ZagierMethod}

In this appendix we will apply a method by Zagier, outlined in \cite{ZagierAppendixTM} in chapter 6 of \cite{Zeidler2006}, to derive the perturbative part of the transseries in the $N=2$ cases. We include it as the method is very efficient. However it doesn't apply to higher $N$, and in this case somewhat hides some of the features of the perturbative series.

\subsection{The method}

Zagier's method works as follows: Suppose we are looking for an asymptotic approximation to a function $g(t)$ of the following form:
\begin{eqnarray}\label{eq:zagierfunction}
g(t)=f(t)+f(2t)+f(3t)+\dots\;.
\end{eqnarray}
Suppose we also have access to a series approximation of $f(t)$ of the form
\begin{eqnarray}
f(t)\sim \sum\limits_{\lambda> -1}b_\lambda t^\lambda
\;.
\end{eqnarray}
Here the sum is over whatever the exponents of $t$ happen to be, real or complex, and not necessarily integer spaced, so long as the real part of all the $\lambda$ are greater than $-1$. From substituting this series expansion into the sum \eqref{eq:zagierfunction} and swapping the sums, we might approximate $g(z)$ as
\begin{eqnarray}
g(t)\sim \sum\limits_{\lambda> -1}b_\lambda \zeta(-\lambda)(t)^\lambda\;.
\end{eqnarray}
But, you may also think to approximate $g(t)$ for small $t$ as a series approximation to the integral
\begin{eqnarray}
I_f&=&\int\limits_0^\infty dt\;f(t)\;,\nonumber\\
g(t)&\sim&\frac{I_f}{t}\;.
\end{eqnarray}
The correct answer is to add them:
\begin{eqnarray}
g(t)\sim\frac{I_f}{t}+\sum\limits_{\lambda> -1}b_\lambda \zeta(-\lambda)(t)^\lambda\;\;\;\;(t\rightarrow0)\;.
\end{eqnarray}
For a formal proof of this see \cite{ZagierAppendixTM}. (Note we can adjust this formula to include terms with real part of $\lambda$ equal to $-1$, but we won't need this so we don't include it here.) Let us now apply this in the weak and strong cases for $SU(2)$.

\subsection{Weak coupling}

For the weak coupling case, the quickest way to apply Zagier's method is to use \eqref{eq:DeltaFunctionIdentity1} in \eqref{eq:SUNPartitionFunctionIntegralRep}, to get
\begin{eqnarray}\label{eq:SU2ft}
Z_{SU(2)}(g,\tilde{h})&=&\sum\limits_{m\in\mathbb{Z}}m^{2-2\tilde{h}}e^{-g m^2/2}\nonumber\\
&=&(1+e^{-2\pi i \tilde{h}}) \sum\limits_{m=1}^{\infty}m^{2-2\tilde{h}}e^{-g m^2/2}\nonumber \\
&=&\frac{(1+e^{-2\pi i \tilde{h}})}{g^{1-\tilde{h}}} \sum\limits_{m=1}^{\infty}(\sqrt{g}m)^{2-2\tilde{h}}e^{- (\sqrt{g}m)^2/2}\;.
\end{eqnarray}
From the last line we see we have
\begin{eqnarray}
f(t)=t^{2-2\tilde{h}}e^{-t^2/2}\;,\;\;\;t=\sqrt{g}\;.
\end{eqnarray}
From this expression we can calculate
\begin{eqnarray}
I_f&=&2^{\frac{1}{2}-\tilde{h}}\Gamma(3/2-\tilde{h})\;,\nonumber\\
f(t)&=&t^{2-2\tilde{h}}\sum\limits_{n=0}^\infty \frac{(-t^2/2)^n}{n!}\;.
\end{eqnarray}
Thus we have
\begin{eqnarray}
Z_{SU(2)}(g,\tilde{h})&=&\frac{(1+e^{-2\pi i \tilde{h}})}{g^{1-\tilde{h}}} \left(\frac{2^{\frac{1}{2}-\tilde{h}}\Gamma(3/2-\tilde{h})}{\sqrt{g}}+\sum\limits_{n=0}^\infty\frac{g^{1+n-\tilde{h}}}{2^n n!}\zeta(-2-2n+2\tilde{h})\right)\;.\nonumber\\
\end{eqnarray}
Using the standard reflection formula for the zeta function we arrive at the following perturbative expansion for the partition function:
\begin{eqnarray}
Z_{SU(2)}^{\mathrm{pert}}(g,\tilde{h})&=&(1+e^{-2\pi i \tilde{h}})2^{1/2-2\tilde{h}}\Gamma(3/2-\tilde{h})g^{-3/2+\tilde{h}} \\
&&\;\;\;\;+\left(e^{\pi i \tilde{h}}\left(1-e^{-4\pi i \tilde{h}}\right)+e^{-\pi i \tilde{h}}\left(1-e^{4\pi i \tilde{h}}\right)\right)\left(\frac{i}{2\pi }\right)^{3-2\tilde{h}}\nonumber\\
&&\;\;\;\;\;\;\;\;\;\;\;\;\times\sum\limits_{a=0}^{\infty}\frac{(g/(8\pi^2))^a}{a!}\zeta(3+2a-2\tilde{h})\Gamma(3+2a-2\tilde{h})\;.\nonumber
\end{eqnarray}
This is not what we found in \eqref{eq:deformedSU2PF}. But the reason is simple. \eqref{eq:deformedSU2PF} was the result of including all the perturbative saddles with $n\geq 0$. The above is the result of including the perturbative saddles for negative $n$ as well, which we shouldn't as they have transseries parameter 0. We see again that knowledge of the saddles and their intersection numbers is important. The above method is very efficient, but the result is an infinite sum of different saddle contributions with the wrong transseries parameters for our purposes. Of course, a resurgence analysis of the above will produce all the correct non-perturbative data in all sectors, after which all there is to do is fix the transseries parameters.

\subsection{Strong coupling}

For the strong case things are slightly more involved. Starting from the integral representation of the partition function given in \eqref{eq:SUNPartitionFunctionIntegralRep}, we can rearrange our expression so it is in the form \eqref{eq:zagierfunction}:
\begin{eqnarray}
Z_{SU(2)}(g,\tilde{h})&=&\sum\limits_{n\in\mathbb{Z}} \int\limits_{-\infty}^{\infty}d\Phi\;\Phi^{2-2\tilde{h}}e^{2\pi i n \Phi-\frac{g}{2}\Phi^2}\nonumber\\
&=&g^{-3/2+\tilde{h}}\sum\limits_{n\in\mathbb{Z}} \int\limits_{-\infty}^{\infty}d\Phi\;\Phi^{2-2\tilde{h}}e^{2\pi i \frac{n}{\sqrt{g}} \Phi-\frac{1}{2}\Phi^2}\\
&=&(1+e^{-2\pi i \tilde{h}})2^{\frac{1}{2}-\tilde{h}}\Gamma(3/2-\tilde{h})g^{-3/2+\tilde{h}}\nonumber\\
&&\;\;\;\;\;\;\;\;\;\;\;+g^{-3/2+\tilde{h}}\sum\limits_{n=1}^{\infty}\int\limits_{-\infty}^{\infty}d\Phi\;\Phi^{2-2\tilde{h}}\left(e^{2\pi i \frac{n}{\sqrt{g}} \Phi}+e^{-2\pi i \frac{n}{\sqrt{g}} \Phi}\right)e^{-\frac{1}{2}\Phi^2}\;\;\nonumber.
\end{eqnarray}
We can now apply Zagier's method to the sum in the final line of the above. This time we have
\begin{eqnarray}
f(t)=\int\limits_{-\infty}^{\infty}d\Phi\;\Phi^{2-2\tilde{h}}\left(e^{2\pi i t \Phi}+e^{-2\pi i t \Phi}\right)e^{-\frac{1}{2}\Phi^2}\;,\;\;\;t=\frac{1}{\sqrt{g}}\;.
\end{eqnarray}
The calculation of $I_f$ goes as follows:
\begin{eqnarray}
I_f&=&\int\limits_0^\infty dt \int\limits_{-\infty}^{\infty}d\Phi\;\Phi^{2-2\tilde{h}}\left(e^{2\pi i t \Phi}+e^{-2\pi i t \Phi}\right)e^{-\frac{1}{2}\Phi^2}\nonumber\\
&=&\int\limits_{-\infty}^\infty dt \int\limits_{-\infty}^{\infty}d\Phi\;\Phi^{2-2\tilde{h}}e^{2\pi i t \Phi-\frac{1}{2}\Phi^2}\nonumber\\
&=&\int\limits_{-\infty}^{\infty}d\Phi\;\Phi^{2-2\tilde{h}}\delta(\Phi)e^{-\frac{1}{2}\Phi^2}\nonumber\\
&=&0\;.
\end{eqnarray}
The calculation of a series expansion for f(t) goes as follows; first we rewrite the integral using $\Phi\rightarrow-\Phi$ for the second term as
\begin{eqnarray}
f(t)&=&\int\limits_{-\infty}^{\infty}d\Phi\;\Phi^{2-2\tilde{h}}\left(e^{2\pi i t \Phi}+e^{-2\pi i t \Phi}\right)e^{-\frac{1}{2}\Phi^2}\nonumber\\
&=&(1+e^{-2\pi i \tilde{h}})\int\limits_{-\infty}^{\infty}d\Phi\;\Phi^{2-2\tilde{h}}e^{2\pi i t \Phi-\frac{1}{2}\Phi^2}\;.
\end{eqnarray}
Then we Taylor expand $e^{2\pi i t \Phi}$ and perform the integral:
\begin{eqnarray}
f(t)&=&(1+e^{-2\pi i \tilde{h}})\sum\limits_{n=0}^\infty \int\limits_{-\infty}^{\infty}d\Phi\;\Phi^{2-2\tilde{h}}\frac{(2\pi i t \Phi)^n}{n!}e^{-\frac{1}{2}\Phi^2}\nonumber\\
&=&2^{\frac{1}{2}-\tilde{h}}(1+e^{-2\pi i \tilde{h}})^2\sum\limits_{n=0}^\infty (2\sqrt{2}\pi i t)^{n}\frac{\Gamma(3/2-\tilde{h}+n/2)}{n!}\;.
\end{eqnarray}
Thus we have
\begin{eqnarray}
g(t)&=&2^{\frac{1}{2}-\tilde{h}}(1+e^{-2\pi i \tilde{h}})^2\sum\limits_{n=0}^\infty (2\sqrt{2}\pi i t)^{n}\frac{\Gamma(3/2-\tilde{h}+n/2)}{n!}\zeta(-n)\nonumber\\
&=&-2^{\frac{1}{2}-\tilde{h}}(1+e^{-2\pi i \tilde{h}})^2\frac{1}{2\pi it}\sum\limits_{n=0}^\infty \left(-8\pi^2 t \right)^{n}\frac{\Gamma(1+n-\tilde{h})B_{2n}}{(2n)!}\;.
\end{eqnarray}
To get to the last line we have used that
\begin{eqnarray}
\zeta(0)=-\frac{1}{2}\;,\;\;\;\zeta(-2n)=0\;,\;\;\;\zeta(-2n+1)=-\frac{B_{2n}}{2n}\;.
\end{eqnarray}
Thus we arrive at an asymptotic strong coupling perturbative expansion
\begin{eqnarray}\label{eq:strongdeformedPF}
Z_{SU(2)}(g,\tilde{h})&=&\frac{(1+e^{-2\pi i\tilde{h}})}{2}2^{1/2-\tilde{h}}\Gamma(3/2-\tilde{h})g^{-3/2+\tilde{h}}\\
&&\;\;\;\;\;+(1-e^{-2\pi i\tilde{h}})\frac{1}{2^{1+\tilde{h}}\pi i}\sum\limits_{a=0}^\infty\frac{B_{2a}(-8\pi^2)^{a}}{(2a)!}\Gamma(1+a-\tilde{h})g^{\tilde{h}-1-a}\nonumber\;.
\end{eqnarray}
This is exactly what we found in \eqref{eq:strongseries}.



\bibliography{bibliography.bib}



\end{document}